\documentclass[]{jfm}
\usepackage{graphicx}
\usepackage{newtxtext}
\usepackage{newtxmath}
\usepackage{natbib}
\usepackage{hyperref}
\hypersetup{
	colorlinks = true,
	urlcolor   = blue,
	citecolor  = blue,
	linkcolor  = blue,
	filecolor  = blue,
}

\newcommand{\RomanNumeralCaps}[1]
\linenumbers

\usepackage{float}
\usepackage{caption}
\usepackage{subcaption}
\usepackage{amssymb}
\usepackage{booktabs}
\usepackage{multirow}
\usepackage[font=normalsize,justification=justified]{caption}
\title{A self propelling clapping body}
\author{Suyog Mahulkar\aff{1}
	\corresp{\email{suyogm@iisc.ac.in}},
	Jaywant Arakeri\aff{1}}

\affiliation{\aff{1}Indian Institute of Science, Banglore, India}

\begin{document}
\maketitle
\begin{abstract}
		We report an experimental study of the motion of a clapping body consisting of two flat plates pivoted at the leading edge by a torsion spring. Clapping motion and forward propulsion of the body are initiated by the sudden release of the plates, initially held apart at an angle $2\theta_0$. Results are presented for the clapping and forward motions, and for the wake flow field for 24 cases, where depth to length ratio ($d^*$ = 1.5,1 and 0.5), spring stiffness per unit depth ($Kt$), body mass ($m_b$), and initial separation angle  ($2\theta_0$ = 45 and 60 deg) are varied.  The body  motion consists of a rapid forward acceleration to a maximum velocity followed by a slow retardation to nearly zero velocity. Whereas the acceleration phase involves a complex interaction between plate and fluid motions, the retardation phase is simply fluid dynamic drag slowing the body. The wake consists of either a single axis-switching elliptical vortex loop (for $d^*$ = 1 and 1.5) or multiple vortex loops (for $d^*$ = 0.5). The  body motion is nearly independent of $d^*$ and most affected by variation in $\theta_0$ and Kt. Using conservation of linear momentum and conversion of spring strain energy into kinetic energy in the fluid and body, we obtain a relation for the translation velocity of the body in terms of the various parameters. Approximately 80\% of the initial stored energy is transferred to the fluid, only 20\% to the body. The experimentally obtained $COT$ lies between 2 and 8.             
\end{abstract}

\begin{section}{Introduction}
	
	  In aquatic habitats, the two common types of propulsion are through the flapping of fins or tails and through pulsed jets. Whereas the former has been extensively studied, pulse jet propulsion has received far less attention. A recent review gives (Gemmell et al.\cite{Gemmel_Jettting}) an overview of the different types of pulsed propulsion found among marine invertebrates and a detailed comparative analysis of their swimming performances. Jellyfishes and squids use the pulse jet propulsion mechanism. In both creatures, contraction of the body cavity produces a jet. Most of the studies on pulsed propulsion have looked at the structure of the wake (for example,  Daribi et al.\cite{Dabiri05}, Daribi et al.\cite{Dabiri06}, Bartol et al.\cite{Bartol2D09}). Bartol et al.\cite{Bartol2D09} have showed the existence of two types of jetting patterns behind a  squid {\it Lolliguncula brevis}: the first one consists of the isolated vortex ring, and the second one consists of the vortex ring followed by a trailing jet. In a later study  Bartol et al.\cite{Bartol3D16}  studied the interaction between fin motion and short pulse jets in the same species. Pulsed jet propulsion has also been used in several aquatic robots, such as Robosqid\cite{robosquid10}, CALAMAR-E\cite{Krieg08}, and flexible robots with eight radial arms\cite{Bujard21}.

   	 Clapping motion provides an alternative way to produce pulse jets, though most studies have been in the context of the flight of butterflies (Brodsky\cite{Brodsky91} and Johansson et al.\cite{Johanassan21}). D. Kim et al.\cite{Kim13} have made a detailed study of the flow created and the thrust generated due to the clapping of two plates in quiescent fluid. In a comparative analysis between flapping and clapping, Martin et al.\cite{Martin17} show that clapping produces a higher thrust, whereas the flapping form of propulsion is more efficient. \par
	
   	The relation between the pulsed jet and the thrust is quite clear when the body is stationary in quiescent fluid. In pulsed jet propulsion systems, however, the body accelerates and moves forward due to the inherently unsteady thrust. The fluid flow affects the body motion, and the body motion, in turn, affects the fluid flow. A computational study of medusae swimming (S. Alben et al.\cite{Alben13}) also shows that the kinematics of the bell radius plays a crucial role in efficient swimming. The nature of the body motion and the effect of the body motion on the pulsed jet itself are important fundamental questions that need to be answered to better understand the propulsion of creatures such as jellyfish and squids. In the present study, we use a simple model of pulsed propulsion to study these issues. We have a body that rapidly moves forward due to the action of a pulsed jet. The self-propelling body, consists of two rigid thin plates, pivoted at the front and held together by a torsion-like spring. In the natural state, both plates touch each other; the interplate angle is zero degrees. Initially, the plates are held at some angle, in the range of 45-60 degrees, in quiescent water. The release of the holding force brings the plates rapidly together, expelling the water between the plates, creating a jet, and propelling the body forward. Figure \hyperref[fig:ProbSchematic]{1} shows a schematic of the setup. Our interest is to study the kinematics of the body motion and the flow and the interaction of the two. \par
	
	In this paper, \S \hyperref[sec:ExptSetup]{2} describes the apparatus design. A brief discussion of the overall analysis \S\hyperref[sec:briefDiscussion]{3} is followed by a detailed of body kinematics \S \hyperref[sec:BodyKinematics]{3.1}, a quantitative description of the wake from 2D-PIV data \S\hyperref[sec:WakeDynamics]{3.2}, and wake energetics \S\hyperref[sec:WakeEnergetics]{3.3}. The concluding remarks are presented in \S\hyperref[sec:Conclusion]{4}.

\end{section}

\begin{section}{Experimental setup}
	\label{sec:ExptSetup}
     The requirements that need to be satisfied for the body to move in a horizontal direction subsequent to the clapping motion are that it has to be neutrally buoyant, that the center of mass COM and buoyancy COB coincide, and that the thrust force passes through the center of mass. In this neutrally buoyant configuration, the total vertical force acting on the body is zero.\par
	%%%%%%%%%%%%%%%%%%%%%%%%%%%%%%%%%%%%%%%%%%%%%%%%%%%%%%%%%%%%%%%%%%%%%%%%%%%%%%%%%%%%%%% 
	\begin{gather}\label{eq:NeutralBuoyancy}
	F_{net}=F_{Buoyancy}-F_{mass} = 0 
	\end{gather} 
	%%%%%%%%%%%%%%%%%%%%%%%%%%%%%%%%%%%%%%%%%%%%%%%%%%%%%%%%%%%%%%%%%%%%%%%%%%%%%%%%%%%%%%%
	
	These requirements required careful design and fabrication. The main components of the body include Balsa wood (SG 0.22 gm/cc), hard plastic (SG 0.89 gm/cc), ‘Bond-Tite’ glue (SG 1.05 gm/cc), fishing thread (SG 1.22 gm/cc), and steel plate (SG 8.09 gm/cc). The balsa wood mainly provides the buoyant force to balance the weight of the steel plates. \par
	%%%%%%%%%%%%%%%%%%%%%%%%%%%%%%%%%%%%%%%%%%%%%%%%%%%%%%%%%%%%%%%%%%%%%%%%%%%%%%%%%%%%%%%
	\begin{figure}
		\centering\
		\begin{subfigure}[b]{0.4\textwidth}
			\includegraphics[width=\textwidth]{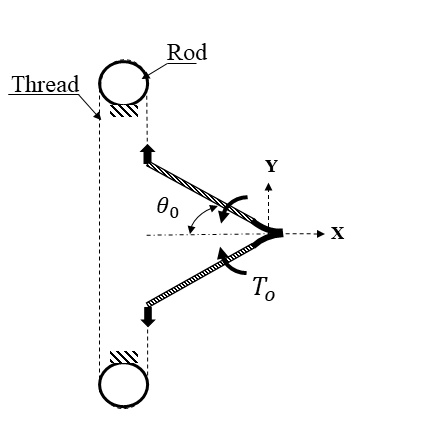}
			\caption{}
			\label{fig:DynProblem_to}
		\end{subfigure}\hspace{5mm}
		\begin{subfigure}[b]{0.4\textwidth}
			\includegraphics[width=\textwidth]{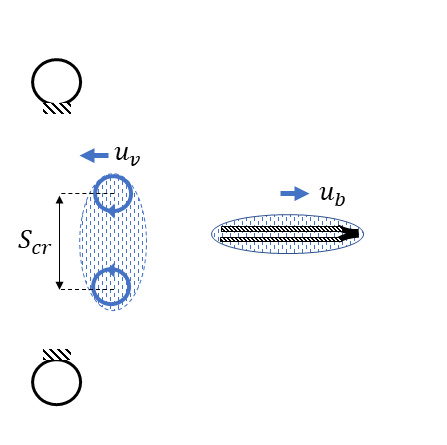}
			\caption{}
			\label{fig:DynProblem_tf}
		\end{subfigure}
	\caption{(a) The clapping body at rest with an initial torque $T_o$ and inter-clap angle $2\theta_o$, held by the fishing thread (0.25mm in diameter and shown by the black dashed line) that loops over two stationary solid rods. The plane Z = 0 is located at the mid-depth of the body. (b) The clapping body moves with a velocity $u_b$  along the X-direction subsequent to cutting of the thread. The vortex pair, with a core separation = $S_{cr}$, moves with a velocity $u_v$. The blue color indicates water masses traveling with the body and the wake vortex pair.}\label{fig:ProbSchematic}	
	\end{figure}

    Figure \hyperref[fig:IsoApparatus_Dyn]{2} shows different views of one of the clapping bodies and its different components. Each arm of the clapping body consists of a steel plate of length $L_{Steel}$ on which a balsa piece having an aerofoil shape is attached at the front end. The steel plate provides the necessary spring action. A rectangular sheet of hard plastic is attached at the back end of the steel plate; another piece of balsa is glued onto the plastic sheet at the back. A canopy made of a thin plastic sheet (0.17mm thickness) is attached at the back to change the body's mass $m_b$. It envelopes the back balsa piece and the rigid plastic plate. A detailed analysis of the distributions of weight and buoyancy is required to ensure the requirements of neutral buoyancy and coincidence of COM and COB to arrive at the final body configuration. The clapping body is created by gluing two identical arms over the front end with 'Bond-Tite'. See figure \hyperref[fig:TV_superimposed_Lines]{2e}. \par
     
    Two of the parameters that change the clapping body configuration are spring stiffness and body mass. Bending of the steel plates over the length $L_e$ gives the spring action. Plates of two different thicknesses (0.14 mm and 0.10 mm) and lengths (60 mm and 55 mm)  were used to make bodies with two stiffnesses per unit depth denoted by $Kt_1$ and $Kt_2$. The Euler-Bernoulli beam theory was used to determine $L_e$ such that the steel plates were still in the elastic limit for an angular deformation of 30 degrees. The canopy was used to increase the mass from the base (no canopy) value. The extra body mass is mainly due to the water that occupies the space between the streamlined plastic canopy and rigid plastic with it of length denoted by $L_{Plastic}$. The ratio of the body mass with canopy to the body mass without canopy is denoted by $M^*$. The construction and shape of the canopy are such that slightly different values of $M^*$ and $Kt$ were obtained for nominally the same configuration. See figures \hyperref[fig:ProbDrawing_Dyn]{2a, 2b, 2c}. The third parameter that was changed was the non-dimensional depth of the body $d^*$ ($=d/L$); bodies with three values of $d^*$ = 1.5, 1.0 and 0.5, were used in the experiments. Table\hyperref[tab:DesignData]{1} lists all relevant data corresponding to the bodies used in the present study. \par
    
	%%%%%%%%%%%%%%%%%%%%%%%%%%%%%%%%%%%%%%%%%%%%%%%%%%%%%%%%%%%%%%%%%%%%%%%%%%%%%%%%%%%%%%%
	\begin{figure}
		\centering\
		\begin{subfigure}[b]{0.7\textwidth}
			\includegraphics[width=\textwidth]{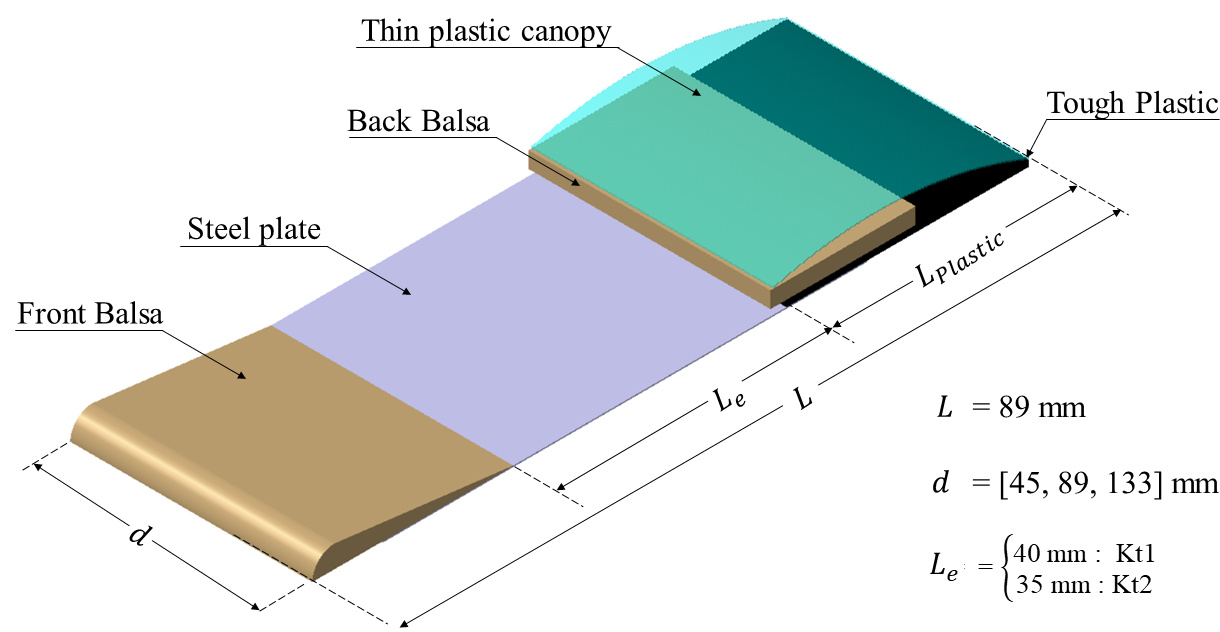}
			\caption{}
			\label{fig:IsoApparatus_Dyn}
		\end{subfigure}
		\begin{subfigure}[b]{0.8\textwidth}
			\includegraphics[width=\textwidth]{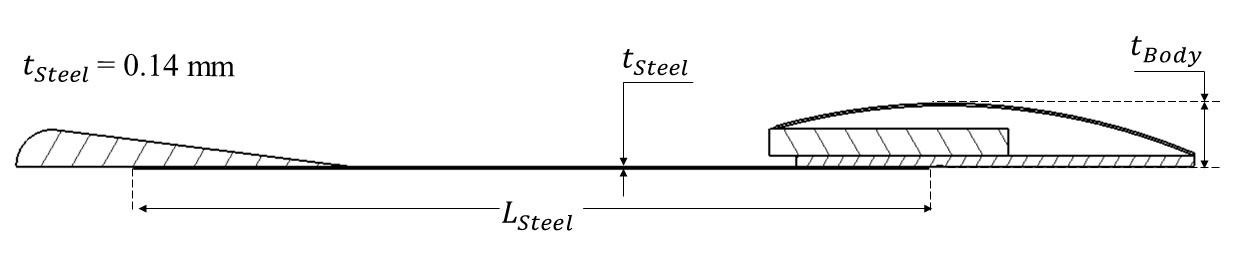}
			\caption{}
			\label{fig:2DApparatus_Kt1_M150_Dyn}
		\end{subfigure}
		\begin{subfigure}[b]{0.8\textwidth}
			\includegraphics[width=\textwidth]{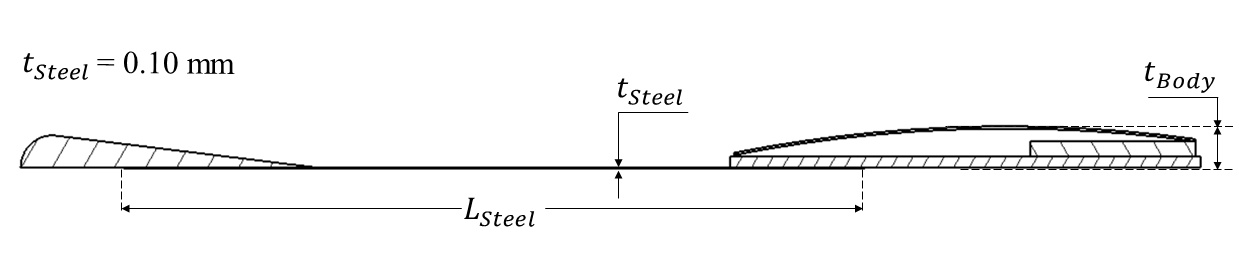}
			\caption{}
			\label{fig:2DApparatus_Kt2_M150_Dyn}
		\end{subfigure}
		\begin{subfigure}[b]{0.4\textwidth}
			\includegraphics[width=\textwidth]{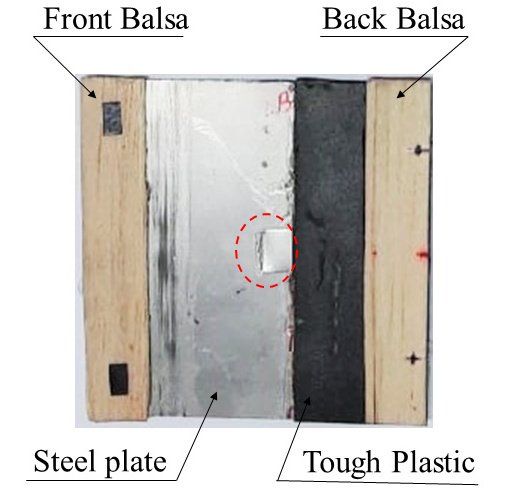}
			\caption{}
			\label{fig:ClappingBody_Kt2_AR100_M150_Dyn}
		\end{subfigure}\hspace{10mm}
		\begin{subfigure}[b]{0.38\textwidth}
		\includegraphics[width=\textwidth]{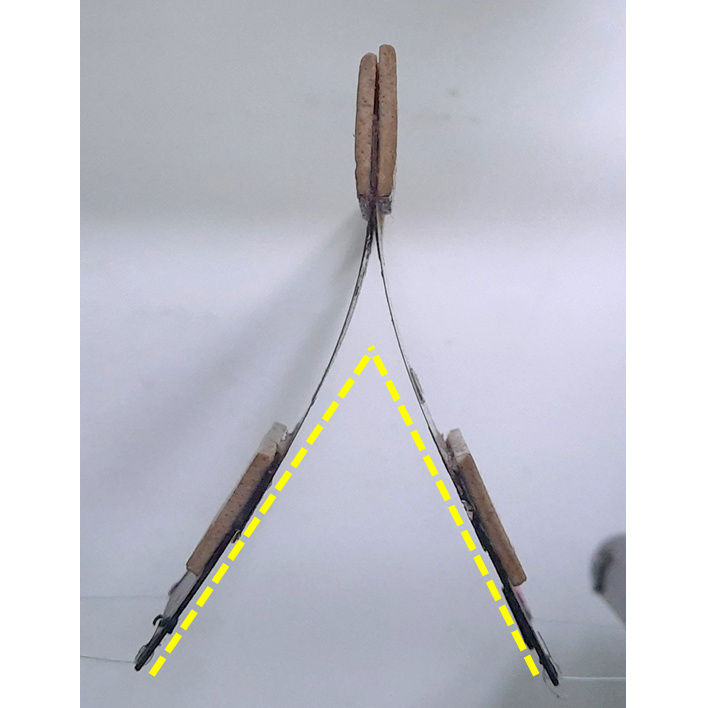}
		\caption{}
		\label{fig:TV_superimposed_Lines}
     	\end{subfigure}
		\caption{(a) Iso-metric view of one of the plates that form the clapping body. (b) Crossection of a plate with $M^*$ = 1.5 and stiffness per unit depth $Kt_1$, and (c) of a clapping plate with $M^*$ = 1.5 and stiffness per unit depth $Kt_2$. (d) Side view of the clapping body with $d^*$ = 1.0, $M^*$ = 1.0, and $Kt_2$. The small piece of steel encircled with a red dashed line indicates the additional mass used for hydrodynamical balancing. (e) Top view of the clapping body with $d^*$ = 0.5, $M^*$ = 1.0, and $Kt_1$ with the plates pulled apart. The leading portion of the body shows steel plates glued together and sandwiched between two balsa-wood aerofoils.}\label{fig:ProbDrawing_Dyn}	
	   \end{figure}
	   %%%%%%%%%%%%%%%%%%%%%%%%%%%%%%%%%%%%%%%%%%%%%%%%%%%%%%%%%%%%%%%%%%%%%%%%%%%%%%%%%%
	  
	  The experiment required the design of an arrangement to hold the arms at an initial clapping angle $2\theta_o$ and a release mechanism to allow the arms to quickly come together to give the clapping action. A fishing thread ({\it Caperlan}) with a diameter of 0.25mm was used to construct a loop connecting both arms, where both arms experience initial effective torque $T_o$ (See figure \hyperref[fig:DynProblem_to]{1a}). The release stand consists of a pair of rigid acrylic circular rods mounted on an aluminum base. The initial clapping angle was adjusted by changing the separation distance between the rods. All experiments were done in quiescent water in a tank of dimension 80cm $\times$ 80cm $\times$ 30cm (height). The clapping body was placed at a depth of 15cm from the water surface. The body was set in motion by cutting the thread using a laparoscopic scissor. The scissor with an arm of 30cm and 5mm diameter minimized disturbance in the water during the cutting. \par 
	   
	   The experimental input parameters are mass ratio $M^*$, initial clapping angle $2\theta_o$, non-dimensional depth $d^*$, and spring stiffness per unit depth $Kt$. We performed the experiments for: $M^*$ = 1.5 and 1; $2\theta_o$ = 45 and 60degrees; $d^*$ = 1.5, 1.0, and 0.5; stiffness per unit depth $Kt_1$ = 0.8-1.1 mJ / mm.rad$^2$ for steel plate of 0.14mm thickness and $Kt_2$ = 0.3-0.6 mJ / mm.rad$^2$ for steel plate of 0.10mm thickness. The body length $L$ is the same in all experiments (= 89mm); three values of depth (d) were used, 45 mm, 89 mm, and 133 mm, giving $d^*$ =0.5, $d^*$ = 1.0, and $d^*$ =1.5. The thicknesses of the various components are: steel = 0.14mm for $Kt_1$ and 0.10mm for $Kt_2$; front balsa aerofoil = 2.7mm for $Kt_1$ and 2.2mm for $Kt_2$; back balsa = 2mm for $Kt_1$ and 1.15 mm for $Kt_2$, rigid plastic = 0.8mm. The thicknesses of the body  $t_{Body}$ before and after canopy addition: 3mm and 4.7 for $Kt_1$; 2mm and 3mm for $Kt_2$.
	   Experiments were repeated three times for each of the 24 cases in the parametric space. The spring stiffness of the steel plate ($Kt_d$) in the clapping body is defined as a proportionality constant correlating the initial strain energy with the initial clapping angle, refer \eqref{eq:SE_Kt}. The details of spring stiffness calculations are discussed in the \S \hyperref[sec:SE_Exp]{5}. The values of $m_b$, $Kt$, $Kt_d$, and centroid of the clapping body $X_c$ are given in the table\hyperref[tab:DesignData]{1}. Note that, $Kt_1$ ranges from 0.8-1.1 $mJ/mm.rad^2$ and $Kt_2$ ranges from 0.3-0.6 mJ / mm.rad$^2$.  \par
       %%%%%%%%%%%% Design data: Calpping body %%%%%%%%%%%%%%%%%%%%%%%%%%%%%%%%%%%%%%%%%%%
		% Table generated by Excel2LaTeX from sheet 'Design'
		\begin{table}
			\centering
		
		\begin{tabular}{cccccccccc}
			\toprule
			\textbf{$t_{Steel}$} & \textbf{$Kt$} & \textbf{$M^*$} & \textbf{$d^*$} & \textbf{$m_b$} & \textbf{$d$} & \textbf{$Kt_d$} & \textbf{$L_{Steel}$} & \textbf{$L_{Plastic}$} & \textbf{$X_c$} \\
			\text{[mm]} & \text{[mJ / mm.rad$^2$]} &       &       & \text{[gm]} & \text{[mm]} & \text{[mJ / rad$^2$]} & \text{[mm]} & \text{[mm]} & \text{[mm]} \\
			\midrule
			\multirow{6}[4]{*}{0.14} & \multirow{6}[4]{*}{0.8-1.1} & \multirow{3}[2]{*}{1.0} & 1.5   & 33.4  & 133   & 114.4 & \multirow{6}[4]{*}{60.0} & \multirow{6}[4]{*}{30.0} & \multirow{3}[2]{*}{$\sim$ 36} \\
			&       &       & 1.0   & 21.2  & 89    & 99.3 &       &       &  \\
			&       &       & 0.5   & 10.6  & 45    & 40.5  &       &       &  \\
			\cmidrule{3-7}\cmidrule{10-10}          &       & \multirow{3}[2]{*}{1.5} & 1.5   & 48.8  & 133   & 111.1 &       &       & \multirow{3}[2]{*}{$\sim$ 37} \\
			&       &       & 1.0   & 31.2  & 89    & 74.4 &       &       &  \\
			&       &       & 0.5   & 15.3  & 45    & 46.6  &       &       &  \\
			\midrule
			\multirow{6}[4]{*}{0.10} & \multirow{6}[4]{*}{0.3-0.6} & \multirow{3}[2]{*}{1.0} & 1.5   & 23.1  & 133   & 36.9  & \multirow{6}[4]{*}{55.0} & \multirow{6}[4]{*}{35.0} & \multirow{3}[2]{*}{$\sim$ 39} \\
			&       &       & 1.0   & 15.0  & 89    & 26.9  &       &       &  \\
			&       &       & 0.5   & 7.5   & 45    & 28.2  &       &       &  \\
			\cmidrule{3-7}\cmidrule{10-10}          &       & \multirow{3}[2]{*}{1.5} & 1.5   & 32.3  & 133   & 55.0  &       &       & \multirow{3}[2]{*}{$\sim$ 41} \\
			&       &       & 1.0   & 22.0  & 89    & 37.5  &       &       &  \\
			&       &       & 0.5   & 11.1  & 45    & 28.7  &       &       &  \\
			\bottomrule
		\end{tabular}%
		\label{tab:DesignData}%
		\caption{Design data of the clapping bodies}
	\end{table}%
	%%%%%%%%%%%%%%%%%%%%%%%%%%%%%%%%%%%%%%%%%%%%%%%%%%%%%%%%%%%%%%%%%%%%%%%%%%%%%%%%%%%%%%%%
	A lot of care was required to achieve neutral buoyancy and coincidence of COM and COB. The neutral buoyancy condition gets easily disturbed even if a small air bubble is present on the body surface. Bubbles form during the insertion of the release stand along with the body into the water. A jet from a syringe was used to remove the bubbles entrapped in the gap between leading balsa plates and the leading edge of steel plates. Balsa wood would absorb water, due to which, over time, the neutral buoyancy condition was disturbed. The balsa wood was coated with ‘Plastik 70’ to overcome this problem. The body mass was verified  before each experiment. Placement of small masses of steel or balsa was required to achieve the requirements listed above(see figure \hyperref[fig:ProbDrawing_Dyn]{2d}). The weights and locations of these masses were identified for the 12 clapping bodies by trial and error method. Each body had to be properly balanced before each experiment. \par
	
	Two-dimensional particle image velocimetry (PIV) was used to measure the flow field in the unsteady wake. The guidelines given by Raffel et al.\cite{Raffel18} were followed. The PIV setup consists of a continuous wave 5W power, 532nm wavelength laser, a high-speed camera, and two plano concave lenses. Two plano concave lenses of radius 6mm are positioned opposite to each other to increase the divergence angle of the laser sheet; the laser sheet thickness was 2-3mm. Silver-coated particles (CONDUCT-O-FIL, Potters Inc) of 10-15 $\mu$m were used as tracers. A high-speed camera (Photron-SA5) with 1024 $\times$ 1024 Pixel$^2$ resolution recorded the flowfield at 1000FPS using a Nikon lens of 105mm focal length. The post-processing of the PIV database was performed using ‘IDT-ProVision’ software. The 2D-PIV measurements were performed on the XY plane at Z =0 and the XZ plane at Y =0. The term ‘Top view PIV’ corresponds to the XY plane, whereas ‘Side view PIV’ corresponds to the XZ plane.\par
	
	In the top view PIV, as shown in figure \hyperref[fig:PIV_Schematic_Dyn]{3}, the laser light sheet is along the mid-XY plane located at half the body depth. Top view of the PIV setup shows a green laser sheet illuminating the interplate cavity, and the black region represents the shadow on the back side of the cavity. For the top view, a mirror at 45 degrees placed on top of the tank was used. The interrogation window size was 24 $\times$ 24 Pixel$^2$, where 1pixel $\approx$ 0.3mm. The region of interest ROI varies between 100$^2$ - 120$^2$ mm$^2$ for the clapping bodies corresponding to $Kt_1$ and $Kt_2$. In the side view PIV, the camera directly recorded the flow field. The interrogation window size was the same as for the one in the top view PIV. The ROI varies between 150 $\times$ 120 mm$^2$ to 120 $\times$ 120mm$^2$ based on the $d^*$ variations. \par
	
	%%%%%%%%%%%%%%%%%%%%%%%%%%%%%%%%%%%%%%%%%%%%%%%%%%%%%%%%%%%%%%%%%%%%%%%%%%%%%%%%%%%%%%
	\begin{figure}
		\centering\
		\begin{subfigure}[b]{0.8\textwidth}
			\includegraphics[width=\textwidth]{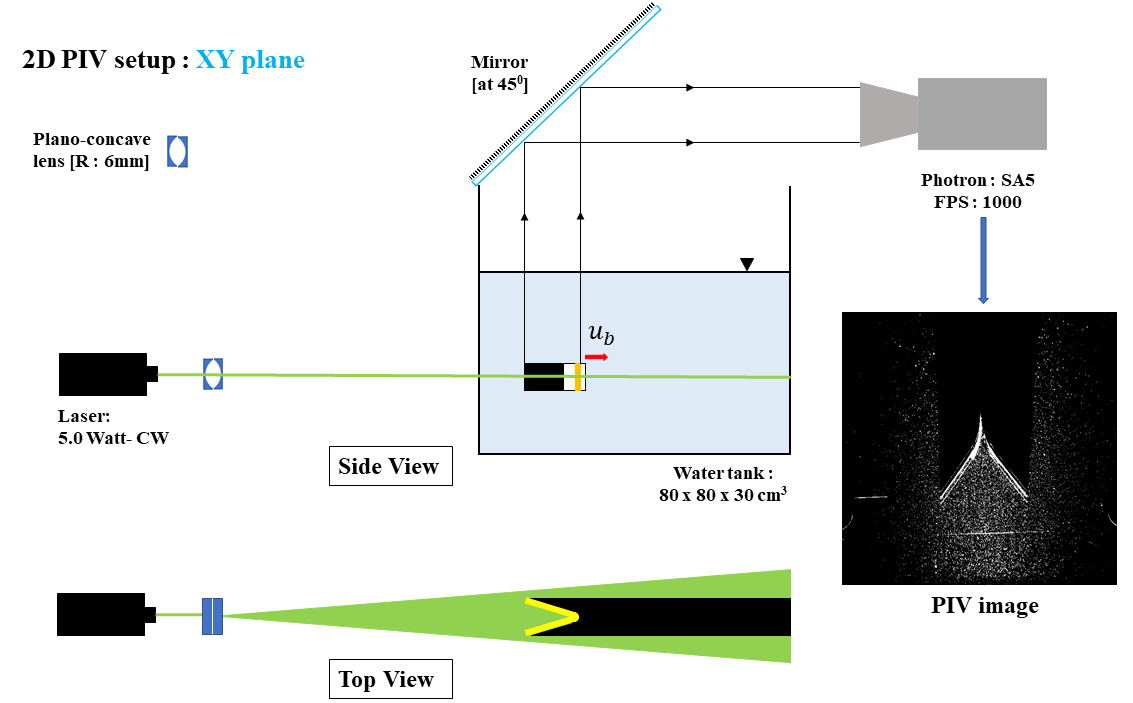}
		\end{subfigure}
		\caption{Schematic of the PIV setup formeasurement of the flow field in the XY plane. Also shown is a photograph of the particles illuminated by the laser sheet.}\label{fig:PIV_Schematic_Dyn}		
	\end{figure}
   %%%%%%%%%%%%%%%%%%%%%%%%%%%%%%%%%%%%%%%%%%%%%%%%%%%%%%%%%%%%%%%%%%%%%%%%%%%%%%%%%%%%%%%
   The flow was visualized on the XY plane using planar laser-induced fluorescence (PLIF). A thin layer of dye paste containing a mixture of Rhodamine B, Acrylic binder(Daler-Rowney Slow Drying gel), and honey was applied along a line at mid-depth on the inside surfaces of the clapping plates; honey provides the required fluidity to the gel. The methodology is adapted from that described in David et al.\cite{JeemReeves18}. RhodamineB emits light at 625nm when excited with the green laser light. The same high-speed camera and CW laser were used for the dye visualization.\par
   
   The body kinematics has been extracted using the Kanade-Lucas-Tomasi (KLT) feature-tracking algorithm in MATLAB. The trajectory of the self-propelling body is recorded both in top and side view (figure \hyperref[fig:PIV_Schematic_Dyn]{3}). The high-speed camera at 1000 FPS is used to record the rapid clapping action from the top view, whereas a regular camera (Nikon-COOLPIX) at 25FPS is used to record the trajectory of the body in the side view until it comes to rest.
   
\end{section}

\begin{section}{Results and discussion}
\label{sec:briefDiscussion}
	First, we give an overview of the body motion and the flow created by the clapping motion of the body following the cutting of the thread. We choose one case with the following  parameters: stiffness per unit depth = $Kt_1$, $M^*$ = 1.0, $2\theta_o$ = 60 deg, and $d^*$ = 0.5. The cutting of the threaded loop initiates the rotation of each plate. Ejection of the fluid from the inter-plate cavity due to the rapid clapping motion creates a transient jet. During this time, the high fluid pressure on the inner surfaces of the two plates provides the propulsive force to the clapping body. The body has two phases of translatory motion: a rapid acceleration during the clapping motion followed by slow retardation. Figures \hyperref[fig:Lin_Velo_Dyn_AR200]{4a} and \hyperref[fig:Lin_Velo_Dyn_AR200]{4b} show the body’s translational velocity $(u_b)$ with time, the former focuses on the initial phase. After attaining a maximum velocity of 0.71m/sec at 50ms, the drag force slowly reduces the body velocity to zero over about 2 seconds. The total distance traveled by the body is approximately 3BL (body lengths). The body is tracked until it is primarily moving in the X direction; when the body speed becomes low, even a slight mismatch between weight and buoyancy force makes the body deviate from the horizontal path. The high-speed camera at 1000FPS records clapping action from the top view, whereas the Nikon camera at 25FPS records the side view. Data is extracted manually during the clapping action, and the KLT tracker is used after the end of the clapping motion until the body comes to rest. During the impulsive phase, images are analyzed at 250FPS instead of 1000FPS, which gives more than one-pixel displacement per frame. A reduction in body velocity during the retardation phase allows the velocity calculation with a time resolution of 125 FPS. Figure \hyperref[fig:Lin_Velo_Dyn_AR200]{4} shows the data points along polynomial fits. \par

	Figure \hyperref[fig:Ang_Pos_Velo_Dyn_AR200]{5a} shows the corresponding variation of semi-clapping angle ($\theta$) with time starting with the initial value of 30 degrees, and figure \hyperref[fig:Ang_Pos_Velo_Dyn_AR200]{5b} shows angular velocity ($\dot{\theta}$) variation. Note that the angular velocity is the rate of change of half of the inter-clap angle. Both clapping plates are set into an impulsive rotation once the thread is cut. The angular velocity attains maxima of 13.1 rad/sec at $\theta$ = 10 degrees, at about 0.02 seconds. The body reaches its maximum translation velocity at approximately when the angular velocity becomes zero. \par
	
	Vorticity is shed from the trailing edges of the plates, culminating in the formation of a three-dimensional vortex loop that appears as two vortex patches in the XY plane. Figure \hyperref[fig:Wake_summary]{6a} shows the PIV velocity and vorticity fields in the central plane. The position of the clapping body is marked with yellow lines, whereas the shadow in the top view PIV image is shown in gray color. There is an indication of the expected bound vortex (shown schematically in figure \hyperref[fig:VortSchem_Velo_Dyn]{7a}) around each plate in figure \hyperref[fig:Wake_summary]{6a}. The bound vortices in the PIV field are not clearly visible due to insufficient spatial resolution and the shadow behind the plates. Figure \hyperref[fig:Wake_summary]{6b} shows the dye initially on the inner sides of the plates, being shed into the wake as the body moves forward. The red circles indicate the starting vortices. \par

	\begin{figure}
		\centering\
		\begin{subfigure}[b]{0.4\textwidth}
			\includegraphics[width=\textwidth]{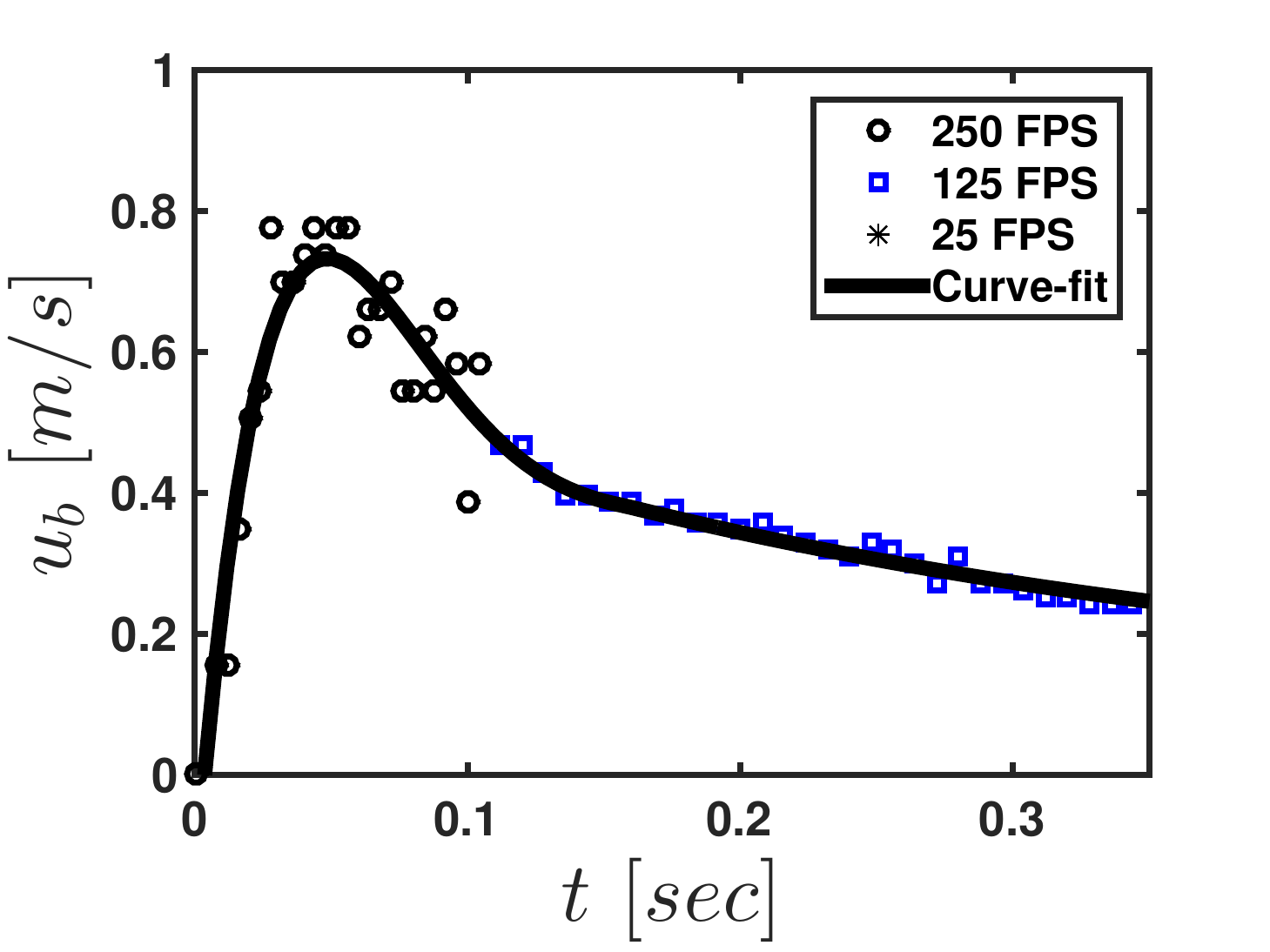}
			\caption{}
			\label{fig:VeloLin_01_Dyn}
		\end{subfigure}\hspace{08mm}
		\begin{subfigure}[b]{0.4\textwidth}
			\includegraphics[width=\textwidth]{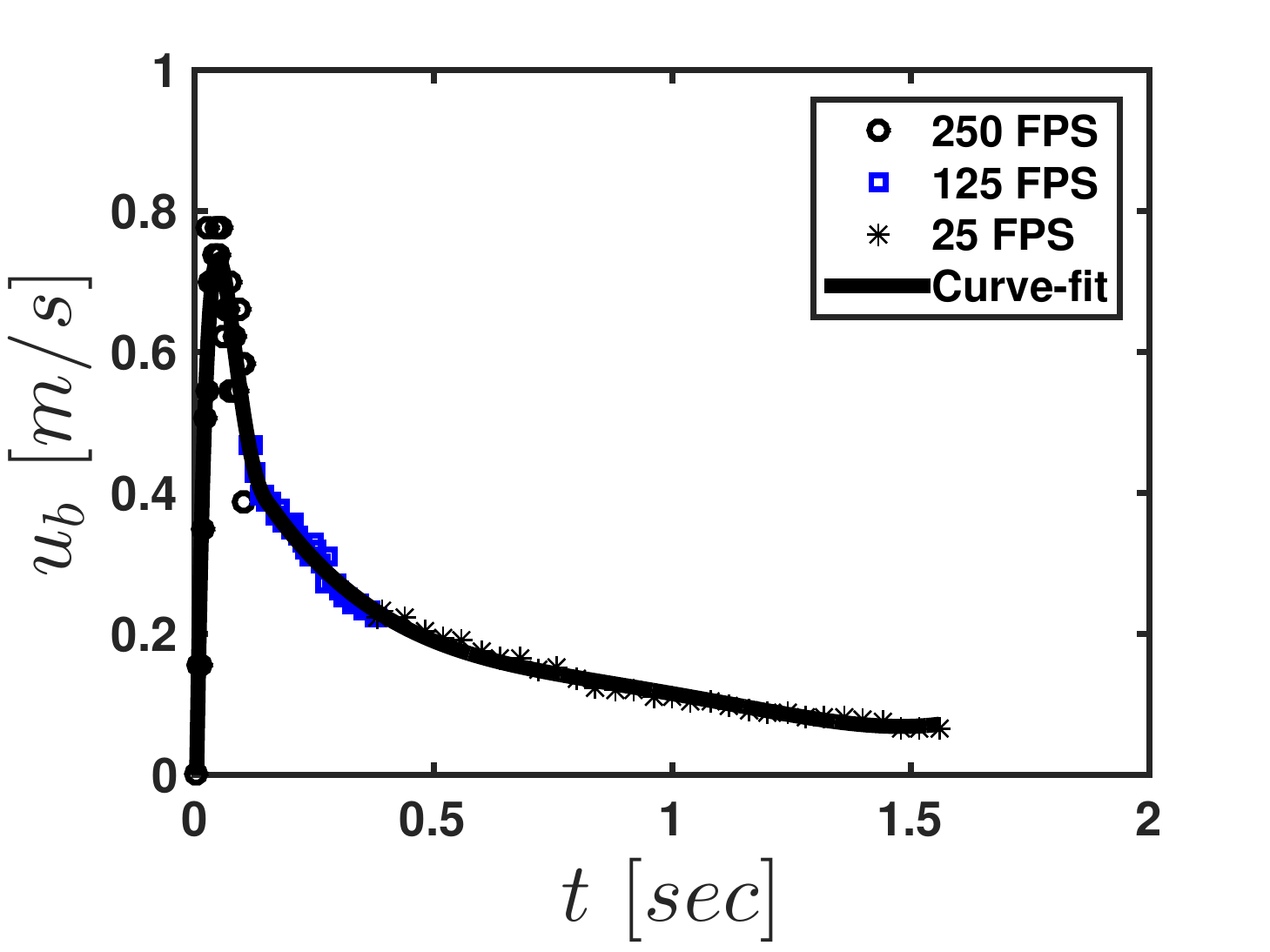}
			\caption{}
			\label{fig:VeloLin_02_Dyn}
		\end{subfigure}
		\caption{(a) Variation of the translational velocity $u_b$ of the body with time upto 0.35sec. (b) Time evolution of $u_b$ till the body translates along the X-direction with negligible displacement in the Z-direction. $M^*$ = 1.0, $2\theta_o$ = 60 deg, and $d^*$ = 0.5.}\label{fig:Lin_Velo_Dyn_AR200}
	\end{figure}

	\begin{figure}
	\centering\
	\begin{subfigure}[b]{0.4\textwidth}
		\includegraphics[width=\textwidth]{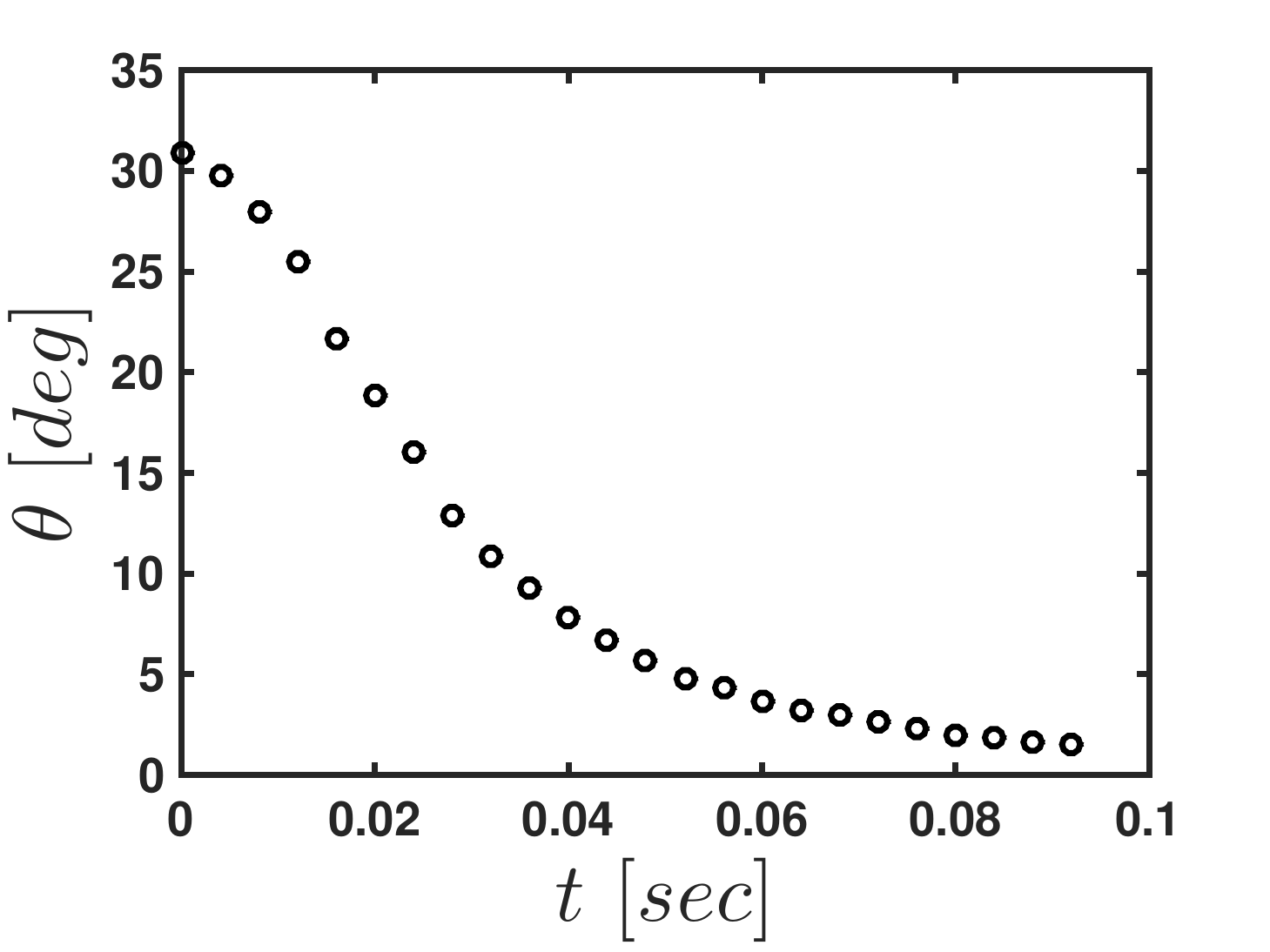}
		\caption{}
		\label{fig:DispAngular_Dyn}
	\end{subfigure}\hspace{08mm}
	\begin{subfigure}[b]{0.4\textwidth}
		\includegraphics[width=\textwidth]{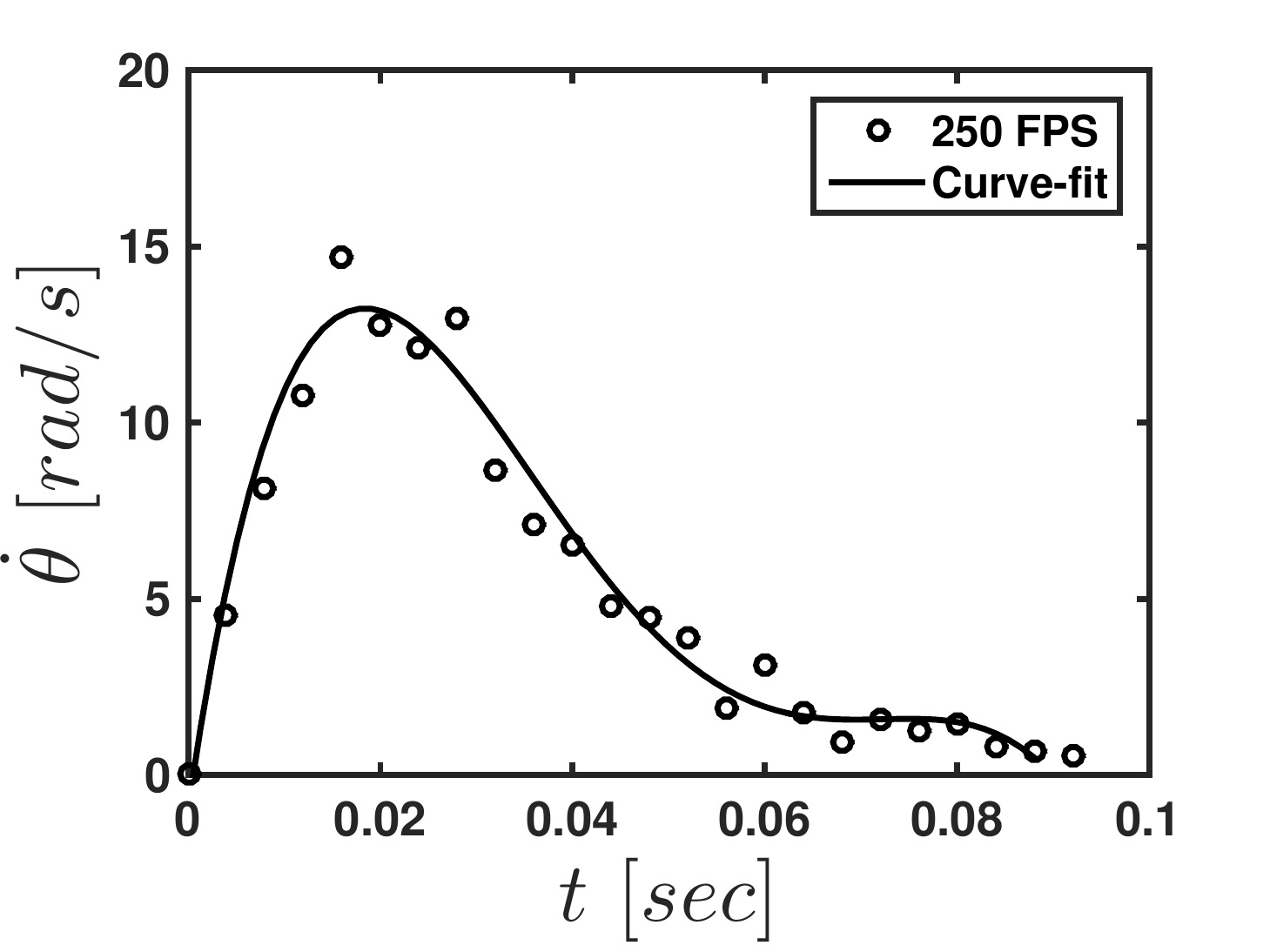}
		\caption{}
		\label{fig:VeloAngular_Dyn}
	\end{subfigure}
	\caption{(a) Variation of semi-clapping angle $\theta$ with time. (b) Variation of angular velocity $\dot{\theta}$ of the clapping plate with time. $M^*$ = 1.0, $2\theta_o$ = 60 deg, and $d^*$ = 0.5.}\label{fig:Ang_Pos_Velo_Dyn_AR200}
	\end{figure}

	\begin{figure}
		\centering\
		\begin{subfigure}[b]{0.4\textwidth}
			\includegraphics[width=\textwidth]{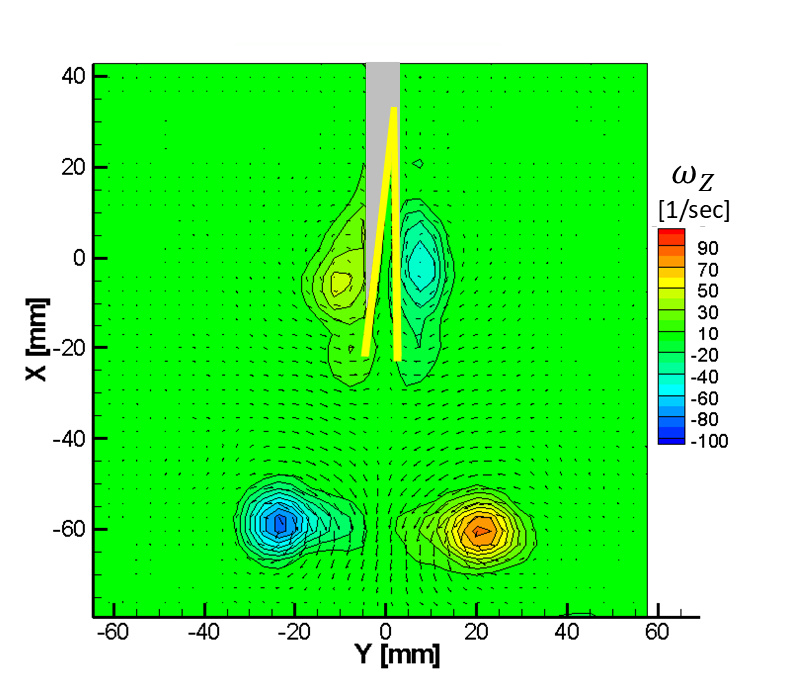}
			\caption{}
			\label{fig:PIV_AR200}
		\end{subfigure}\hspace{08mm}
		\begin{subfigure}[b]{0.4\textwidth}
			\includegraphics[width=\textwidth]{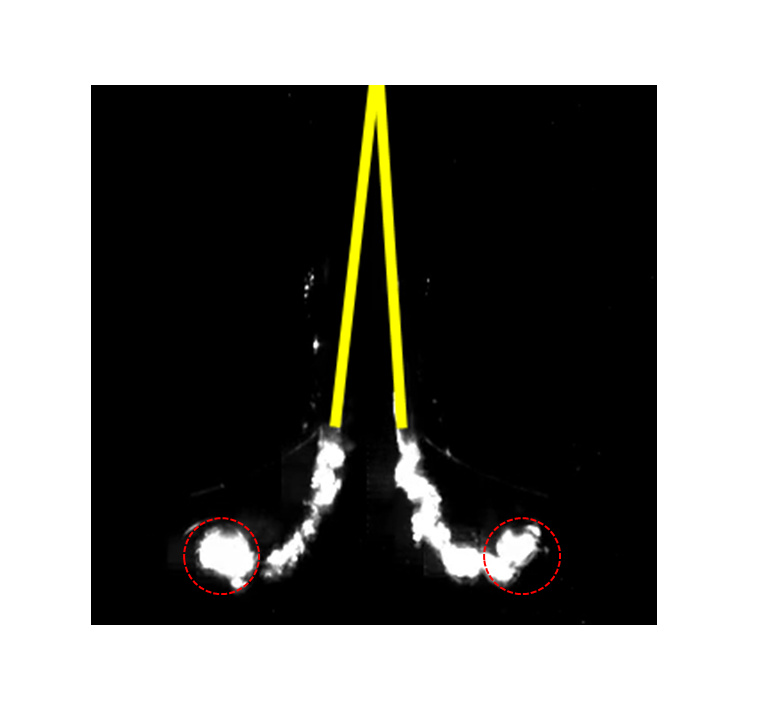}
			\caption{}
			\label{fig:Dye_AR200}
		\end{subfigure}
		\caption{(a) Z-vorticity field $\omega_Z$ show the starting vortices at the end of the clapping motion, t = 60 ms. The two yellow lines show the superimposed clapping body, and the gray color shows a shadow of the body. (Figure \hyperref[fig:TV_superimposed_Lines]{2e} show the correspondence between the  yellow lines and the clapping body.) (b) The wake visualized using the PLIF shows at 45ms. The starting vortices are marked with red dashed circles. $M^*$ = 1.0, $2\theta_o$ = 60 deg, and $d^*$ = 0.5.}\label{fig:Wake_summary}
	\end{figure}

	\begin{figure}
		\centering\
		\begin{subfigure}[b]{0.29\textwidth}
			\includegraphics[width=\textwidth]{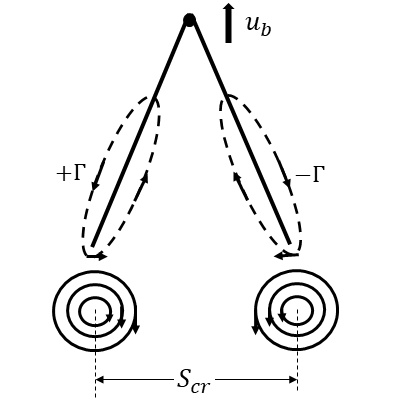}
			\caption{}
			\label{fig:Vort_Schm_Dyn}
		\end{subfigure}\hspace{10mm}
		\begin{subfigure}[b]{0.4\textwidth}
			\includegraphics[width=\textwidth]{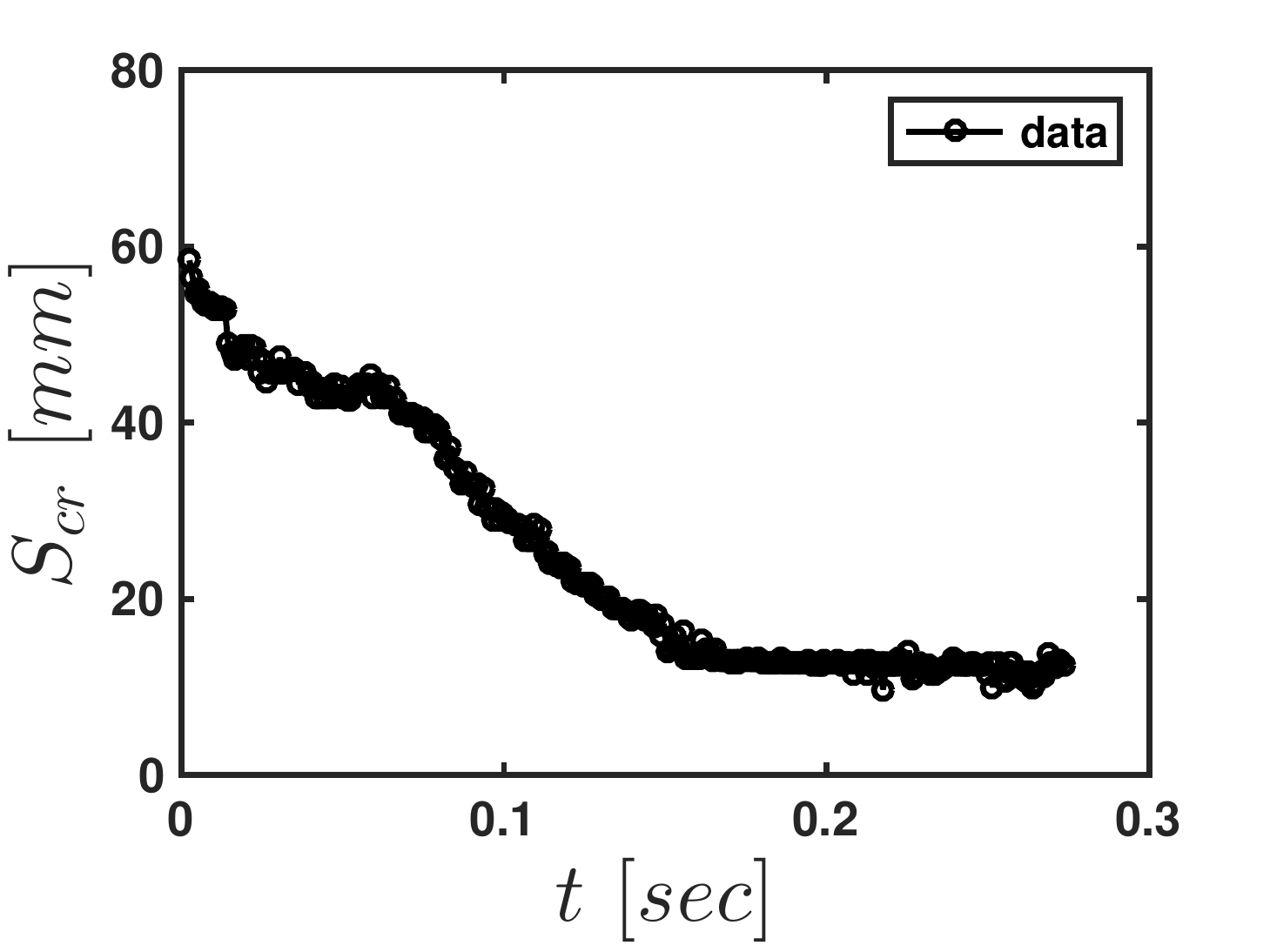}
			\caption{}
			\label{fig:coreSep_AR200}
		\end{subfigure}
		\caption{(a) Schematic showing starting (black circles) and bound vortices (dashed line). (b) Core separation $S_{cr}$ as a function of time. $M^*$ = 1.0, $2\theta_o$ = 60 deg, and $d^*$ = 0.5.}\label{fig:VortSchem_Velo_Dyn}
	\end{figure}

	The fluid velocity near the trailing edge of the plate during the clapping phase (figure \hyperref[fig:velo_x_profile]{8a}) shows a jet-like flow between the plates and a signature of the two vortices being formed. Towards the end of the clapping phase (t $\sim$ 61 ms), when the plates are almost touching, a small amount of fluid is trapped between the plates and dragged behind by the body, resulting in a wake-like velocity profile (figure \hyperref[fig:velo_x_profile]{8b}). \par
	\begin{figure}
		\centering\
		\begin{subfigure}[b]{0.4\textwidth}
			\includegraphics[width=\textwidth]{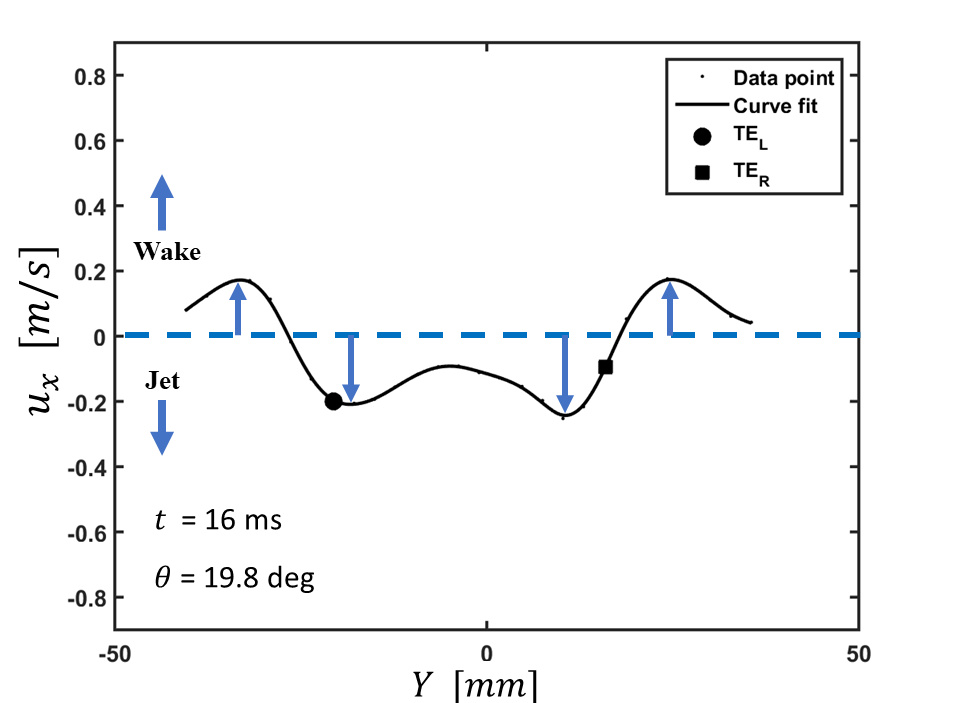}
			\caption{}
			\label{fig:VeloProf_01_Dyn}
		\end{subfigure}\hspace{08mm}
		\begin{subfigure}[b]{0.4\textwidth}
			\includegraphics[width=\textwidth]{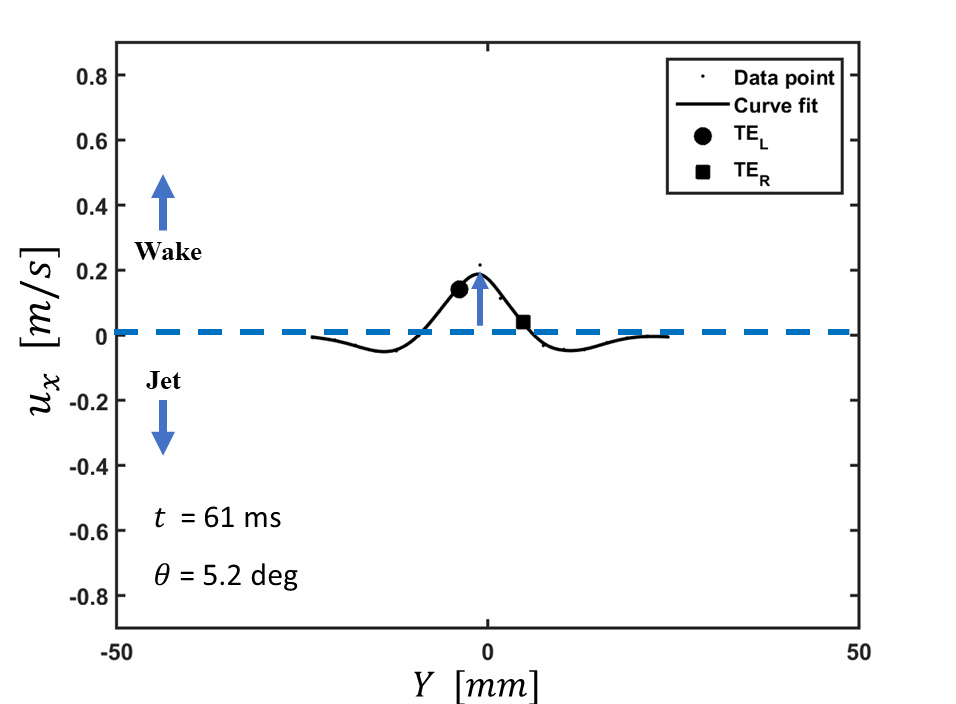}
			\caption{}
			\label{fig:VeloProf_02_Dyn}
		\end{subfigure}
		\caption{ Profiles of X-component of the fluid velocity, $u_x$, across the line joining the trailing edges : (a) at the mid-clapping phase (b) after the end of the clapping phase. $\mathrm{TE_L}$ and $\mathrm{TE_R}$ indicate the trailing edges of the left and right clapping plates, and $\theta$ is the instantaneous semi-clapping angle. $M^*$ = 1.0, $2\theta_o$ = 60 deg, and $d^*$ = 0.5.}\label{fig:velo_x_profile}
	\end{figure}
	
	The circulation around each vortex is calculated using equation \eqref{eq:circulation} where $A_c$ is the area enclosed by contour whose vorticity is $\omega\geq 0.05\ \omega_{max}$. The circulaiton evolution is shown in figure \hyperref[fig:Core_cir_disp]{9a}. 

	\begin{gather}\label{eq:circulation}
	{\Gamma}=\int{\omega}dA_c 
	\end{gather}
	\begin{figure}
		\centering\
		\begin{subfigure}[b]{0.4\textwidth}
			\includegraphics[width=\textwidth]{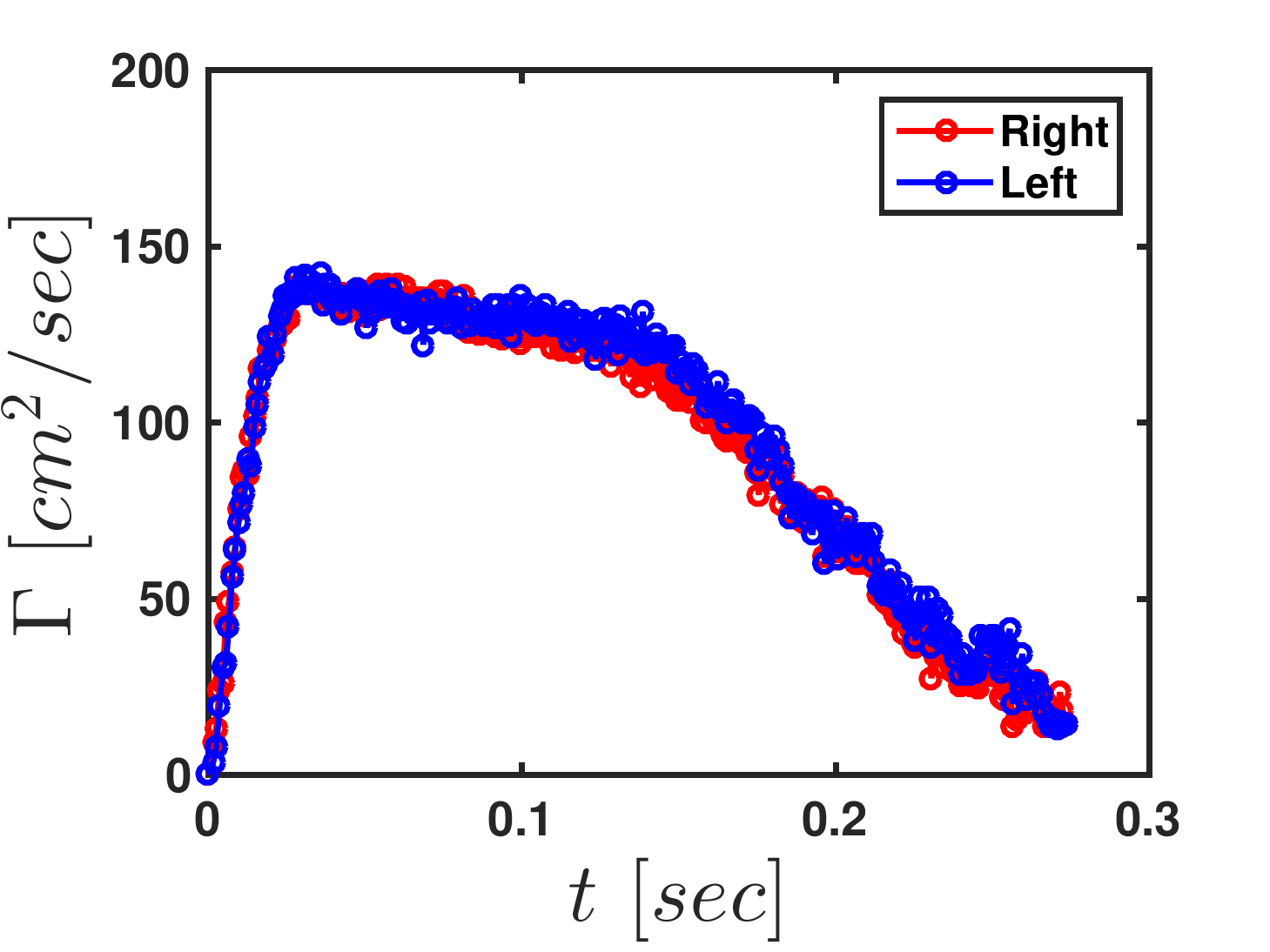}
			\caption{}
			\label{fig:Circualtaion_Dyn}
		\end{subfigure}\hspace{08mm}
		\begin{subfigure}[b]{0.4\textwidth}
			\includegraphics[width=\textwidth]{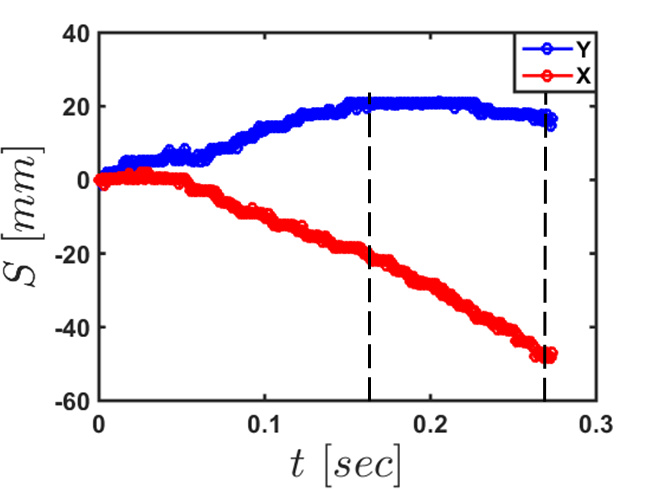}
			\caption{}
			\label{fig:CoreDsip_Dyn}
		\end{subfigure}
		\caption{(a) Circulation magnitudes of the left and right starting vortices. (b) The displacements ($S$) in X and Y directions of the left vortex. The black dashed lines indicate the time interval during which we calculate the steady wake vortex velocity. $M^*$ = 1.0, $2\theta_o$ = 60 deg, and $d^*$ = 0.5.}\label{fig:Core_cir_disp}
	\end{figure}

	The circulation values in the two vortices are nearly identical and show a gradual reduction with time till about 150 ms, after which the change is more rapid. The distance between the vortices $S_{cr}$ (see figure \hyperref[fig:VortSchem_Velo_Dyn]{7b}) also gradually reduces up to 150 ms. After which, it becomes nearly constant when the vortices touch each other, resulting in the cancellation of vorticity and consequent reduction in circulation, as seen in figure \hyperref[fig:Core_cir_disp]{9a}. The displacement of the vortex $S$ in the X direction also shows two phases (see figure \hyperref[fig:CoreDsip_Dyn]{9b}), an initial one with a lower velocity followed by one with a higher velocity when the two vortices are touching each other. The time corresponding to the later phase is identified using black dashed lines, where the Y-displacement is constant(slope $\approx$ 0). In this phase, vortex velocity is 0.18m/sec, four times lower than the maximum body velocity of 0.71 m/s. A detailed discussion on wake evolution is given in \S\hyperref[sec:WakeDynamics]{3.2}. \par

	 In the following sections, we look at how various parameters (non-dimensional depth $d^*$, spring stiffness per unit depth $Kt$, initial clapping angle $2 \theta_{o}$, and mass ratio $M^*$) influence the body kinematics (\S\hyperref[sec:BodyKinematics]{3.1}) and the flow, in particular in the wake (\S\hyperref[sec:WakeDynamics]{3.2}).\par

\begin{subsection}{Body kinematics}
\label{sec:BodyKinematics}
	The two main kinematic parameters are the body translation velocity and the angular velocity of the clapping plates.

\begin{subsubsection}{Translational velocity of the body}

	The body velocity ($u_b$), throughout the parametric space, exhibits the same behavior of rapid increase followed by a slow reduction. For each case in the parametric space, the translational velocity curve is averaged over three experiments; the average of standard deviations in $u_b$ is less than 6\% of the maximum body velocity. The maximum body velocity ($u_m$) lies in the range 0.16m/s to 0.73m/s (See \hyperref[tab:ubm]{Table 2}). The acceleration phase is in  the range of 50-110ms, whereas the deceleration phase continues for more than 1000ms. The acceleration of the body is between 0.2g and 1.5g, where g is the acceleration due to gravity. In all the experimental cases, $u_b$ attains the maximum close to the end of the clapping motion when the clapping angle is 6-8 degrees. \par
	
 \begin{figure}
		\centering\
		\begin{subfigure}[b]{0.4\textwidth}
			\includegraphics[width=\textwidth]{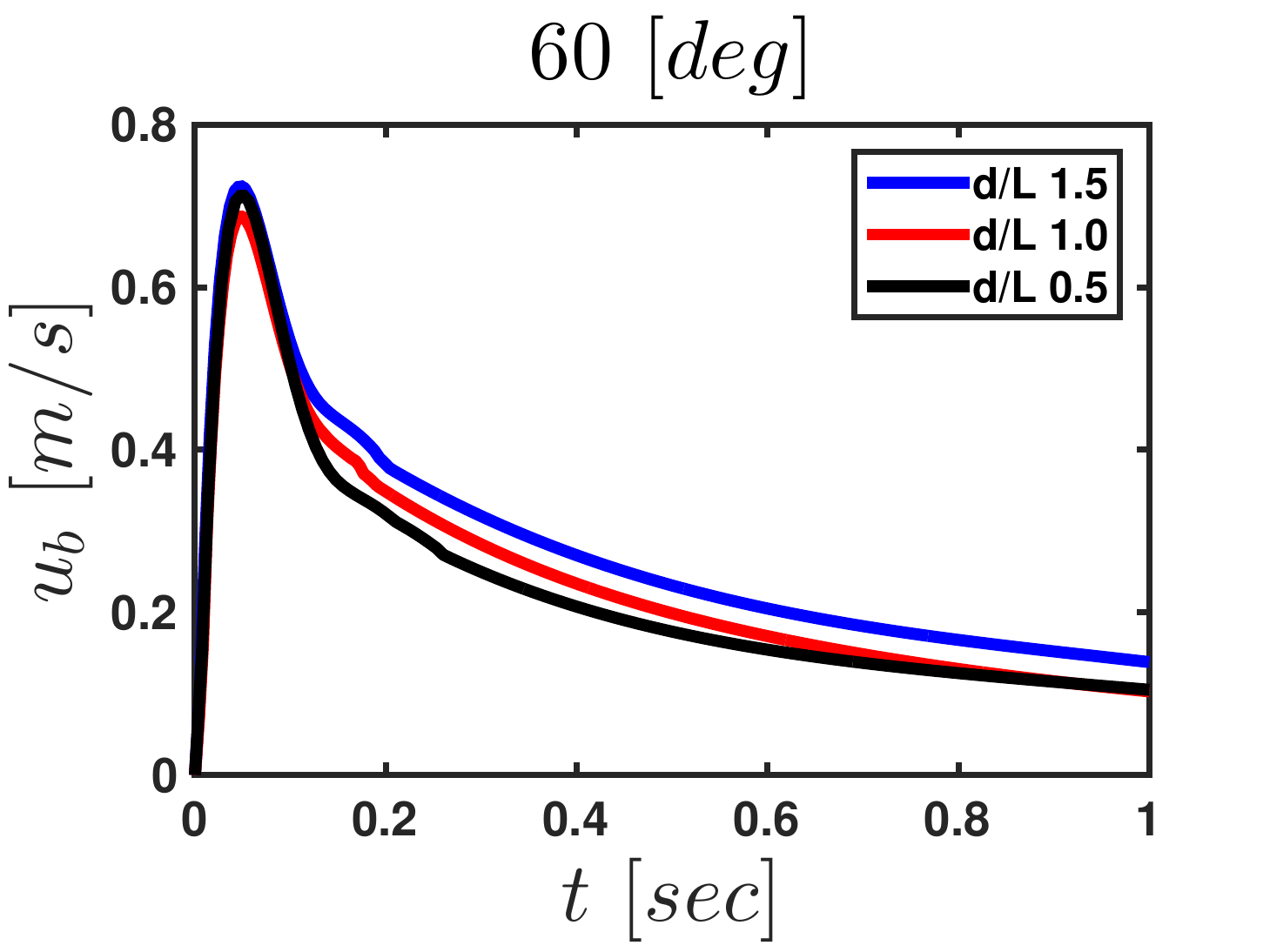}
			\caption{}
			\label{fig:ub_ARconst_Dyn}
		\end{subfigure}\hspace{08mm}
		\begin{subfigure}[b]{0.4\textwidth}
			\includegraphics[width=\textwidth]{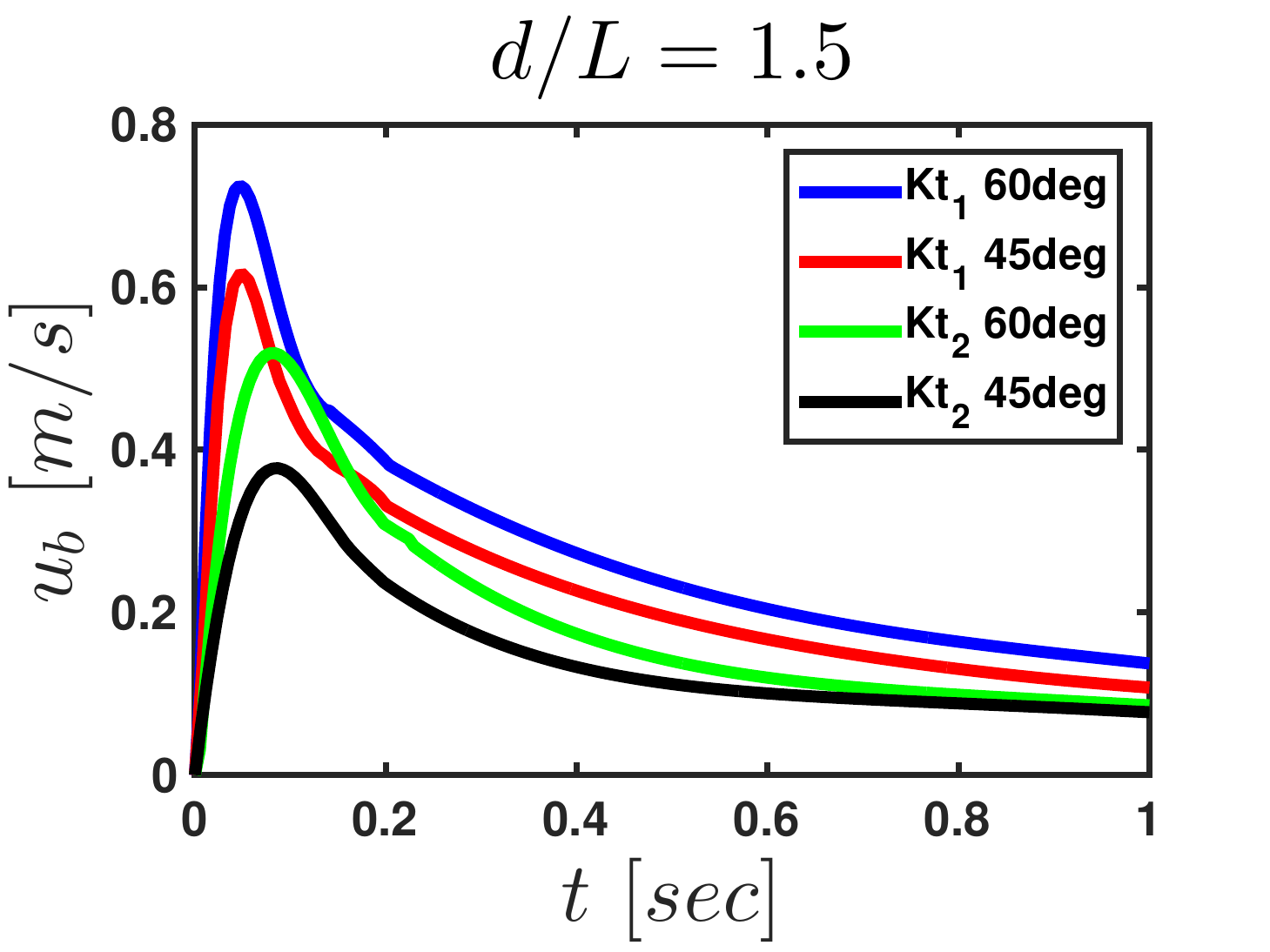}
			\caption{}
			\label{fig:ub_Kt_angle_Dyn}
		\end{subfigure}\vspace{05mm}
		\begin{subfigure}[b]{0.4\textwidth}
			\includegraphics[width=\textwidth]{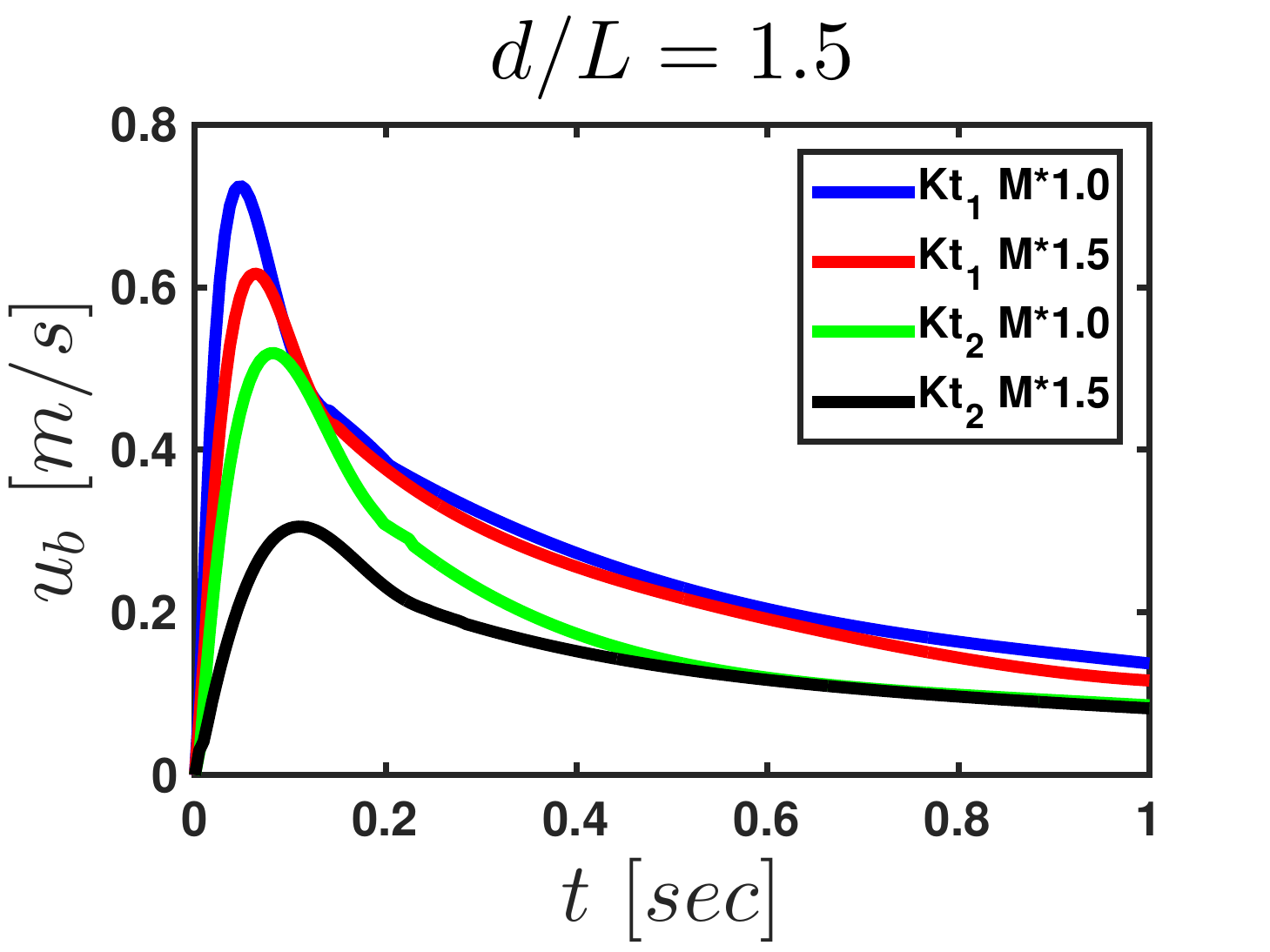}
			\caption{}
			\label{fig:ub_Mstr_angle_Dyn}
		\end{subfigure}
		\caption{Variation of translational velocity $u_b$ with time, (a) for different $d^*$ values, and with stiffness per unit depth = $Kt_1$, $2 \theta_o$= 60 deg and $M^*$= 1.0. (b) for different values of $Kt$ and clapping angle $2 \theta_o$, and with $d^*$= 1.5 and $M^*$= 1.0. (c) for different values of $M^*$ and $Kt$, and with $d^*$= 1.5, $2 \theta_o$= 60 deg.}\label{fig:ub_ar_mstr_angle}
	\end{figure}

	\begin{figure}
		\centering\
		\begin{subfigure}[b]{0.45\textwidth}
			\includegraphics[width=\textwidth]{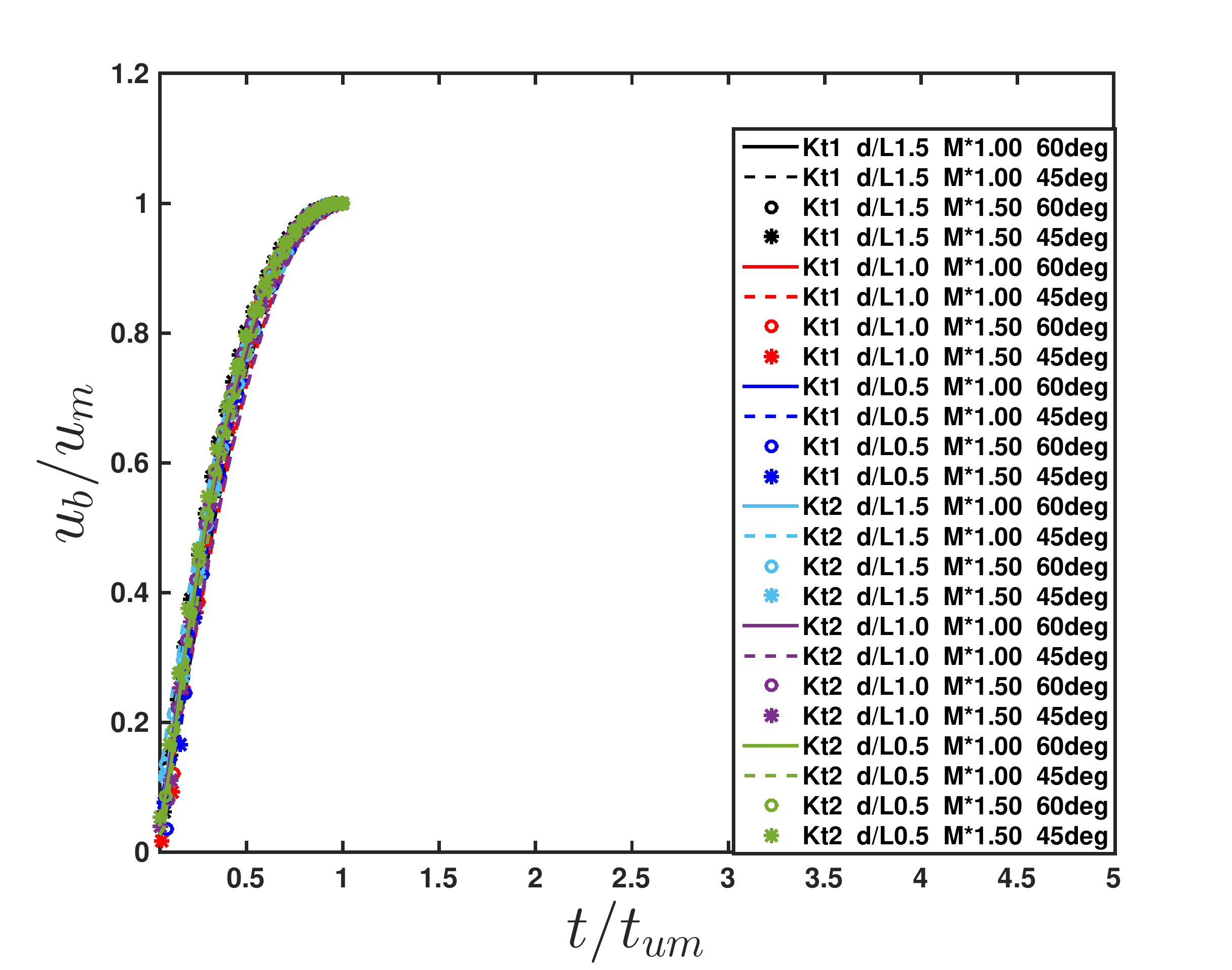}
			\caption{}
			\label{fig:ub_Acc_NonDim}
		\end{subfigure}\hspace{02mm}
		\begin{subfigure}[b]{0.52\textwidth}
			\includegraphics[width=\textwidth]{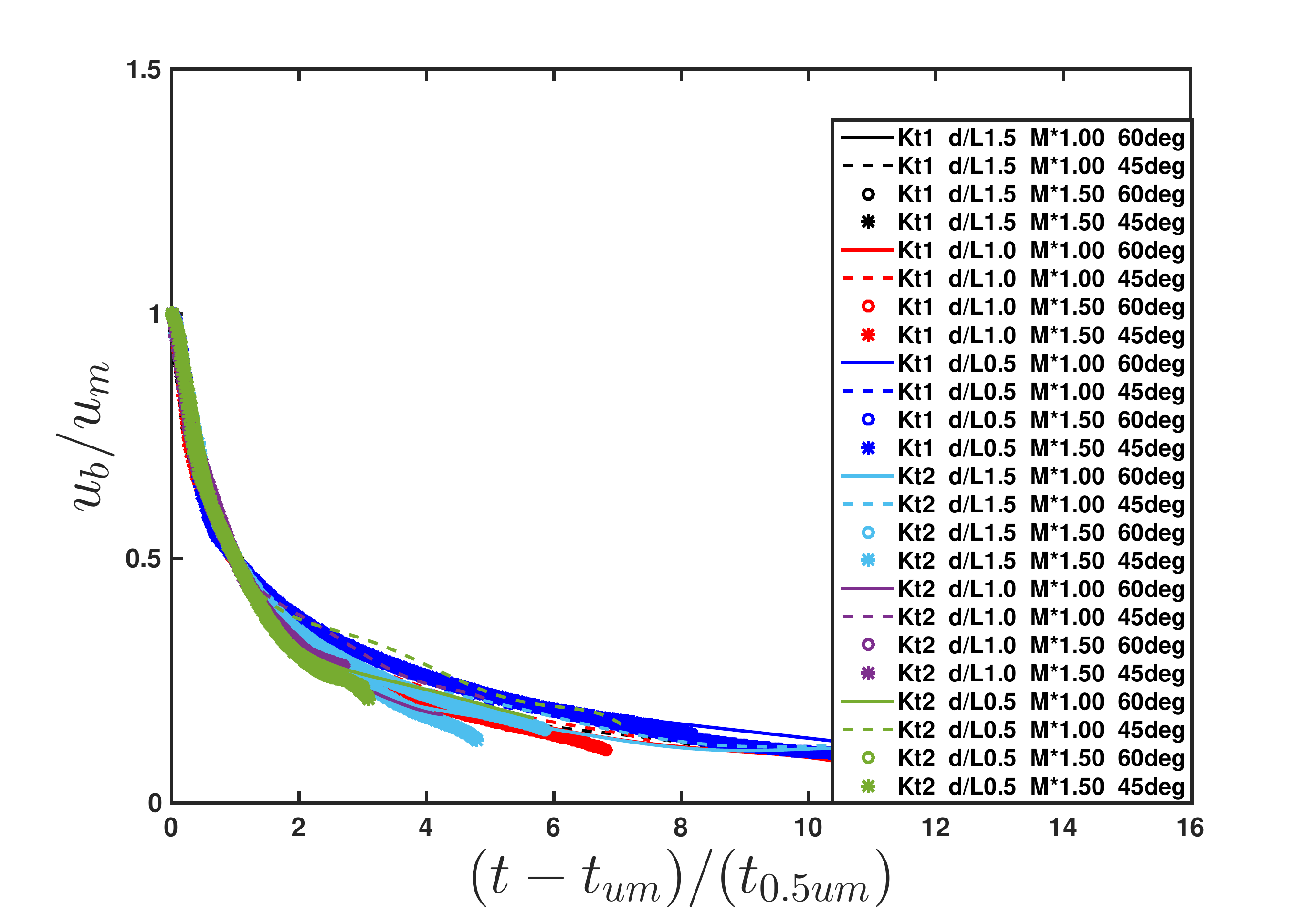}
			\caption{}
			\label{fig:ub_DAcc_NonDim}
		\end{subfigure}\hspace{02mm}
		\caption{Normalized body velocity versus normalized time for all 24 cases during (a) acceleration phase and (b) retardation phase : $u_m$ is the maximum body velocity, $t_{um}$ is the time when $u_b = u_m$, and $t_{0.5\ um}$ is the time when $u_b =u_m /2$ during the retardation phase.}\label{fig:ub_nonDim}
	\end{figure}
	
	In general we found change in the body aspect ratio $d^*$, does not produce noticeable change in the body velocity profile with time. Figure\hyperref[fig:ub_ARconst_Dyn]{10a} illustrates this observation. We see that  profiles for the three values of $d^*$, with the other  parameters fixed, are nearly identical, especially during the acceleration phase.  The relative standard deviation (RSD) in $u_m$ due to $d^*$ variations is less than 10\% except for  $Kt_2$ and $M^* =$ 1.5 case where RSD of 17\% is observed. \hyperref[tab:ubm]{Table 2} lists the $u_m$ values for all the different conditions. \par
	
	Equation \eqref{eq:thrust_eqm} is the equation of motion for the body in the accelerating phase.
	\begin{gather}\label{eq:thrust_eqm}
	\left(m_b +m_{add}\ \right)\frac{d u_b}{dt}= F_T 
	\end{gather}
	The net thrust force $F_T$ (= Thrust - Drag) acting on the body is balanced by an inertial force, where $m_{add}$ is the additional mass of fluid that accelerates with the body. The estimation of $m_{add}$ is difficult to assess due to the complex motion of fluid during clapping.\par

%%%%%%%%%%%%%%%%%%%%%%%%%% Ub_max  %%%%%%%%%%%%%%%%%%%%%%%%%%%%%%%%%%%%%%%%%%%%%%%%%%%%	
% Table generated by Excel2LaTeX from sheet 'Variables'
\begin{table}
	\centering
	
	\begin{tabular}{cccccc}
		\toprule
		\multicolumn{6}{c}{\text{ $u_m$  [m/s]}} \\
		\midrule
		\textbf{$Kt$} & \multicolumn{1}{c}{\multirow{2}[4]{*}{\text{$M^*$}}} & $2\theta_{o}$ & \multicolumn{3}{c}{\text{$d^*$}} \\
		\cmidrule{4-6}    \text{[mJ / mm.rad$^2$]} &       & \multicolumn{1}{p{3em}}{\text{[Deg]}} & \textbf{1.5} & \textbf{1.0} & \textbf{0.5} 	\\	
		\midrule
		\multirow{4}[4]{*}{$Kt_1$: \textbf{0.8-1.1}} & \multirow{2}[2]{*}{\textbf{1.0}} & \textbf{60} & 0.73  & 0.69  & 0.71 \\
		&       & \textbf{45} & 0.62  & 0.54  & 0.57 \\
		\cmidrule{2-6}          & \multirow{2}[2]{*}{\textbf{1.5}} & \textbf{60} & 0.62  & 0.61  & 0.54 \\
		&       & \textbf{45} & 0.38  & 0.40  & 0.42 \\
		\midrule
		\multirow{4}[4]{*}{$Kt_2$: \textbf{0.3-0.6}} & \multirow{2}[2]{*}{\textbf{1.0}} & \textbf{60} & 0.52  & 0.48  & 0.43 \\
		&       & \textbf{45} & 0.38  & 0.33  & 0.27 \\
		\cmidrule{2-6}          & \multirow{2}[2]{*}{\textbf{1.5}} & \textbf{60} & 0.31  & 0.31  & 0.28 \\
		&       & \textbf{45} & 0.18  & 0.19  & 0.16 \\
		\bottomrule
	\end{tabular}%
	
	\label{tab:ubm}%
	\caption{Maximum translational velocity of the body}	
	
\end{table}%
	
%%%%%%%%%%%% Time corresponds to ub max %%%%%%%%%%%%%%%%%%%%%%%%%%%%%%%%%%%%%%%%%%%%%%
% Table generated by Excel2LaTeX from sheet 'Variables'
\begin{table}
	\centering
	\begin{tabular}{cccccc}
		\toprule
		\multicolumn{6}{c}{\text{ $t_{um}$   [ms]}} \\
		\midrule
		\textbf{$Kt$} & \multicolumn{1}{c}{\multirow{2}[4]{*}{\textbf{$M^*$}}} & $2\theta_{o}$ & \multicolumn{3}{c}{\textbf{$d^*$}} \\
		\cmidrule{4-6}    \text{[mJ / mm.rad$^2$]} &       & \multicolumn{1}{p{3em}}{\text{[Deg]}} & \textbf{1.5} & \textbf{1.0} & \textbf{0.5} \\
		\midrule
		\multirow{4}[4]{*}{\textbf{0.8-1.1}} & \multirow{2}[2]{*}{\textbf{1.0}} & \textbf{60} & 47.0  & 50.5  & 49.2 \\
		&       & \textbf{45} & 48.2  & 51.3  & 48.2 \\
		\cmidrule{2-6}          & \multirow{2}[2]{*}{\textbf{1.5}} & \textbf{60} & 62.7  & 59.5  & 52.9 \\
		&       & \textbf{45} & 70.2  & 61.5  & 57.4 \\
		\midrule
		\multirow{4}[4]{*}{\textbf{0.3-0.6}} & \multirow{2}[2]{*}{\textbf{1.0}} & \textbf{60} & 83.4  & 86.2  & 79.6 \\
		&       & \textbf{45} & 87.2  & 87.8  & 76.0 \\
		\cmidrule{2-6}          & \multirow{2}[2]{*}{\textbf{1.5}} & \textbf{60} & 108.3  & 105.5  & 91.7 \\
		&       & \textbf{45} & 120.6  & 103.1  & 96.5 \\
		\bottomrule
	\end{tabular}%
	\label{tab:tum}%
	\caption{Time corresponds to $u_m$}
\end{table}% 
	In the accelerating phase, independence of $u_b$ with $d^*$ implies that the  thrust force per unit depth, and $m_{add}$ per unit depth must be approximately constant. As it to be expected, both the spring stiffness per unit depth (Kt) and the initial clapping angle ($2 \theta_o$) influence the body velocity. The increase in spring stiffness per unit depth from $Kt_2$ to $Kt_1$, increases the $u_m$ by 1.4-2 times, whereas the time corresponding to the velocity maximum ($t_{um}$) shows a reduction from 76-121 ms to 47-70 ms (See \hyperref[tab:tum]{Table 3} and figures \hyperref[fig:ub_Kt_angle_Dyn]{10b},\hyperref[fig:ub_Kt_angle_Dyn]{10c}). A higher initial clapping angle or a lower body mass results in a larger body velocity, though the time to reach maximum velocity does not change much (see figure\hyperref[fig:ub_Kt_angle_Dyn]{10b},\hyperref[fig:ub_Kt_angle_Dyn]{10c}, and tables \hyperref[tab:tum]{3} and \hyperref[tab:ubm]{2}). A higher body mass reduces the maximum body velocity. The maximum distance covered by the bodies along the X-direction is approximately: 3 BL for spring stiffness per unit depth =  $Kt_1$ and $M^* =$ 1; 2-3BL for stiffness per unit depth = $Kt_1$ and $M^* =$ 1.5; 1.5 -2 BL for stiffness per unit depth = $Kt_2$ and $M^* =$ 1; 1-1.5BL for stiffness per unit depth = $Kt_2$ and $M^* =$ 1.5. \par

The translational velocity profiles during the acceleration phase are similar: data from all 24 cases, when plotted as $u_b/u_m$ versus $t/t_{um}$, collapse onto a single curve (figure  \hyperref[fig:ub_Acc_NonDim]{11a}). During the retardation phase, reasonable collapse is obtained when $u_b$ is scaled with $u_m$, and time is scaled with time corresponding to when the body velocity has reduced by half from its maximum value (see figure \hyperref[fig:ub_DAcc_NonDim]{11b}). 
\end{subsubsection}

\begin{subsubsection}{Angular velocity of the clapping plates}
\begin{figure}
	\centering\
	\begin{subfigure}[b]{0.4\textwidth}
		\includegraphics[width=\textwidth]{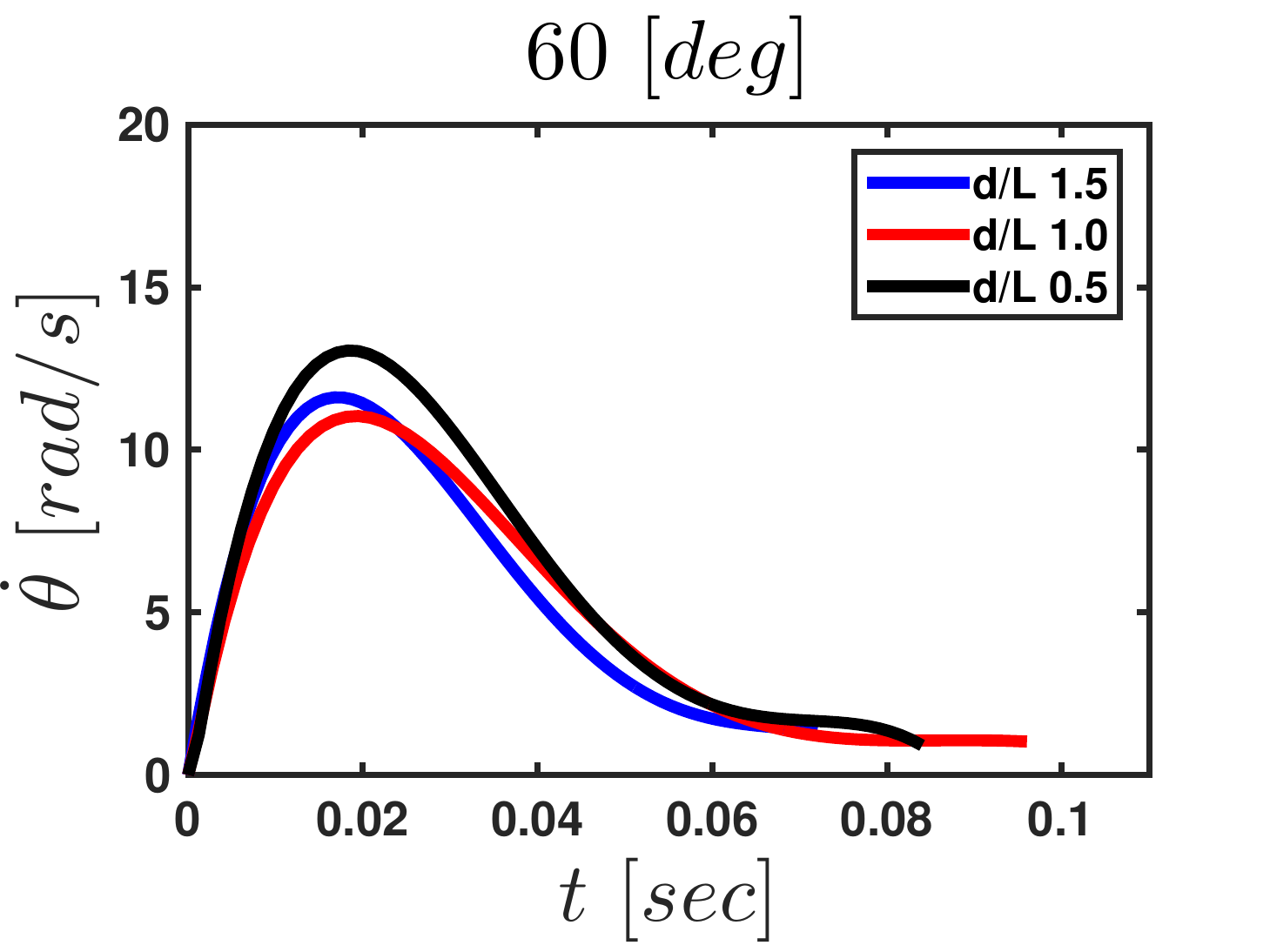}
		\caption{}
		\label{fig:AngVelo_ARconst_Dyn}
	\end{subfigure}\hspace{08mm}
	\begin{subfigure}[b]{0.4\textwidth}
		\includegraphics[width=\textwidth]{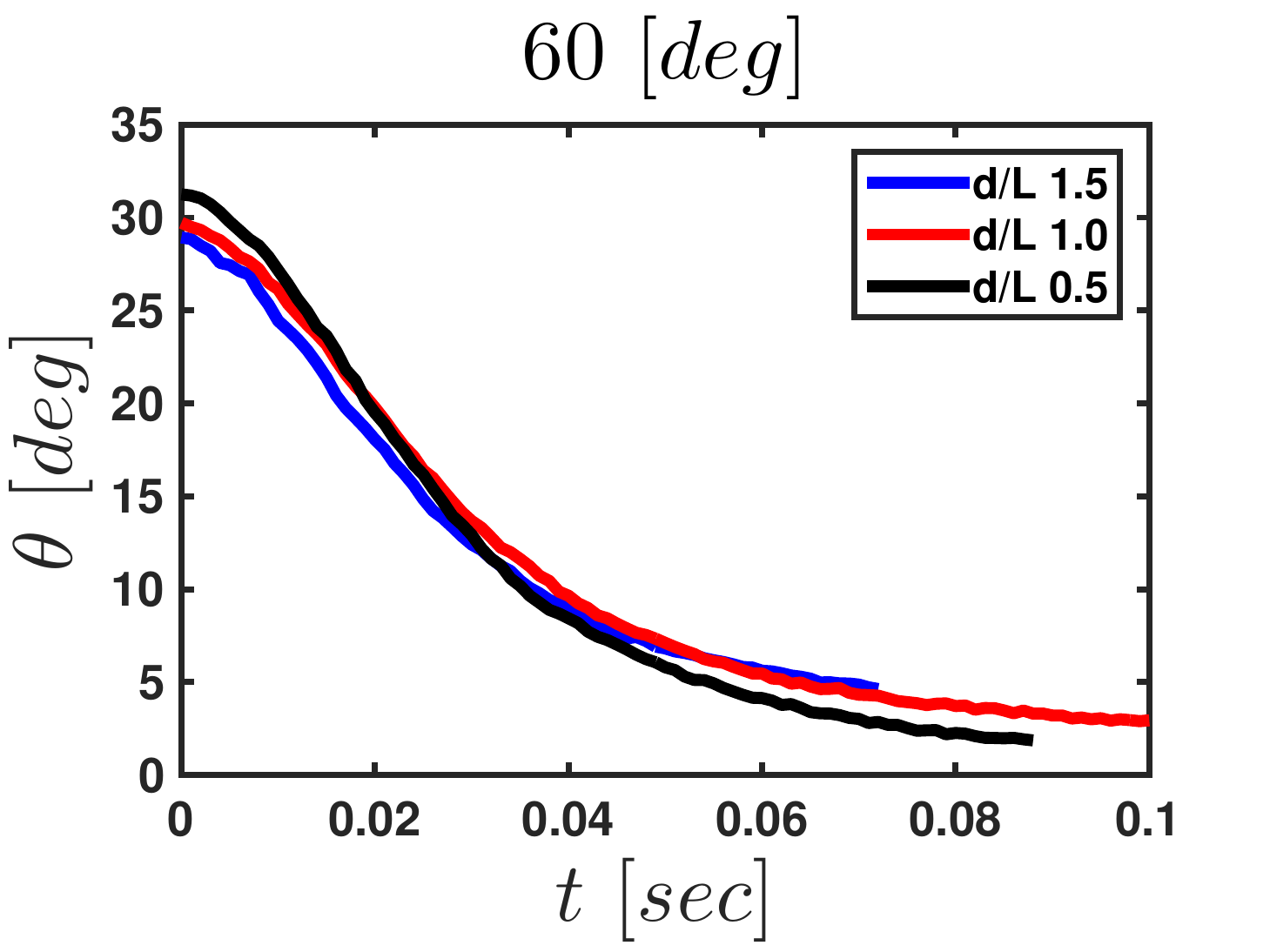}
		\caption{}
		\label{fig:Clapping angle evolution}
	\end{subfigure}\vspace{05mm}
	\begin{subfigure}[b]{0.4\textwidth}
		\includegraphics[width=\textwidth]{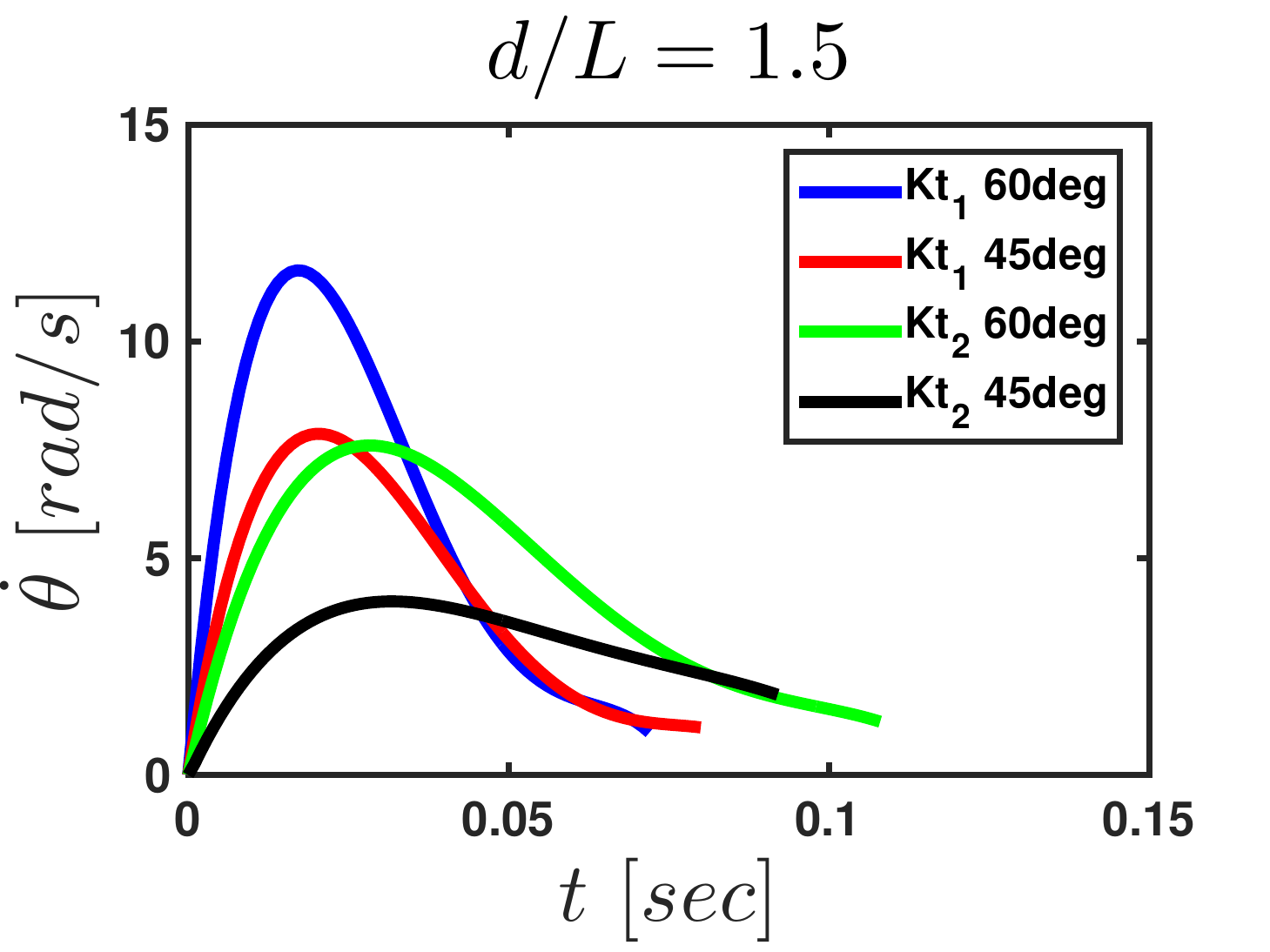}
		\caption{}
		\label{fig:AngVelo_Kt_angle_Dyn}
	\end{subfigure}\hspace{08mm}
	\begin{subfigure}[b]{0.4\textwidth}
		\includegraphics[width=\textwidth]{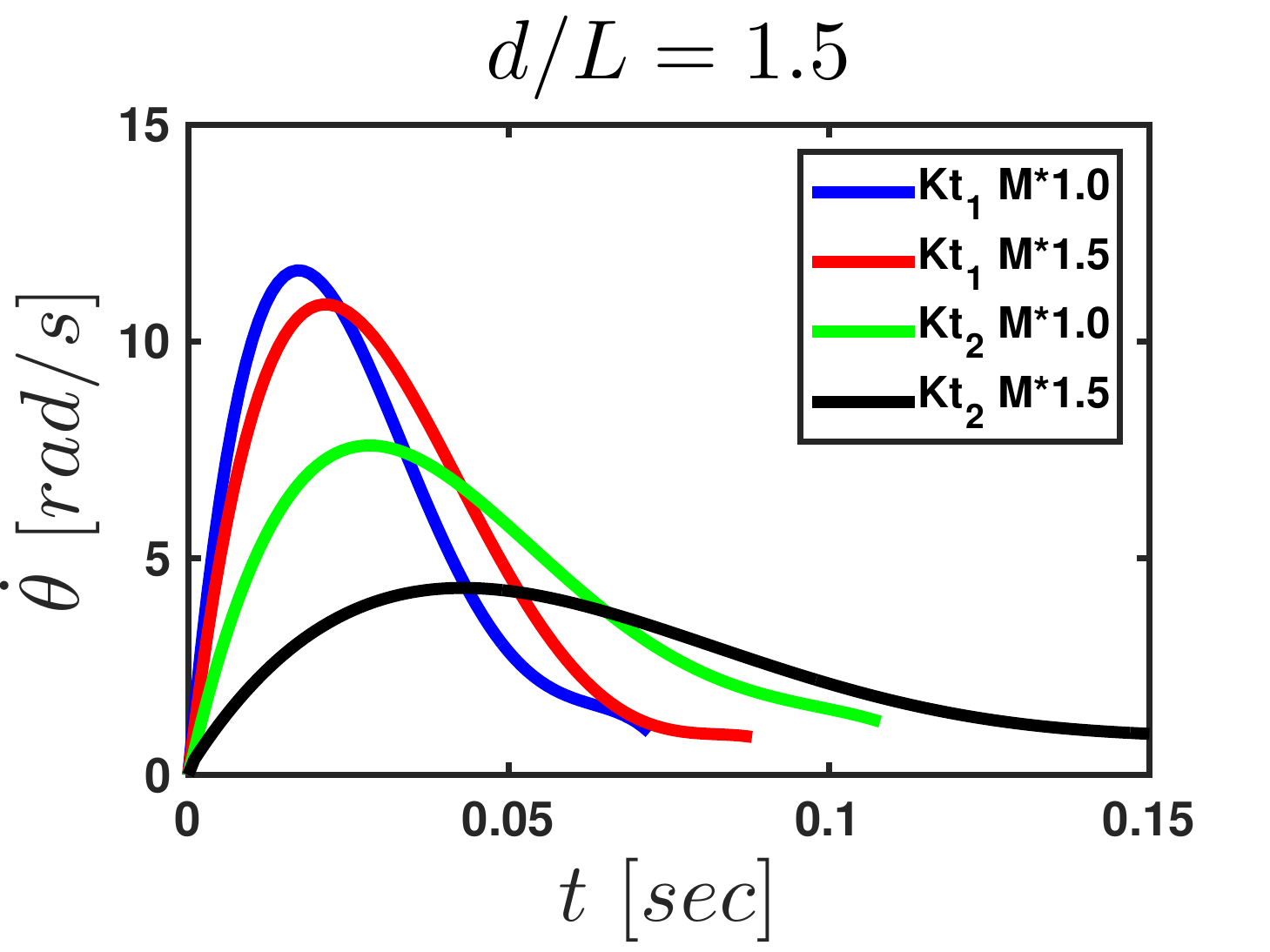}
		\caption{}
		\label{fig:AngVelo_Mstr_angle_Dyn}
	\end{subfigure}
	\caption{ (a) Variation of angular velocity $\dot{\theta}$ with time, (a) for different $d^*$ values, and with stiffness per unit depth = $Kt_1$, $2 \theta_o$= 60 deg and $M^*$= 1.0. (b) Variation in $\theta$ with time, for different $d^*$ values, and with stifness per unit depth $Kt_1$, $2 \theta_o$= 60 deg and $M^*$= 1.0. Angular velocity $\dot{\theta}$ variation with time: (c) for different values of $Kt$ and clapping angle $2 \theta_o$, and with $d^*$= 1.5 and $M^*$= 1.0. (d) for different values of $M^*$ and $Kt$, and with $d^*$= 1.5, $2 \theta_o$= 60 deg.}\label{fig:thdot_ar_mstr_angle}
\end{figure}
\begin{figure}
	\begin{subfigure}[b]{0.51\textwidth}
		\includegraphics[width=\textwidth]{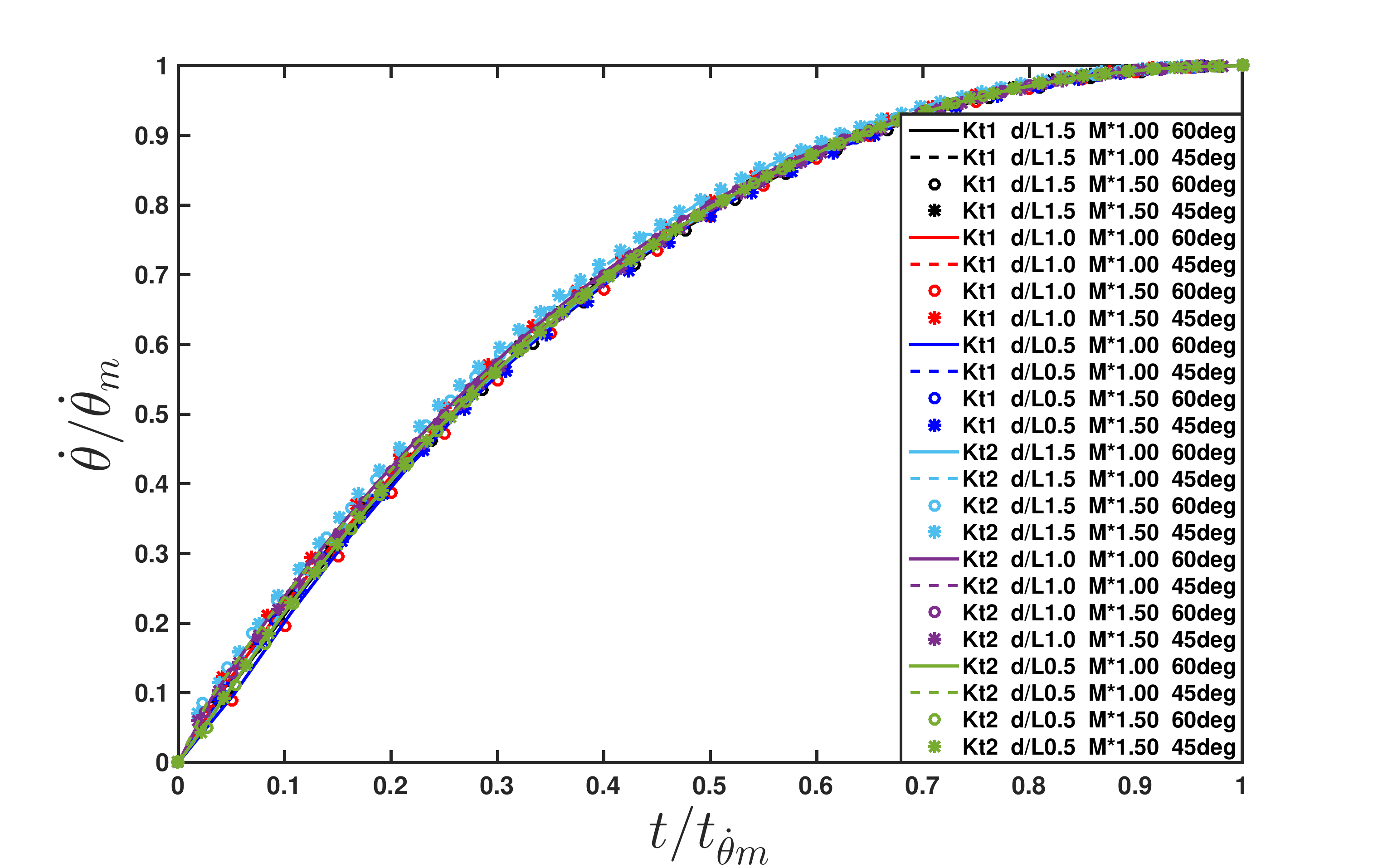}
		\caption{}
		\label{fig:thdot_Acc_NonDim}
	\end{subfigure}\hspace{02mm}
	\begin{subfigure}[b]{0.51\textwidth}
		\includegraphics[width=\textwidth]{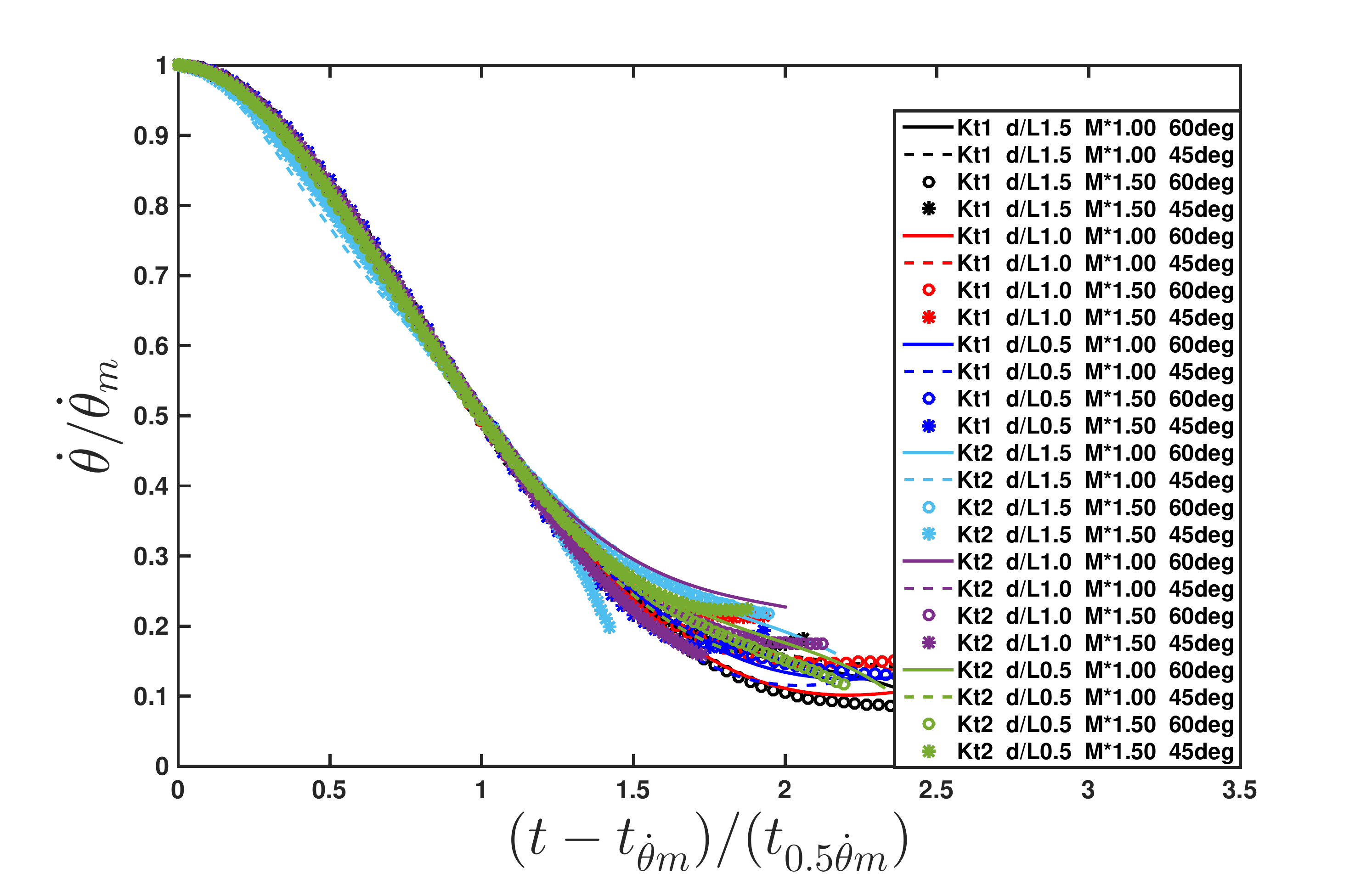}
		\caption{}
		\label{fig:thdot_DAcc_NonDim}
	\end{subfigure}\hspace{02mm}
	\caption{Normalized angular velocity versus normalized time for all 24 cases during (a) acceleration phase and (b) retardation phase : $\dot{\theta}_m$ is the maximum angular velocity, $t_{\dot{\theta}m}$ is the time when $\dot{\theta} = \dot{\theta}_m$, and $t_{0.5 \dot{\theta}m}$ is the time when $\dot{\theta} = \dot{\theta}_m /2$ during the retardation phase.}\label{fig:thdot_nonDim}
\end{figure}
 The angular velocity shows a rapid increase till it reaches a maximum ($\dot{\theta}_m$) at the time, $t_{\dot{\theta}m}$, and slower reduction to zero as the two plates come close to each other. The influence of change of the various parameters on the $\dot{\theta}$ versus time profile is similar to that observed for the $u_b$ versus time profile: the $\dot{\theta}$ curves do not change with change in $d^*$; reduction in the maximum value of angular velocity $(\dot{\theta}_m)$ is significant when $Kt$ is reduced or when $M^*$ is increased. Figures \hyperref[fig:AngVelo_ARconst_Dyn]{12a},\hyperref[fig:AngVelo_ARconst_Dyn]{12c}, and \hyperref[fig:AngVelo_ARconst_Dyn]{12d} show data on angular velocity from a selected few cases that illustrate these features. \hyperref[tab:thdot_m]{Table 4} lists the values of $(\dot{\theta}_m)$ for all the 24 cases. As in the case of $u_b$ data, the angular velocity data for each case is averaged over three experiments. The average of standard deviations in $\dot{\theta}$ is less than 7\% of the maximum angular velocity. In all the experiments, we observed symmetric clapping, both plates had the same angular velocity. The maximum angular velocity $(\dot{\theta}_m)$ ranges from 2 rad/sec to 13 rad/sec (see \hyperref[tab:thdot_m]{Table 4}). \par
 
 The rotational equilibrium of each plate is given by \eqref{eq:rot_eqm}. In the equation, applied torque $(T)$ is proportional to spring stiffness $(Kt_d)$ and semi-clapping angle $(\theta)$. Reactive torque can be given as the product angular acceleration $(\ddot{\theta})$ and total rotational inertia $(I_t)$ which is the sum of mass moment of inertia of the  plate $(I_m)$ and added inertia of water $(I_{add})$. The clapping motion involves complex three-dimensional unsteady flow due to the simultaneous translation and rotation of the plates.  $I_{add}$ is not negligible in such a flow field, but analytical expressions for it are unavailable.  \par

\begin{gather}\label{eq:rot_eqm}
 T = (I_b + I_{add}) \ \ddot{\theta}
\end{gather}

%%%%%%%%%%%%%%%%%%%%%%%%%%%% Angular velocity %%%%%%%%%%%%%%%%%%%%%%%%%%%%%%%%%%%%%%%%%%%%%
% Table generated by Excel2LaTeX from sheet 'Variables'
\begin{table}
	\centering
	
	\begin{tabular}{cccccc}
		\toprule
		\multicolumn{6}{c}{\text{$\dot{\theta}_{m}$  [rad/s]}} \\
		\midrule
		\textbf{$Kt$} & \multicolumn{1}{c}{\multirow{2}[4]{*}{\textbf{$M^*$}}} & $2\theta_{o}$ & \multicolumn{3}{c}{\textbf{$d^*$}} \\
		\cmidrule{4-6}    \text{[mJ / mm.rad$^2$]} &       & \multicolumn{1}{p{3em}}{\text{[Deg]}} & \textbf{1.5} & \textbf{1.0} & \textbf{0.5} \\
		\midrule
		\multirow{4}[4]{*}{\textbf{0.8-1.1}} & \multirow{2}[2]{*}{\textbf{1.0}} & \textbf{60} & 11.61 & 11.04 & 13.05 \\
		&       & \textbf{45} & 7.89  & 6.98  & 9.32 \\
		\cmidrule{2-6}          & \multirow{2}[2]{*}{\textbf{1.5}} & \textbf{60} & 10.86 & 11.49 & 10.83 \\
		&       & \textbf{45} & 5.70  & 6.52  & 7.86 \\
		\midrule
		\multirow{4}[4]{*}{\textbf{0.3-0.6}} & \multirow{2}[2]{*}{\textbf{1.0}} & \textbf{60} & 7.17  & 6.51  & 7.02 \\
		&       & \textbf{45} & 4.07  & 3.42  & 3.80 \\
		\cmidrule{2-6}          & \multirow{2}[2]{*}{\textbf{1.5}} & \textbf{60} & 4.31  & 5.33  & 6.11 \\
		&       & \textbf{45} & 2.18  & 2.75  & 3.18 \\
		\bottomrule
	\end{tabular}%
	\label{tab:thdot_m}%
	\caption{The maximum angular velocity of the clapping plate}
\end{table}%

The angular velocity curves are approximately independent of $d^*$ (see figure\hyperref[fig:AngVelo_ARconst_Dyn]{12a}), implying the added moment of inertia per unit depth must not vary with $d^*$ as applied torque per unit depth and plate inertia per unit depth do not vary with d*. The differences in $\dot{\theta}_m$ with $d^*$ are due to the slight and inevitable variations in $Kt_d$ in the different models (\hyperref[tab:DesignData]{Table 1}) . See \hyperref[tab:thdot_m] {Table 4}. Figure \hyperref[fig:Clapping angle evolution]{12b} shows  $\theta - t$ variations for the same three cases as in figure \hyperref[fig:AngVelo_Mstr_angle_Dyn]{12a} and show near collapse. There is some variation in the initial clapping angle, of the order of 3 degrees. \par

Change in $Kt$ produces a noticeable change in $\dot{\theta}$. The increase in spring stiffness per unit depth from $Kt_2$ to $Kt_1$ increases $\dot{\theta}_m$ by a factor of 1.6 – 2.5 and reduces the time scale over which  $\dot{\theta}$ reduces to zero. Similarly, the higher initial clap angle results in a higher  $\dot{\theta}_m$ ; the increase in $\dot{\theta}_m$ is 1.5-2 times when $\theta_o$ changes from 45 to 60 degrees (refer \hyperref[tab:thdot_m]{table 4}, figure \hyperref[fig:AngVelo_ARconst_Dyn]{12c}). \par

The influence of body mass on $\dot{\theta}_m$ is marginal, and some the variation can be attributed to the differences in the actual stiffness value for the same steel plate thickness (See figure \hyperref[fig:AngVelo_Mstr_angle_Dyn]{12d}). The maximum value of angular velocity is observed at an angular displacement of 8-11 degrees for $2\theta_o = $60 degrees, and 5-9 degrees for $2\theta_o = $45 degrees. The time ($t_{\dot{\theta}m}$) when angular velocity reaches its maximum value is most influenced by the spring stiffness per unit depth $Kt$  and not so much by in $M^*$, $d^*$, and $\theta_o$; for $Kt_1$, the $t_{\dot{\theta}m}$ lies in the range 19-28 ms, and for $Kt_2$, it is 28-53 ms. \par

As in the case of $u_b$, the $\dot{\theta}$-t profiles are similar in the acceleration and in the deceleration phases, and collapse when suitably scaled (figures \hyperref[fig:thdot_Acc_NonDim]{13a} and \hyperref[fig:thdot_DAcc_NonDim]{13b}). In the angular acceleration and retardation phase, the $\dot{\theta}$ is scaled by its maxima $\dot{\theta}_m$. For the acceleration phase time is scaled with time when angular velocity$(t_{\dot{\theta}m})$ reaches the maximum value and for the retardation phase by the time$(t_{0.5\dot{\theta} m})$ when $\dot{\theta}$ reaches half its maximum value. 
\end{subsubsection}

\begin{subsubsection}{Retardation phase}
In this phase, the clapping motion is completed, and both plates are in close contact with each other, and the drag force slows down the body. The motion of the body is modeled as a submerged body undergoing retardation due to a net drag force: 
\begin{gather}\label{eq:cd_eqm}
\left(m_b +m_{add}\ \right)\frac{d u_b}{dt_r}=-F_D 
\end{gather}

Here $m_b$ and $m_{add}$ represent the body mass and any added mass of water that is carried along with the body, which we will assume to be zero, as the plates are in close contact; $t_r$ ($=t-t_{um} \ ;\ t_r \geq 0$) represents time during the retardation phase. We may write,
\begin{gather}\label{eq:cd_Fd}
F_D=0.5\ C_d\ \rho\ A_p \ u_b^2 
\end{gather}

where by $A_p$ and $\rho$ are planform area and fluid density and $C_d$ is the drag coefficient. The solution to equations \eqref{eq:cd_eqm} and \eqref{eq:cd_Fd} subjected to the condition, $u_b=u_m$ at the start of retardation phase ($t_r$=0) is 
\begin{gather}\label{eq:ub_model}
u_b=\ \left(\phi\ t_r+\ \frac{1}{u_{m}}\right)^{-1}
\end{gather}
\begin{gather}\label{eq:phi_exp}
\phi=\frac{0.5\ C_d\ \rho A_p}{(m_b + m_{add})} 
\end{gather}

 Due to the assumption of $m_{add} \sim 0$, $C_d$ derived from \eqref{eq:phi_exp} is given as 
\begin{gather}\label{eq:cd_final}
\boxed{C_d=\frac{2\phi \  m_b}{\rho A_p}}
\end{gather}
 
  \begin{figure}
 	\centering
 	\begin{subfigure}[b]{0.4\textwidth}
 		\includegraphics[width=\textwidth]{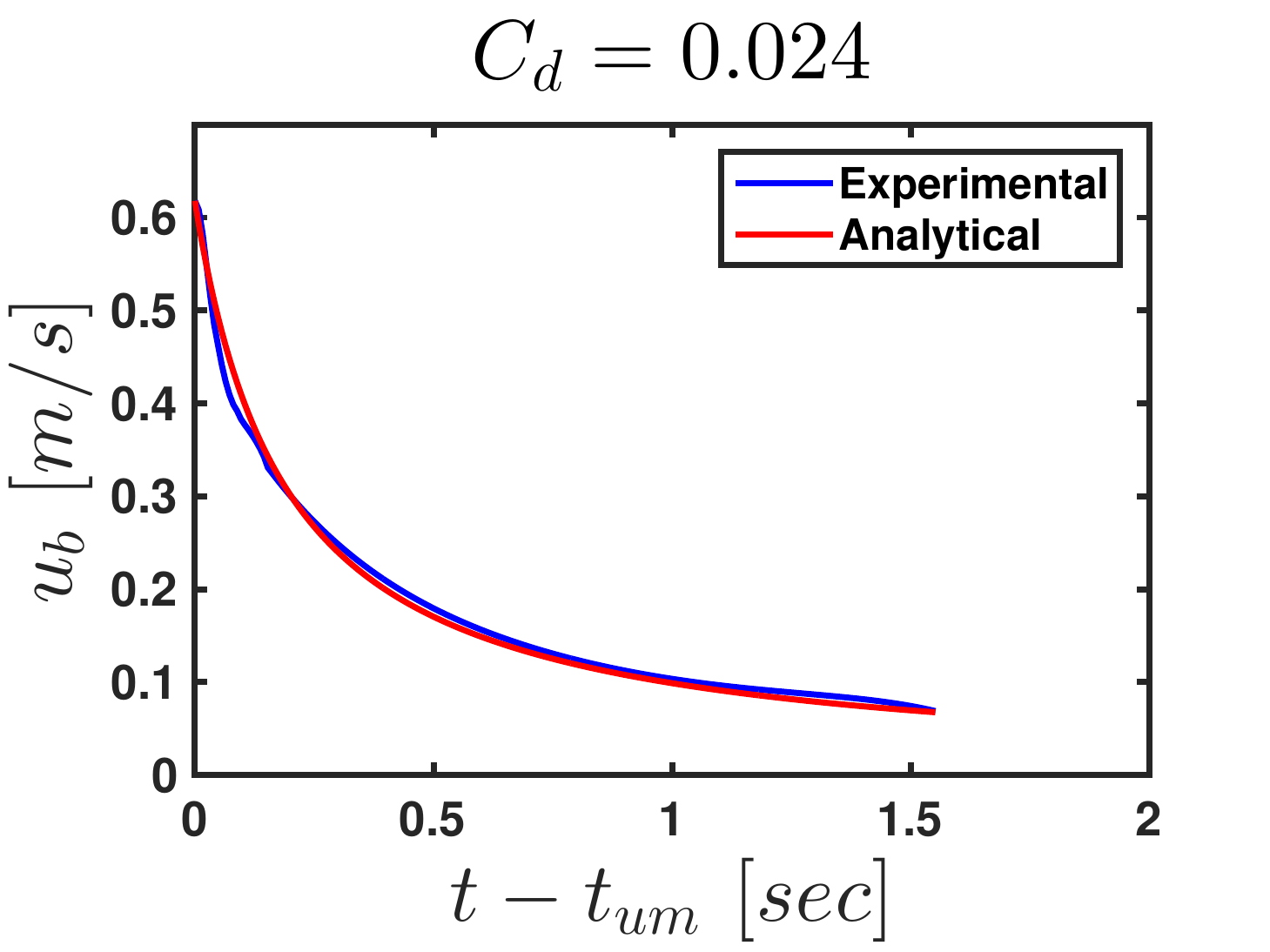}
 	\end{subfigure}
 	\caption{ The translational velocity of the body in the retardation phase for the case with $d^* =$1.5 , $Kt_1$, $M^*$ =1.0 and $2\theta_o$  = 45 deg. Blue curve represents a fit to  the experimental data, and the red curve is data obtained using equation \eqref{eq:ub_model}; $\phi = 8.5 $ for this case.}\label{fig:Cd_retardation}
 \end{figure}
 
 \begin{figure}
 	\centering
 	\begin{subfigure}[b]{0.48\textwidth}
 		\includegraphics[width=\textwidth]{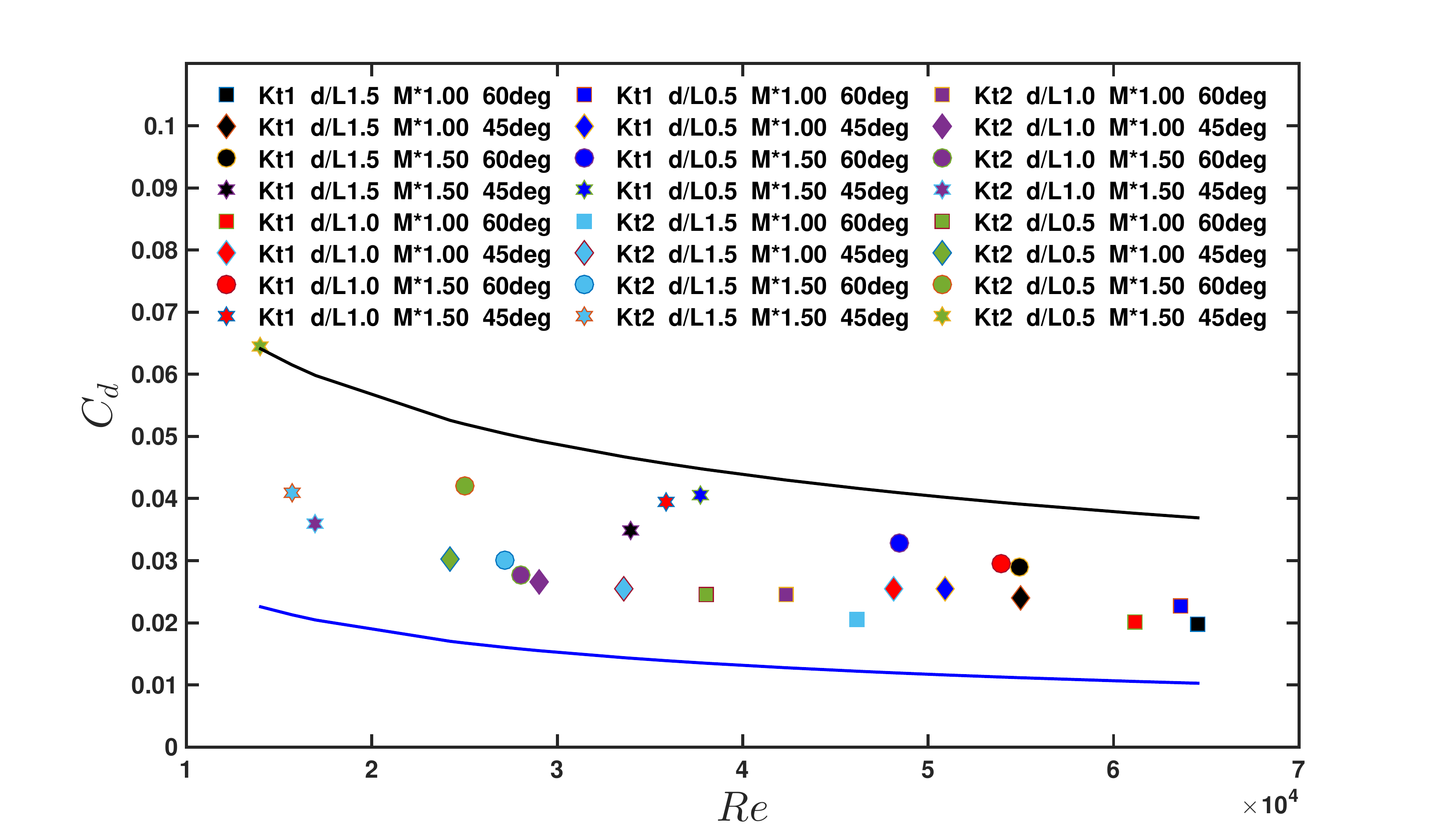}
 	\end{subfigure}
 	\caption{ The calculated values of $C_d$ plotted versus Reynolds number $Re$ for the 24 cases. The blue line represents $C_d$ for a flat plate, and the black line for 18\% thick symmetric aerofoil at zero angles of attack. The $C_d$ data for the aerofoil and the flat plate are extracted from  Munson et al.\cite{Munson13}.}\label{fig:Re_Cd}
 \end{figure}

 %%%%%%%%%%%%%%%%%%%%%%%%%%%%%%%% Drag coefficient %%%%%%%%%%%%%%%%%%%%%%%%%%%%%%%%%%%%%%%
 % Table generated by Excel2LaTeX from sheet 'Variables'
 \begin{table}
 	\centering
 	\begin{tabular}{cccccc}
 		\toprule
 		\multicolumn{6}{c}{$C_d$} \\
 		\midrule
 		\textbf{$Kt$} & \multicolumn{1}{c}{\multirow{2}[4]{*}{\textbf{$M^*$}}} & $2\theta_{o}$ & \multicolumn{3}{c}{\textbf{$d^*$}} \\
 		\cmidrule{4-6}    \text{[mJ / mm.rad$^2$]} &       & \multicolumn{1}{p{3em}}{\text{[Deg]}} & \textbf{1.5} & \textbf{1.0} & \textbf{0.5} \\	
 		\midrule
 		\multirow{4}[4]{*}{\textbf{0.8-1.1}} & \multirow{2}[2]{*}{\textbf{1.0}} & \textbf{60} & 0.020  & 0.020  & 0.023 \\
 		&       & \textbf{45} & 0.024  & 0.025  & 0.025 \\
 		\cmidrule{2-6}          & \multirow{2}[2]{*}{\textbf{1.5}} & \textbf{60} & 0.029  & 0.030  & 0.033 \\
 		&       & \textbf{45} & 0.035  & 0.039  & 0.041 \\
 		\midrule
 		\multirow{4}[4]{*}{\textbf{0.3-0.6}} & \multirow{2}[2]{*}{\textbf{1.0}} & \textbf{60} & 0.021  & 0.025  & 0.025 \\
 		&       & \textbf{45} & 0.025  & 0.0272  & 0.030 \\
 		\cmidrule{2-6}          & \multirow{2}[2]{*}{\textbf{1.5}} & \textbf{60} & 0.030  & 0.028  & 0.042 \\
 		&       & \textbf{45} & 0.041  & 0.036  & 0.064 \\
 		\bottomrule
 	\end{tabular}%
 	\label{tab:Cd}%
 	\caption{Drag coefficient of the body in the retardation phase}
 \end{table} 
 
For each case, the value of $\phi$ is obtained by fitting the experimental data of $u_b$ versus $t_r$ in equation \eqref{eq:ub_model}, from which value of $C_d$ is calculated using equation  \eqref{eq:cd_final}. A sample case ($d^* =$1.5 , $Kt_1$, $M^*$ =1.0 and $2\theta_o$  = 45 deg) is shown in figure \hyperref[fig:Cd_retardation]{14}, which shows equation \eqref{eq:ub_model} is a good model; for this case, we get $\phi$= 8.5, and $C_d =$0.024. The Reynolds number $Re$ ($=L \ u_m / \nu$) for this case is 5.5 $\times 10^4$ . For comparison, at the same $Re$, $C_d$ for a flat plate is 0.011  and that for a symmetric aerofoil with 18\% thickness to chord ratio is 0.039 ; the $C_d$ value of the body lies between these two values. 
\hyperref[tab:Cd]{Table 5} lists the $C_d$ values, and figure \hyperref[fig:Re_Cd]{15} shows the $Cd$ versus $Re$ for the 24 cases. We notice that the $C_d$ values for $M^*$ = 1.5 bodies are higher due to the presence of the canopy, and closer to the $C_d$ value for an aerofoil. The $C_d$ values lie between 0.02 to 0.04 for the entire parametric space, with the exception of the body corresponding to $M^* =$ 1.5, spring stiffness per unit depth = $Kt_2$ and $2\theta_o =$ 45degrees, which shows $C_d =$ 0.06; a slight bulge in the canopy and increased frontal area explains the higher $C_d$.    
 
\end{subsubsection}
\end{subsection}

\begin{subsection}{Wake dynamics}
\label{sec:WakeDynamics}
	In this section, we characterize the wake of the clapping body from dye visualizations in the XY plane and from the 2D PIV data obtained in the XY and XZ planes. The PIV data from the mutually perpendicular planes allows us to get the approximate structure of three-dimensional vortex loops, which are discussed in \S\hyperref[sec:Vorticity_Y]{3.2.2}. The effect of parametric variations on vorticity field, core separation, and circulation is presented in the following sub-sections.
	
\begin{subsubsection}{Vorticity fields in XY plane: $\omega_z$}
	\label{sec:Vorticity_Z}
	The PIV fields at different instants after the start of clapping correspond to the body with the higher stiffness per unit depth ($Kt_1$), $M^*=$ 1, $d^*=$1.5 and $2\theta_o$=60 deg are shown in the figure \hyperref[fig:VortZ_Dyn]{16}, where gray color is used to mark shadow, and yellow is used to identify the rotating portion of the clapping plate. The red and blue patches identify the region with non-zero vorticity, whereas green represents regions with approximately zero vorticity. \par 
	
	Opposite-signed vortices begin to form at the trailing edges of the two plates soon after the motion starts; these are clearly seen at t = 0.0195 s (figure \hyperref[fig:VortZ_Dyn]{16b}). As the body propels forward, the vortices detach and are left behind. As we shall see below, the circulation around each vortex increases rapidly during the initial time, and by t = 0.0395 s (figure \hyperref[fig:VortZ_Dyn]{16c}) it would have reached its peak value. During clapping and just after, as the body moves forward, there is hardly any change in the location and strength of the vortex; at t = 0.1395 s, the body has moved about 75 mm, whereas the vortex pair movement is only about 5 mm (figure \hyperref[fig:VortZ_Dyn]{16i}). The large difference in the velocities of the body and vortex-pair is observed across all the cases in the parametric space.  \par
	
	\begin{figure}
		\centering
		\begin{subfigure}[b]{0.25\textwidth}
			\includegraphics[width=\textwidth]{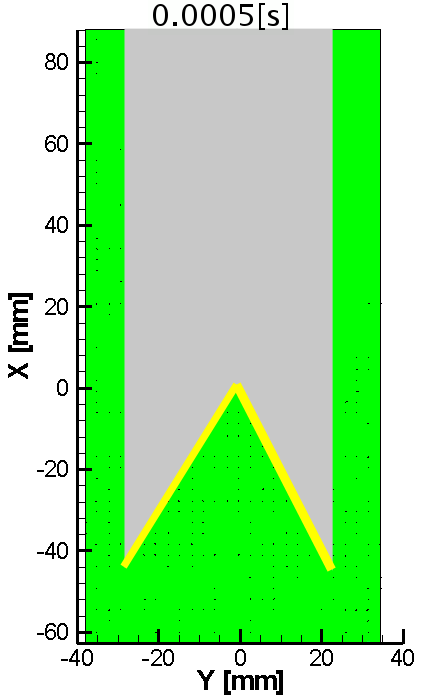}
			\caption{}	
		\end{subfigure}\hspace{05mm}
		\begin{subfigure}[b]{0.25\textwidth}
			\includegraphics[width=\textwidth]{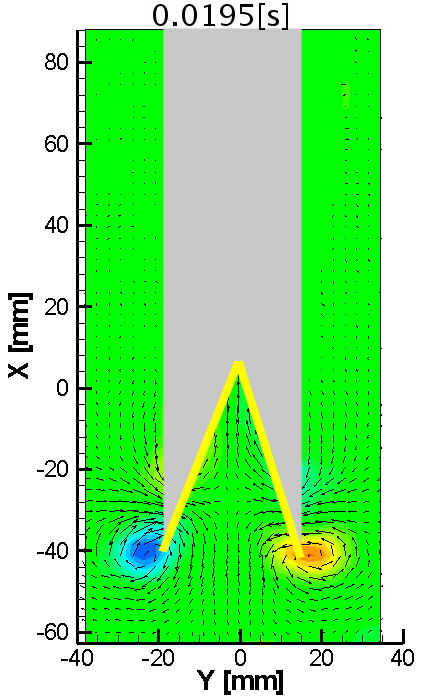}
			\caption{}
		\end{subfigure}\hspace{05mm}
		\begin{subfigure}[b]{0.25\textwidth}
			\includegraphics[width=\textwidth]{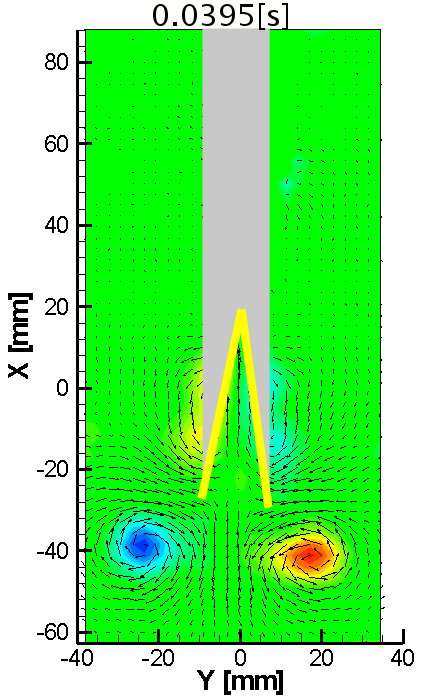}
			\caption{}
		\end{subfigure}\vspace{03mm}
		\begin{subfigure}[b]{0.25\textwidth}
			\includegraphics[width=\textwidth]{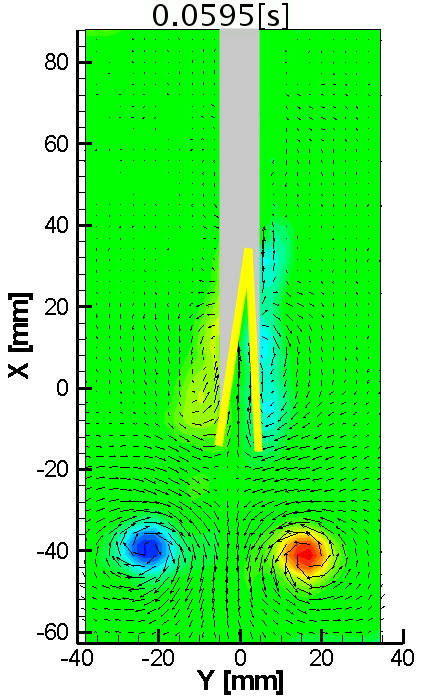}
			\caption{}
		\end{subfigure}\hspace{05mm}
		\begin{subfigure}[b]{0.25\textwidth}
			\includegraphics[width=\textwidth]{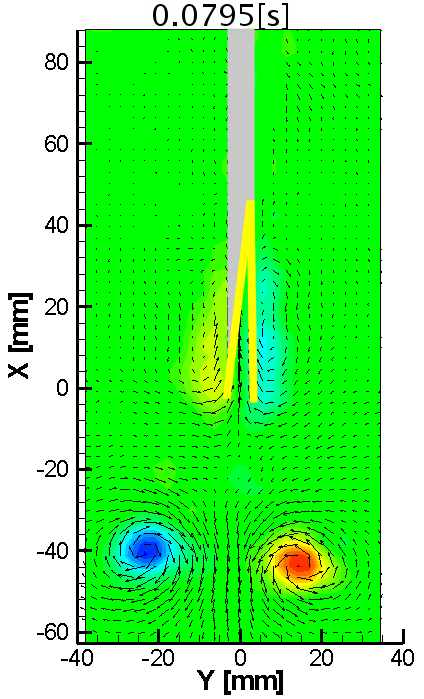}
			\caption{}
		\end{subfigure}\hspace{05mm}
		\begin{subfigure}[b]{0.25\textwidth}
			\includegraphics[width=\textwidth]{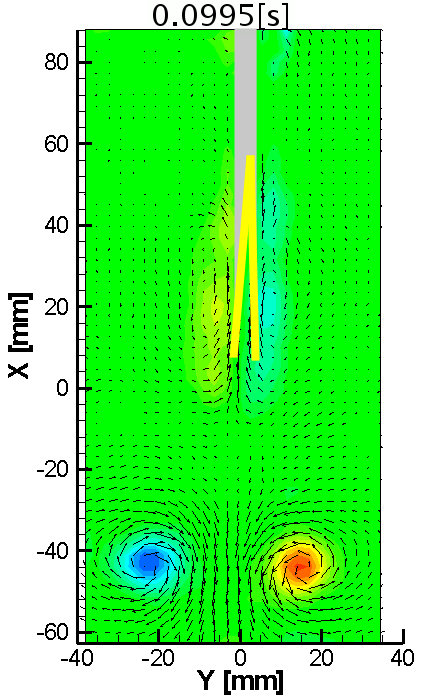}
			\caption{}
		\end{subfigure}\vspace{03mm}
		\begin{subfigure}[b]{0.25\textwidth}
			\includegraphics[width=\textwidth]{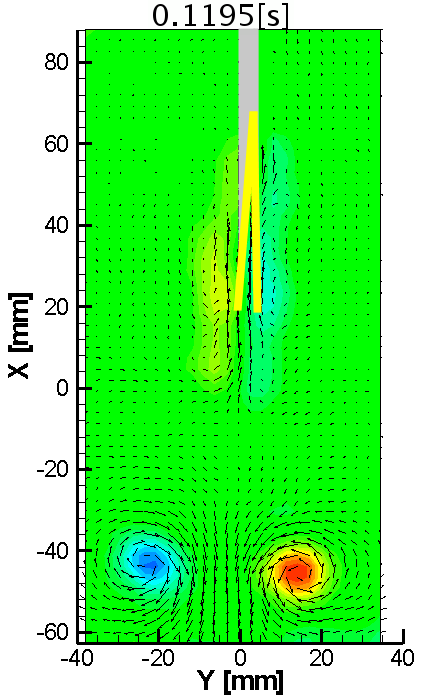}
			\caption{}
		\end{subfigure}\hspace{05mm}
		\begin{subfigure}[b]{0.25\textwidth}
			\includegraphics[width=\textwidth]{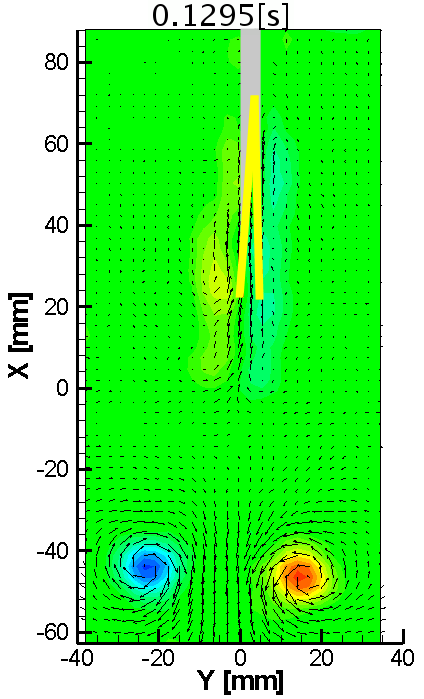}
			\caption{}
		\end{subfigure}\hspace{05mm}
		\begin{subfigure}[b]{0.25\textwidth}
			\includegraphics[width=\textwidth]{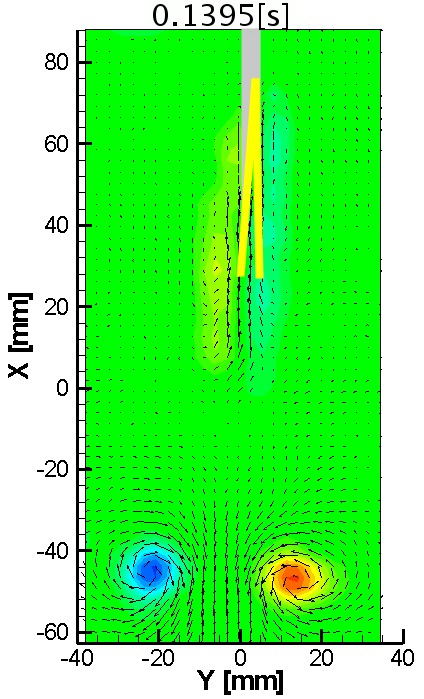}
			\caption{}
		\end{subfigure}\vspace{03mm}
		\begin{subfigure}[b]{0.25\textwidth}
			\includegraphics[width=\textwidth]{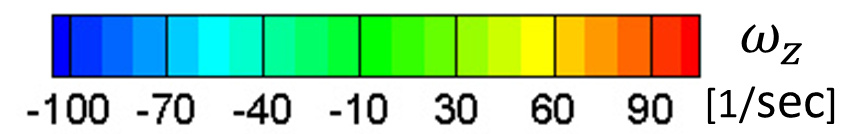}
		\end{subfigure}	
		\caption{Z-component vorticity fields at different time instants for the body with stiffness per unit depth = $Kt_1$, $M^*$ = 1.0, $2\theta_o$ = 60 deg and $d^* =$1.5. }\label{fig:VortZ_Dyn}
	\end{figure}

	The velocity field shows some important features. Two distinct regions are seen during the initial clapping phase (figures \hyperref[fig:VortZ_Dyn]{16b, c}), one where the fluid has an x-velocity component in the forward direction and the other where the fluid, which is in between the vortices, is moving in the opposite direction. At later times, when the vortex pair has separated from the body, fluid between the plates essentially moves with the body. A distinct wake also can be seen just behind the body (figures \hyperref[fig:VortZ_Dyn]{16f -i}). However, the vortex patches are isolated with no trailing jet connected to the body. Such type of wake has also been observed in fast-swimming jellyfish by by Dabiri et al.\cite{Dabiri06} and in the squid by Bartol et al.\cite{Bartol2D09}. Most of the flow features during and just after the clapping described for the case shown in figure \hyperref[fig:VortZ_Dyn]{16} are observed for other cases in the parametric space. Main differences arise in the evolution of the vortex loops, significantly when the body aspect ratio ($d^*$) changes.\par

	At the end of the clapping action, the momentum of the body and of the fluid moving forward with it must be equal to the momentum of the fluid moving in the opposite direction in the wake. If we consider the time when the body has reached maximum velocity, we get \eqref{eq:linMom}.
	
	\begin{gather}\label{eq:linMom}
	(m_b + m_{add}) \ u_m = m_v \ u_v
	\end{gather}
	In the above equation, the $m_{add}$ is the added mass associated with the body, and $m_v$ is the mass of the vortex pair. In all the cases, we found $u_m$ is always higher than $u_v$, implying that  $m_v$ is always higher than total body mass($= m_b + m_{add}$). Refer \hyperref[tab:ubm]{Table 2} and \hyperref[tab:uv]{Table 6}. The steady velocity $u_v$ is in the range 12 - 21 cm/s for $Kt_1$ and 4 - 9 cm/s for $Kt_2$. See \hyperref[tab:uv]{Table 6}. The velocity is calculated during the time period when the displacement of each vortex has negligible displacement in the lateral (Y) direction; this time period is marked by two dashed lines in figure \hyperref[fig:CoreDsip_Dyn]{9b}. The influence of change in $d^*$, $2\theta_o$, and $M^*$  on $u_v$ is not much.\par
	
	%%%%%%%%%%%%%%%%%%%%%%%%%%%%%%%%%%%% Uv max  %%%%%%%%%%%%%%%%%%%%%%%%%%%%%%%%%%%%%%%%%%%%
	% Table generated by Excel2LaTeX from sheet 'Variables'
	
	\begin{table}
		\centering
		
		\begin{tabular}{cccccc}
			\toprule
			\multicolumn{6}{c}{\text{ $u_v$   [m/s]}} \\
			\midrule
			\textbf{$Kt$} & \multicolumn{1}{c}{\multirow{2}[4]{*}{\textbf{$M^*$}}} & $2\theta_{o}$ & \multicolumn{3}{c}{\textbf{$d^*$}} \\
			\cmidrule{4-6}    \text{[mJ / mm.rad$^2$]} &       & \multicolumn{1}{p{3em}}{\text{[Deg]}} & \textbf{1.5} & \textbf{1.0} & \textbf{0.5} \\	
			\midrule
			\multirow{4}[4]{*}{\textbf{0.8-1.1}} & \multirow{2}[2]{*}{\textbf{1.0}} & \textbf{60} & 0.14  & 0.15  & 0.18 \\
			&       & \textbf{45} & 0.13  & 0.12  & 0.19 \\
			\cmidrule{2-6}          & \multirow{2}[2]{*}{\textbf{1.5}} & \textbf{60} & 0.17  & 0.19  & 0.20 \\
			&       & \textbf{45} & 0.12  & 0.15  & 0.21 \\
			\midrule
			\multirow{4}[4]{*}{\textbf{0.3-0.6}} & \multirow{2}[2]{*}{\textbf{1.0}} & \textbf{60} & 0.08  & 0.08  & 0.08 \\
			&       & \textbf{45} & 0.04  & 0.07  & 0.08 \\
			\cmidrule{2-6}          & \multirow{2}[2]{*}{\textbf{1.5}} & \textbf{60} & 0.07  & 0.09  & 0.09 \\
			&       & \textbf{45} & 0.06  & 0.07  & 0.08 \\
			\bottomrule
		\end{tabular}%
		\label{tab:uv}%
		\caption{Steady translational velocity of wake votices.}
	\end{table}%
	
\end{subsubsection}
 
\begin{subsubsection}{Vorticity field in XZ plane: $\omega_y$}
	\label{sec:Vorticity_Y}
	
	The flow field in the XZ plane at Y = 0 reveals the three-dimensional structure of the vortex loop. The clapping action results in a high-pressure region between the plates that produces not only a downstream jet but also jets from the top and bottom sides of the inter-plate cavity. Vorticity shed from the top and bottom edges of the plates finally reconnect to form vortex loops whose configurations depend mainly on the aspect ratio $d^*$. The starting vortices seen in the XY plane are cross-sections of these vortex loops. Figures \hyperref[fig:3D Vortex loop_d*1.5]{17}, \hyperref[fig:3D Vortex loop_d*1.0]{18} and \hyperref[fig:3D Vortex loop_d*0.5]{19}
	show schematics of the vortex loops that form for the bodies with the three different aspect ratios. These schematics are based on the PIV measurements in both planes and the dye visualizations. In the case of $d^*$ = 1.5, the length of the major axis and the minor axis is approximately 100mm and 50mm at 400ms (figure \hyperref[fig:3D Vortex loop_d*1.5]{17a}); at 1300ms, axis switching is complete with the major axis in the horizontal direction and the minor axis in the vertical direction (figure \hyperref[fig:3D Vortex loop_d*1.5]{17b}). For $d^*$ = 1.0 case, at 400ms, the wake vortex loop has major and minor axes of 75mm and 35mm (figure \hyperref[fig:3D Vortex loop_d*1.0]{18a}), and at 700ms, the minor axis (= 45mm) is in the vertical direction and the major axis (= 75mm) is horizontal (figure \hyperref[fig:3D Vortex loop_d*1.0]{18b}). Axis switching is a well-known phenomenon (see for example, Dhanak et al.\cite{Dhanak81} and Cheng et al.\cite{MCheng16}). The wake corresponding to $d^*$ = 0.5 case shows a more complex structure: at 200ms, six prominent vortex loops (ringlets) are observed (figure \hyperref[fig:3D Vortex loop_d*0.5]{19a}), of which only two vortex loops existed futher (figure \hyperref[fig:3D Vortex loop_d*0.5]{19b}). In all three $d^*$ configurations, during the clapping action flow ejected in $\pm$Z direction results vortex loop being formed on both the top and bottom sides of the body, see figures \hyperref[fig:3D Vortex loop_d*1.5]{17a}, \hyperref[fig:3D Vortex loop_d*1.0]{18a}, and \hyperref[fig:3D Vortex loop_d*0.5]{19a}. \par
	 
		\begin{figure}
		\centering
		\begin{subfigure}[b]{0.28\textwidth}
			\includegraphics[width=\textwidth]{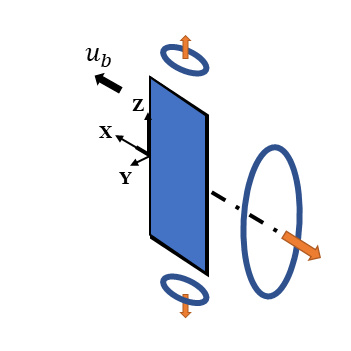}
			\caption{}
			\label{fig:3DVort_AR067_Dyn_01}
		\end{subfigure}\hspace{20mm}
		\begin{subfigure}[b]{0.28\textwidth}
			\includegraphics[width=\textwidth]{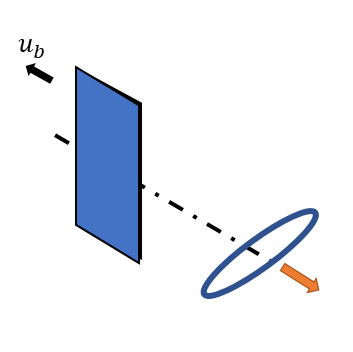}
			\caption{}
			\label{fig:3DVort_AR067_Dyn_02}
		\end{subfigure}
		\caption{Schematic showing a  elliptical vortex loop behind the $d^*$ = 1.5 body switching axis as it moves downstream. Figure (a) and (b) shows wake at 400ms and 1300ms respectively.}\label{fig:3D Vortex loop_d*1.5}
	\end{figure}

	\begin{figure}
		\centering
		\begin{subfigure}[b]{0.28\textwidth}
			\includegraphics[width=\textwidth]{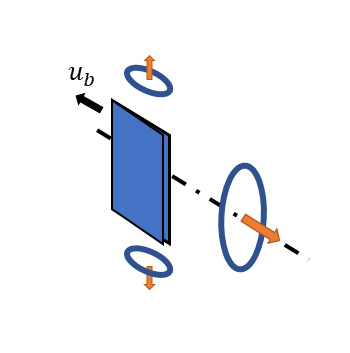}
			\caption{}
			\label{fig:3DVort_AR100_Dyn_01}
		\end{subfigure}\hspace{15mm}
		\begin{subfigure}[b]{0.28\textwidth}
			\includegraphics[width=\textwidth]{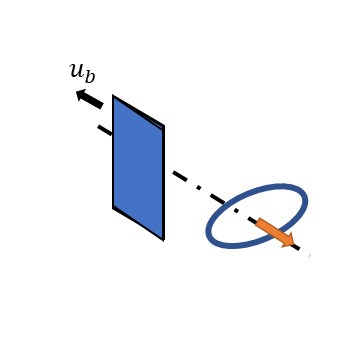}
			\caption{}
			\label{fig:3DVort_AR100_Dyn_02}
		\end{subfigure}
		\caption{The wake vortex loop for the body with $d^*$ = 1.0 at (a) 400ms and (b) 700ms. Here also axis switching is observed.}\label{fig:3D Vortex loop_d*1.0}
	\end{figure}
	\begin{figure}
		\centering
		\begin{subfigure}[b]{0.28\textwidth}
			\includegraphics[width=\textwidth]{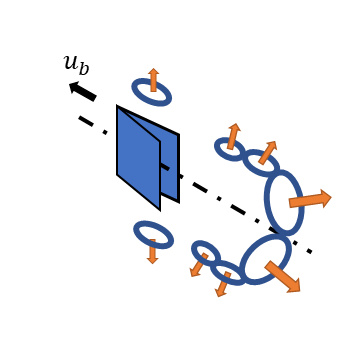}
			\caption{}
			\label{fig:3DVort_AR200_Dyn_01}
		\end{subfigure}\hspace{15mm}
		\begin{subfigure}[b]{0.28\textwidth}
			\includegraphics[width=\textwidth]{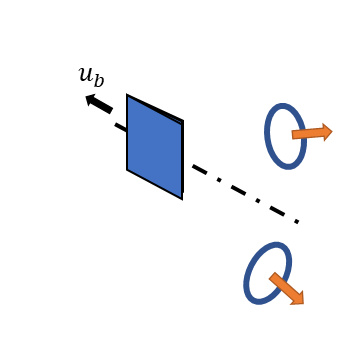}
			\caption{}
			\label{fig:3DVort_AR200_Dyn_02}
		\end{subfigure}
		\caption{The wake for the body with $d^*$ = 0.5 has multiple ringlets (a) at 200 ms (b) out of which only two survive at a later time = 360 ms.}\label{fig:3D Vortex loop_d*0.5}
	\end{figure}

	Figure \hyperref[fig:Vorticity-Y]{20} shows the flow fields at different instants in the XZ plane for $d^*$ = 1.5 (\hyperref[fig:Vorticity-Y]{20a,d}), for $d^*$ = 1.0 (\hyperref[fig:Vorticity-Y]{20b,e}) and for $d^*$ = 0.5 (\hyperref[fig:Vorticity-Y]{20c,f}). The gray thick vertical line represents a stationary rod of 6mm diameter, which is part of the release stand. The $d^*$ = 1.5 case corresponds to the flow field shown in figure \hyperref[fig:VortZ_Dyn]{16} in the XY plane. At t = 0.3175 s (Figure \hyperref[fig:Vorticity-Y]{20a}), when clapping action is complete, we see two oppositely signed vorticity patches on either side of a high-velocity region; at this time, the body is out of the picture, at x $\approx$ +14 cm. These two vortex patches connect to the vortices in the XY plane as shown in figure \hyperref[fig:VortZ_Dyn]{16} to form an elliptical vortex loop. The loop essentially forms after the reconnection of independent vortex loops associated with each clapping plate. A detailed discussion on vortex reconnection is found in Kida et al.\cite{Kida89} and Melander et al.\cite{Melander89}.  The DDPIV for stationary pair clapping plates performed by Kim et al.\cite{Kim13} also shows similar vortex reconnection. For $d^*$ =1.5 case, due to the  axis switching (figure  \hyperref[fig:3D Vortex loop_d*1.5]{17a,b}), the vortex patches in Y-Z plane move closer with time  (figure \hyperref[fig:Vorticity-Y]{20a}) and (figure \hyperref[fig:Vorticity-Y]{20d}). \par
		
	\begin{figure}
		\centering
		\begin{subfigure}[b]{0.32\textwidth}
			\includegraphics[width=\textwidth]{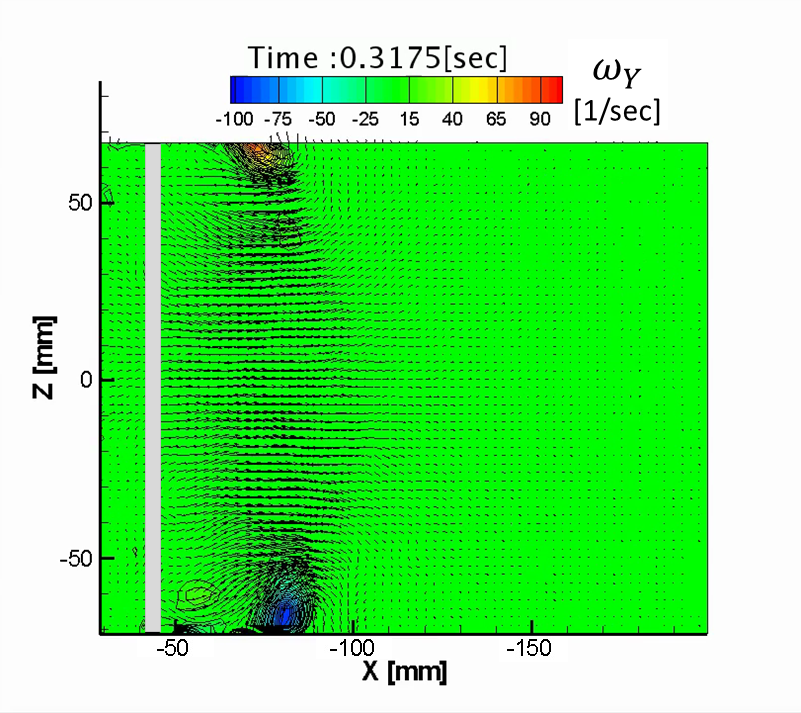}
			\caption{}
			\label{fig:VortY_AR067_Dyn_01}
		\end{subfigure}\hspace{0mm}
		\begin{subfigure}[b]{0.32\textwidth}
			\includegraphics[width=\textwidth]{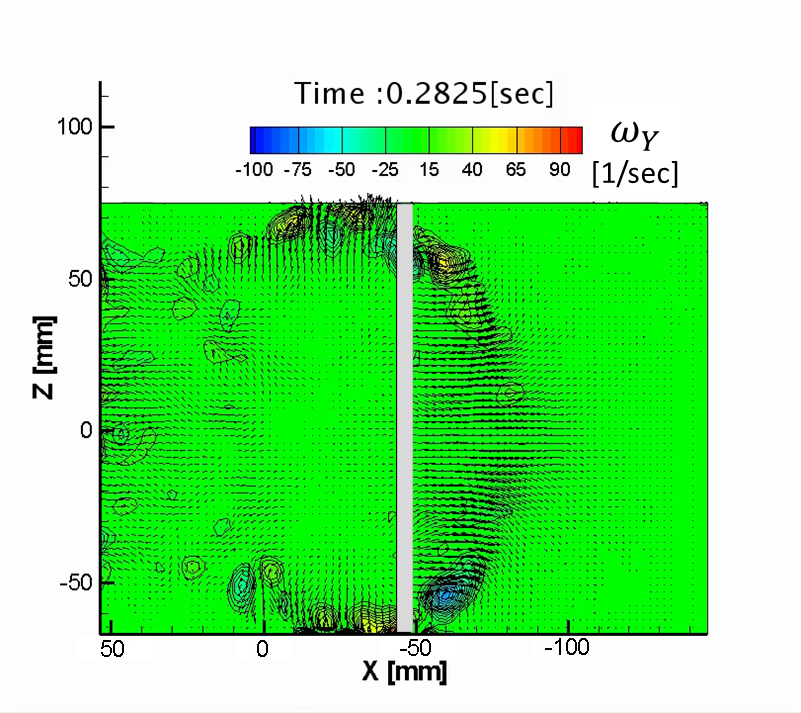}
			\caption{}
			\label{fig:VortY_AR100_Dyn_01}
		\end{subfigure}\hspace{0mm}
		\begin{subfigure}[b]{0.32\textwidth}
			\includegraphics[width=\textwidth]{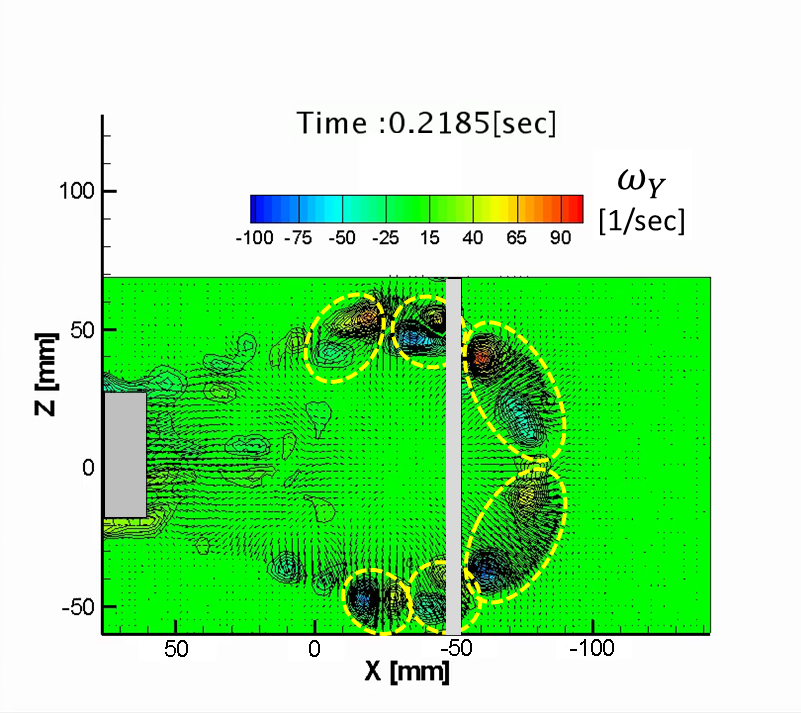}
			\caption{}
			\label{fig:VortY_AR200_Dyn_01}
		\end{subfigure}\vspace{05mm}
		\begin{subfigure}[b]{0.32\textwidth}
			\includegraphics[width=\textwidth]{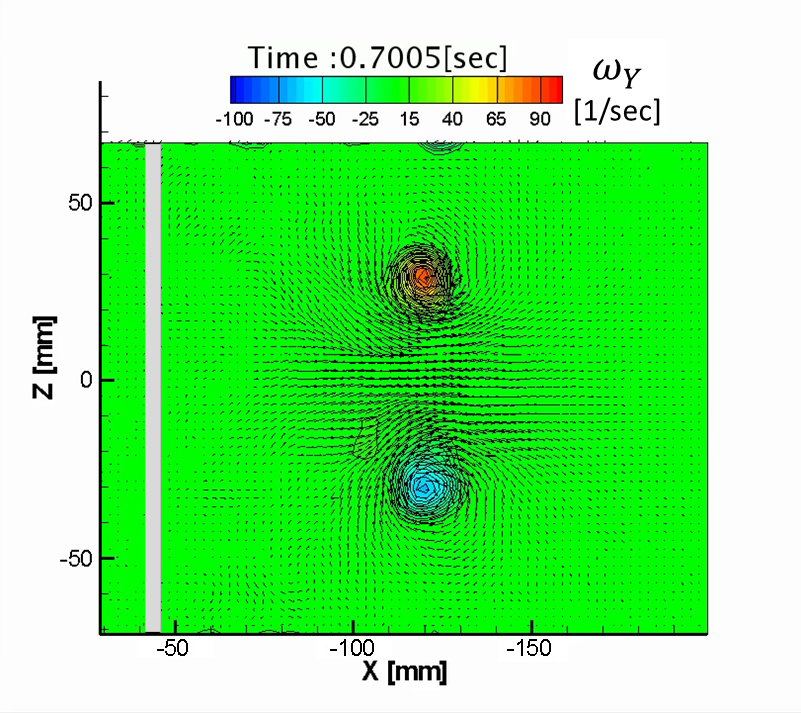}
			\caption{}
			\label{fig:VortY_AR067_Dyn_02}
		\end{subfigure}\hspace{0mm}
		\begin{subfigure}[b]{0.32\textwidth}
			\includegraphics[width=\textwidth]{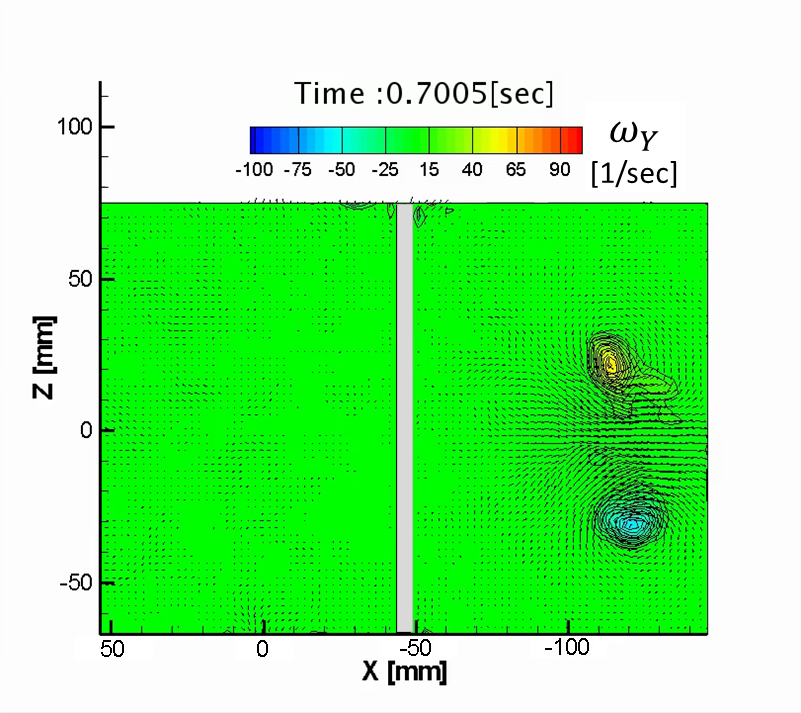}
			\caption{}
			\label{fig:VortY_AR100_Dyn_02}
		\end{subfigure}\hspace{0mm}
		\begin{subfigure}[b]{0.32\textwidth}
			\includegraphics[width=\textwidth]{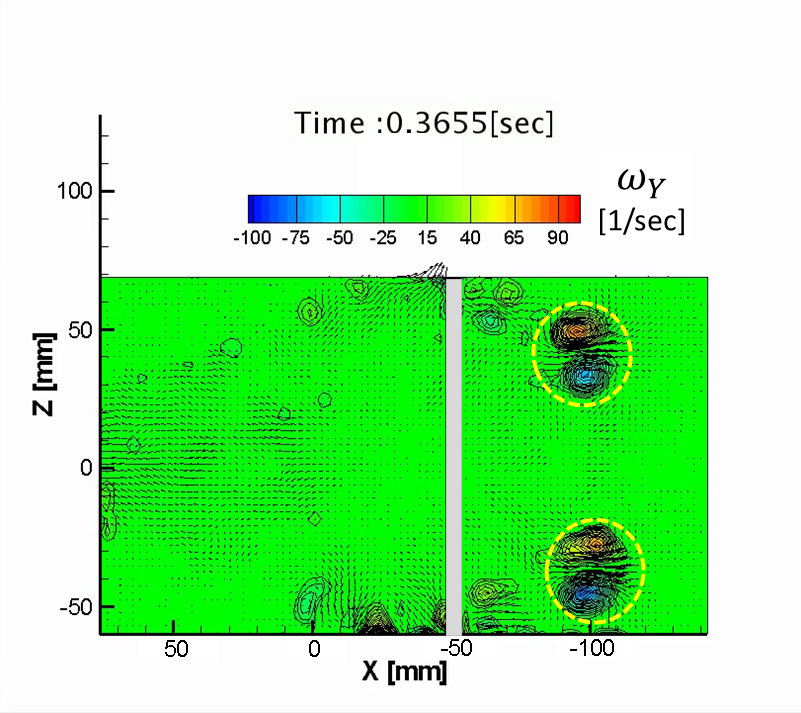}
			\caption{}
			\label{fig:VortY_AR200_Dyn_02}
		\end{subfigure}
		\caption{Flow fields in XZ plane at different instants corresponding to the body with $d^*$ = 1.5 (figures a-d), $d^*$ = 1.0 (figures b-e), and $d^*$ = 0.5 (figures c-f). Other  parameters:  stiffness per unit depth = $Kt_1$, $M^*$ = 1.0, and $2\theta_o$ = 60 deg.}\label{fig:Vorticity-Y}
	\end{figure}

	The flow in the wake for the $d^*$ = 1.0 case (figure \hyperref[fig:Vorticity-Y]{20b}) shows the strong backward flow and two vortex patches that are less distinct than for the $d^*$ = 1.5 case. In addition, we can observe outward flow in the Z-direction, created at the top and bottom edges. Towards the left of the picture, forward flow associated with the immediate wake of the body is seen. At a later time (figure \hyperref[fig:Vorticity-Y]{20e}), two distinct vortex patches are seen, which is the cross-section in the XZ plane of the elliptical loop with horizontal major axis as shown in figure \hyperref[fig:3D Vortex loop_d*1.0]{18b}. \par
	
	For the lowest body depth ($d^*$ = 0.5), the initial flow behind the body shows the six vortex-patch pairs (ringlets) (figure \hyperref[fig:Vorticity-Y]{20c}). Later only two vortex ringlets (figure \hyperref[fig:3D Vortex loop_d*0.5]{19}) remain, seen as two pairs of vortex patches in the XZ plane (figure \hyperref[fig:Vorticity-Y]{20f}). Similar vortex ringlets are  observed in the studies of the head-on collision of two vortex rings (Lim et al.\cite{Lim92} and Cheng et al.\cite{MCheng18}). \par

\end{subsubsection}

\begin{subsubsection}{Strength of the starting vortices: }
	\label{sec:circulation}
	In this section, we present the data on circulation obtained from PIV for the various cases. Circulation is calculated using \eqref{eq:circulation}, with cutoff values $\omega\leq 0.05\ \omega_{max}$ for $Kt_1$ and $\omega\leq 0.07\ \omega_{max}$ for $Kt_2$. For each case in parametric space, the circulation is averaged over three experiments, where the time average of standard deviation is less than 12\% of the maximum circulation($\Gamma_m$). It helps to view the evolution of circulation keeping in mind the vortex structures depicted in figures \hyperref[fig:3D Vortex loop_d*1.5]{17-19}. \par 
	
		\begin{figure}
		\centering
		\begin{subfigure}[b]{0.4\textwidth}
			\includegraphics[width=\textwidth]{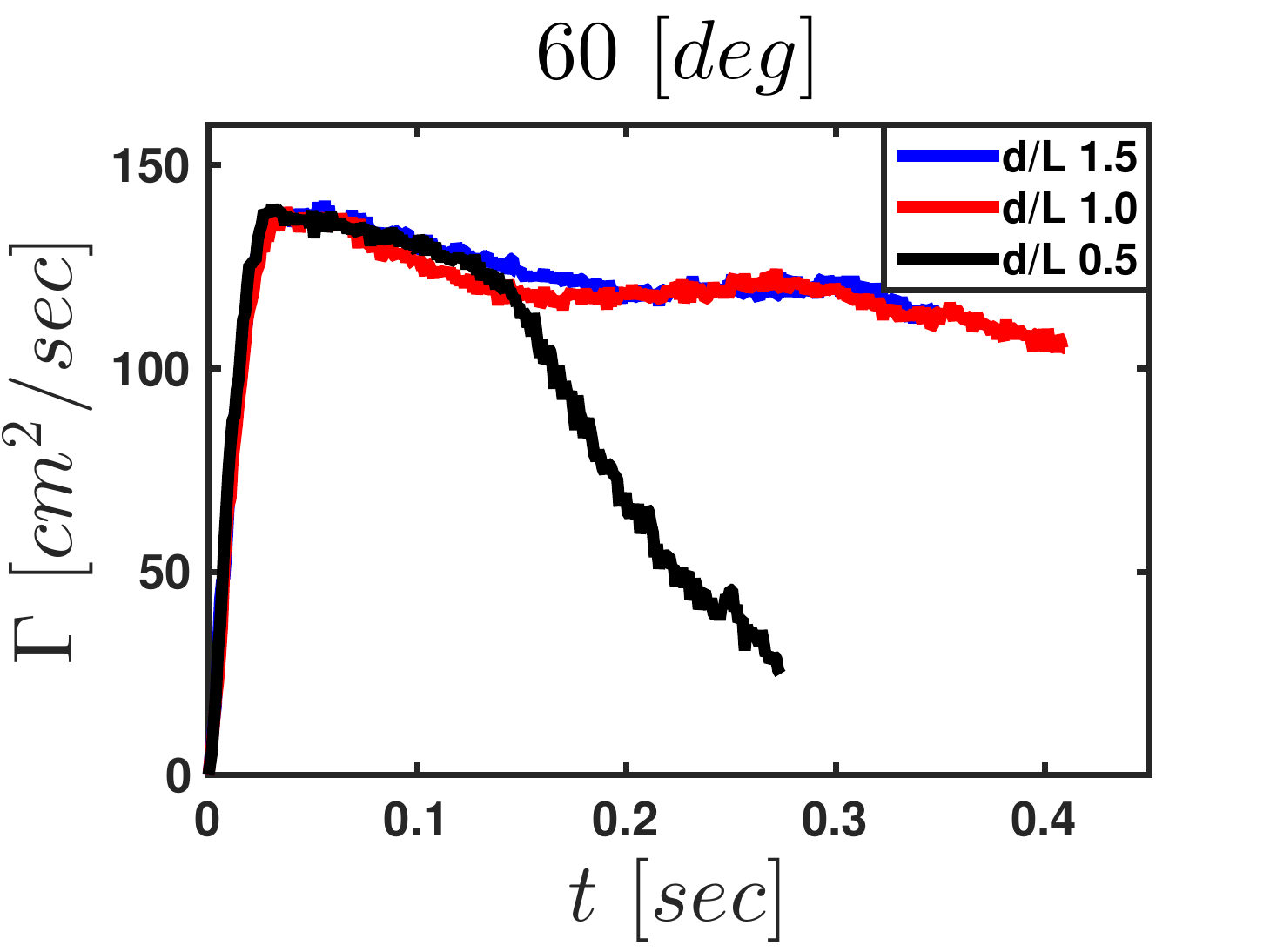}
			\caption{}
			\label{fig:Circulation_AR067_Dyn}
		\end{subfigure}\hspace{05mm}
		\begin{subfigure}[b]{0.4\textwidth}
			\includegraphics[width=\textwidth]{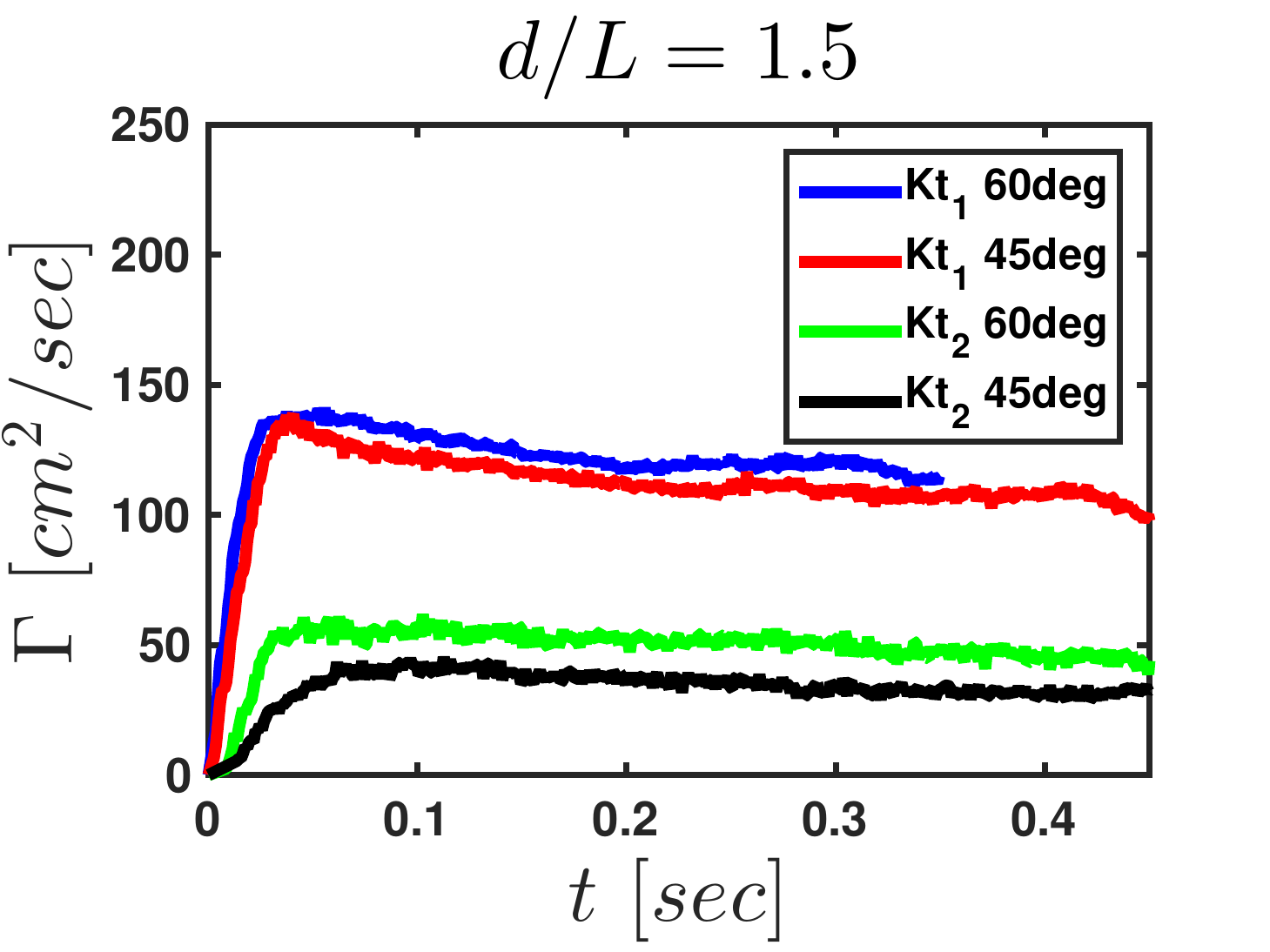}
			\caption{}
			\label{fig:Circulation_AR100_Dyn}
		\end{subfigure}\vspace{05mm}
		\begin{subfigure}[b]{0.4\textwidth}
			\includegraphics[width=\textwidth]{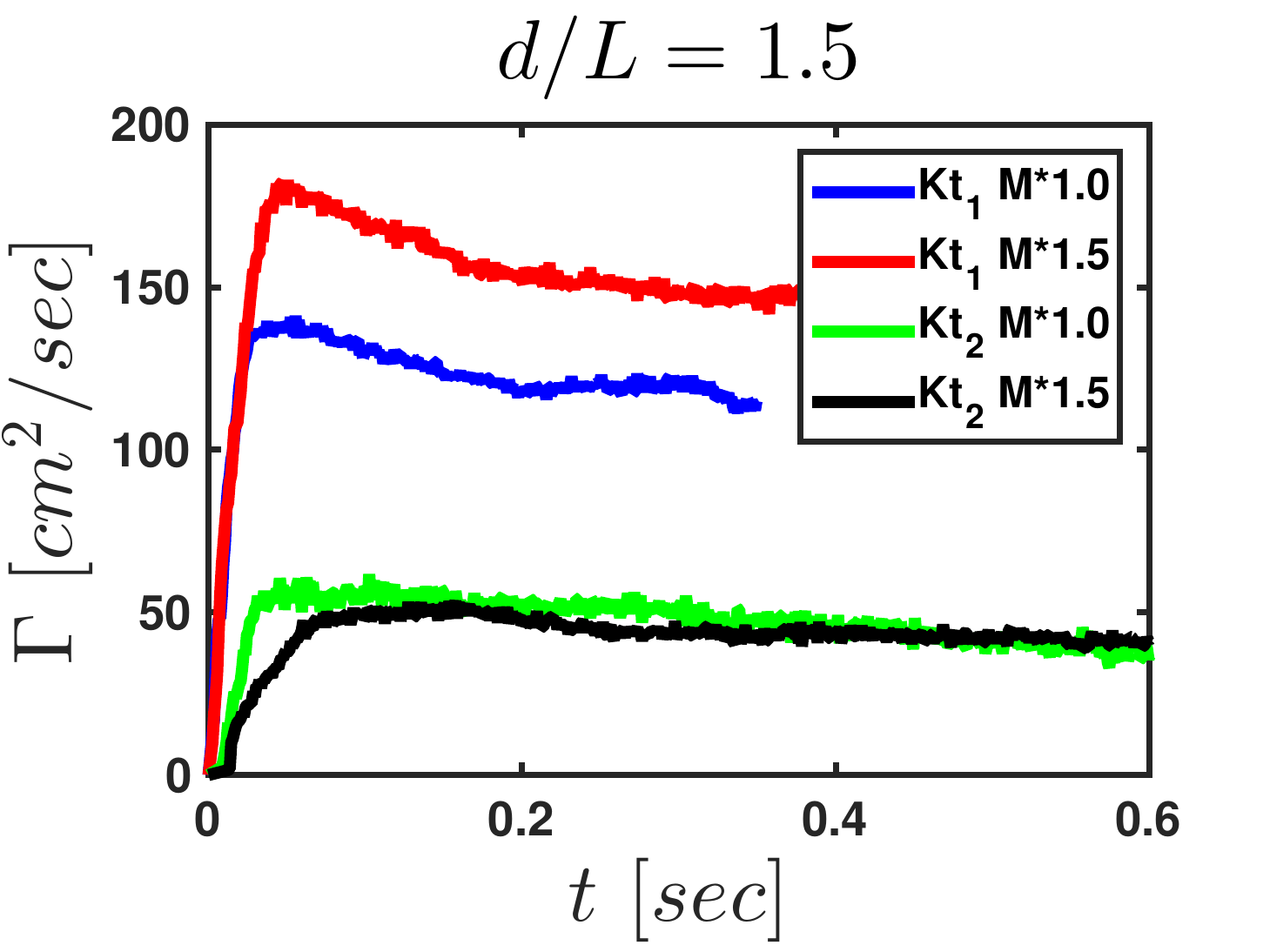}
			\caption{}
			\label{fig:Circulation_AR200_Dyn}
		\end{subfigure}\hspace{05mm}
		\begin{subfigure}[b]{0.4\textwidth}
			\includegraphics[width=\textwidth]{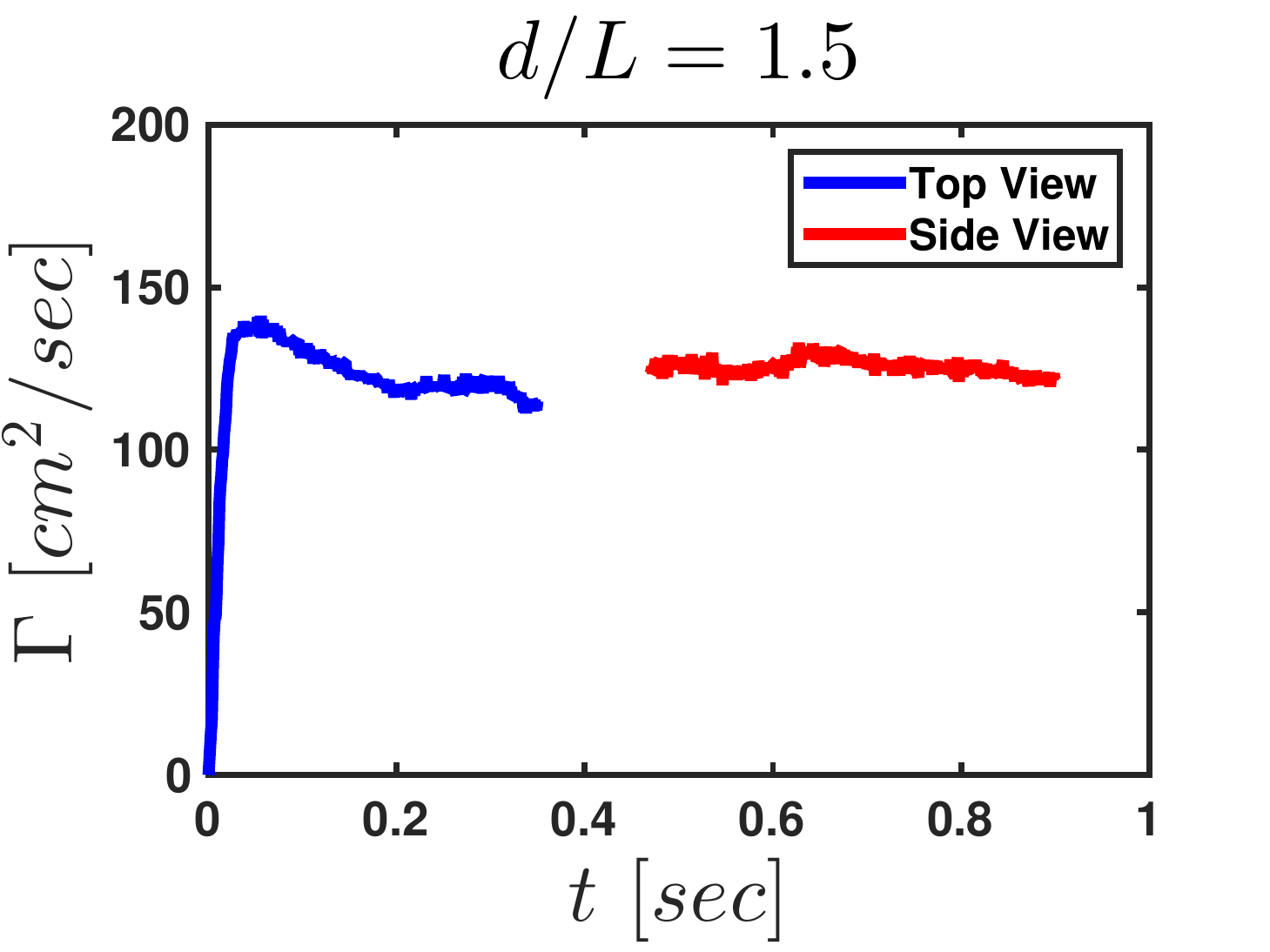}
			\caption{}
			\label{fig:Circulation_AR067_TV_SV_Dyn}
		\end{subfigure}
		\caption{Variation of circulation ($\Gamma$) in the starting vortices with time, (a) for different $d^*$ values, and with stiffness per unit depth = $Kt_1$, $2 \theta_o$= 60 deg and $M^*$= 1.0. (b) for different values of $Kt$ and clapping angle $2 \theta_o$, and with $d^*$= 1.5 and $M^*$= 1.0. (c) for different values of $M^*$ and $Kt$, and with $d^*$= 1.5, $2 \theta_o$= 60 deg. For (a), (b) and (c), circulation is from flow fields measured in the XY plane. (d) The time evolution of $\Gamma$ of the vortices in the XY (top view) and XZ (side view) planes during different time periods. Other  parameters: $Kt_1$, $2\theta_o$ = 60 deg, $M^*$ = 1.0.}\label{fig:CirculationCurves_dyn}
	\end{figure}
	
	In all cases, till the $\dot{\theta}$ of the plates becomes zero, the $\Gamma$ increases rapidly and reaches a maximum value ($\Gamma_m$) and its further evolution largely depends on $d^*$. The nature of circulation variation  till it reaches the maximum is nearly independent of $d^*$ (see figure \hyperref[fig:CirculationCurves_dyn]{21} and \hyperref[tab:gamma_m]{Table 7}). The RSD in $\Gamma_m$ due to $d^*$ variations is less than 12\% for the bodies with $Kt_1$ and less than 25\% for $Kt_2$. In the case of the body with $d^*$ = 1.5 and 1.0, the circulation becomes steady after reaching the maximum. In the case of the body with $d^*$ = 0.5, the circulation shows rapid reduction (see figure \hyperref[fig:CirculationCurves_dyn]{21a}) after it reaches the maximum because cancellation of vorticity as the two vortices come close to each other (see figure \hyperref[fig:CoreSep_PIV]{22c}). \par
	
	The higher spring stiffness results in higher $\Gamma_m$, bodies with spring stiffness per unit depth $Kt_1$ have 2.4-3.4 times higher $\Gamma_m$ than those with $Kt_2$  for $M^*$ = 1.0; for $M^*$ = 1.5  this ratio is 3.0-4.1 . The time to reach $\Gamma_m$  also reduces with an increase in stiffness (see figure \hyperref[fig:CirculationCurves_dyn]{21b}). An increase in $\theta_o$ increases the circulation slightly, this increase being much less than that due to variation in $Kt$ (see figure \hyperref[fig:CirculationCurves_dyn]{21b}). The time corresponding to $\Gamma_m$ is approximately independent of clapping angle. Effect of $M^*$ on $\Gamma_m$ is not clear. The increase in $M^*$ increases circulation in the case of bodies with stiffness per unit depth = $Kt_1$, whereas this effect is negligible for $Kt_2$ cases (figure \hyperref[fig:CirculationCurves_dyn]{21c}). \par

% Table generated by Excel2LaTeX from sheet 'Variables'
\begin{table}
	\centering
	
	\begin{tabular}{cccccc}
		\toprule
		\multicolumn{6}{c}{\text{ $ \Gamma_{m}$  [cm$^2$/s]}} \\
		\midrule
		\textbf{$Kt$} & \multicolumn{1}{c}{\multirow{2}[4]{*}{\textbf{$M^*$}}} & $2\theta_{o}$ & \multicolumn{3}{c}{\textbf{$d^*$}} \\
		\cmidrule{4-6}    \text{[mJ / mm.rad$^2$]} &       & \multicolumn{1}{p{3em}}{\text{[Deg]}} & \textbf{1.5} & \textbf{1.0} & \textbf{0.5} \\	
		\midrule
		\multirow{4}[4]{*}{\textbf{0.8-1.1}} & \multirow{2}[2]{*}{\textbf{1.0}} & \textbf{60} & 137   & 138   & 139 \\
		&       & \textbf{45} & 136   & 109   & 128 \\
		\cmidrule{2-6}          & \multirow{2}[2]{*}{\textbf{1.5}} & \textbf{60} & 181   & 189   & 160 \\
		&       & \textbf{45} & 130   & 143   & 136 \\
		\midrule
		\multirow{4}[4]{*}{\textbf{0.3-0.6}} & \multirow{2}[2]{*}{\textbf{1.0}} & \textbf{60} & 58    & 51    & 45 \\
		&       & \textbf{45} & 42    & 32    & 26 \\
		\cmidrule{2-6}          & \multirow{2}[2]{*}{\textbf{1.5}} & \textbf{60} & 51    & 60    & 54 \\
		&       & \textbf{45} & 32    & 37    & 25 \\
		\bottomrule
	\end{tabular}%
	\label{tab:gamma_m}%
	\caption{Maximum circulation attained by starting vortices}
\end{table}%

	 Circulation of vortex patches in the side view (XZ plane) has been calculated for the cases with stiffness per unit depth = $Kt_1$, $2\theta_o$ = 60 deg, $M^*$ = 1.0 and $d^*$ = 1.5, 1.0 and 0.5. In this plane, vortices appear only after the initial vortex reconnection. Figure \hyperref[fig:Circulation_AR067_TV_SV_Dyn]{21d} shows the circulation in the side view from 400ms onwards, and is almost same as the circulation in the top view. In the case of the bodies with $d^*$ = 1.5 and 1.0, the magnitude of circulation in the top view and side view is approximately the same, implying negligible circulation reduction after the reconnection of vortex loops.\par

\end{subsubsection}

\begin{subsubsection}{Core separation between the starting vortices}
	\label{sec:Scr_3D loop}
	Analysis of core separation distance($S_{cr}$) between vortex cores provides an insight into the wake dynamics and confirms the general picture of vortex evolution for the three aspect ratio bodies as depicted in figures \hyperref[fig:3D Vortex loop_d*1.5]{17}, \hyperref[fig:3D Vortex loop_d*1.0]{18}, and \hyperref[fig:3D Vortex loop_d*0.5]{19}. For each case in parametric space, the core separation curve is averaged over three experiments, where the time average of standard deviation is less than 6\% of the initial tip separation.The $S_{cr}$ follows the reduction in the tip distance till the time corresponding to the end of clapping motion: 60-80ms for $Kt_1$ and 80-110 ms for $Kt_2$. After that, the evolution of $S_{cr}$ depends on $d^*$. Figures \hyperref[fig:CoreSep_PIV]{22a,b,c} show the vortex pairs some time after completion of clapping for the three values of $d^*$, for stiffness per unit depth = $Kt_1$, $M^*$ = 1.0, and $2\theta_o$= 60deg. Figure \hyperref[fig:Dye_vis_dyn]{23} shows PLIF based dye visualization with the laser sheet in the XY plane for the same parameter values as in figure \hyperref[fig:CoreSep_PIV]{22}. Much of the dye is in the vortices, with some remaining in a trail connected to the body. \par
	
	Once the clapping motion is initiated, the starting vortices appear, with the initial separation slightly less than the initial distance between the tips. The subsequent evolution of $S_{cr}$ depends most strongly on $d^*$. Figure \hyperref[fig:CoreSepCurves_dyn]{24a} shows, obtained from PIV images, the variation of $S_{cr}$ with time for the three $d^*$ values, with all other parameters being same. For the same parameter values figure \hyperref[fig:CoreSepCurves_Dye_vis]{25} shows $S_{cr}$ variation obtained from dye visualization images where the vortices could be tracked over longer distances.\par   
	
	Up to 100ms, core separation reduces rapidly with time and is almost the same for all the $d^*$ cases; after that, the evolution of $S_{cr}$ with time shows a strong dependence on $d^*$. For $d^*$ = 0.5 case, the vortices rapidly approach each other, and later, they disappear due to vorticity cancellation. The close approach of the vortices for $d^*$ = 0.5 may be seen in the PIV and dye visualization pictures (figures \hyperref[fig:CoreSep_AR200_Dyn]{22c}, \hyperref[fig:Dye_vis_dyn]{23c}). This vortex pair seen in the XY plane corresponds to one of the several vortex loops that form for $d^*$= 0.5 (figure \hyperref[fig:3D Vortex loop_d*0.5]{19a}). For $d^*$=1.0, there is some up and down variation in Scr before a rapid increase at around 0.5 s, reaching a maximum value of 86 mm at 0.7 s followed by a drop (Figure \hyperref[fig:CoreSepCurves_Dye_vis]{25}); this variation is consistent with the axis switching of the vortex loop shown in figure \hyperref[fig:3D Vortex loop_d*1.0]{18}. For $d^*$= 1.5 cases also axis switching is observed;  core separation is almost constant shows till 200ms, after which there is a period of slow increase followed by a rapid one reaching a maximum value of 126 mm at 1.4 s. Note that for both these cases, the maximum vortex separation is close to the depth of the body ($d$). \par
		
	Change in the initial clapping angle changes initial vortex spacing but the nature of the subsequent evolution in $S_{cr}$ is similar for both the $2\theta_0$ values (Figure \hyperref[fig:CoreSep_AR067_TV_SV_Dyn]{24b}). Reduction of the spring stiffness per unit depth from $Kt_1$ to $Kt_2$ slows down the swithcing process (Figure \hyperref[fig:CoreSep_AR067_TV_SV_Dyn]{24b}), as is to be expected because both body velocity and vortex propagation speed reduce. The time corresponds to this constant separation phase increases with a reduction in the stiffness per unit depth : 150ms for $Kt_1$ and 450ms for $Kt_2$, see figures \hyperref[fig:CoreSep_AR067_TV_SV_Dyn]{24b} and \hyperref[fig:CoreSep_AR067_TV_SV_Dyn]{24c}.The mass of the body seems to have a negligible influence on the evolution of the vortex spacing (figure \hyperref[fig:CoreSep_AR067_TV_SV_Dyn]{24c}) compared to change in the stiffness value.\par
				
	\begin{figure}
		\centering
		\begin{subfigure}[b]{0.32\textwidth}
			\includegraphics[width=\textwidth]{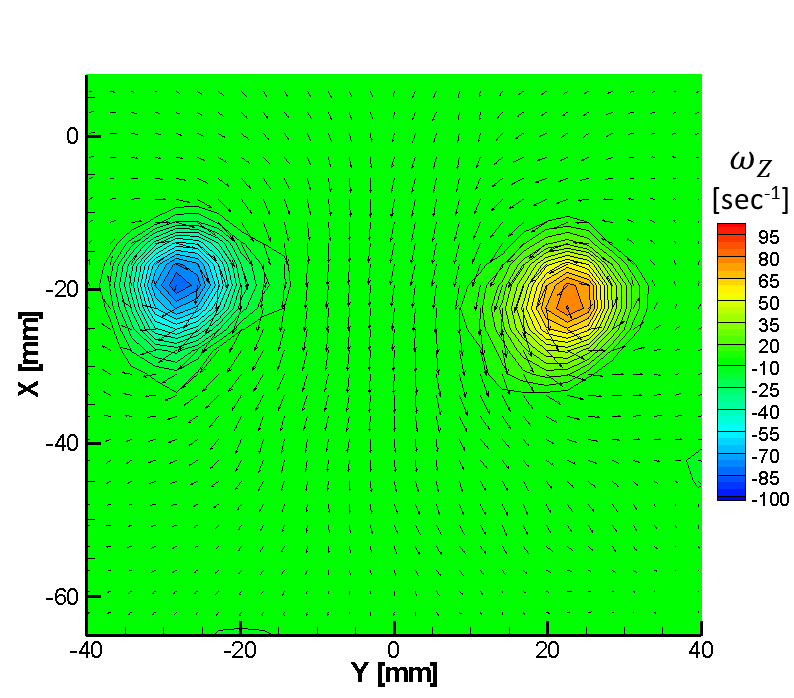}
			\caption{}
			\label{fig:CoreSep_AR067_Dyn}
		\end{subfigure}\hspace{02mm}
		\begin{subfigure}[b]{0.32\textwidth}
			\includegraphics[width=\textwidth]{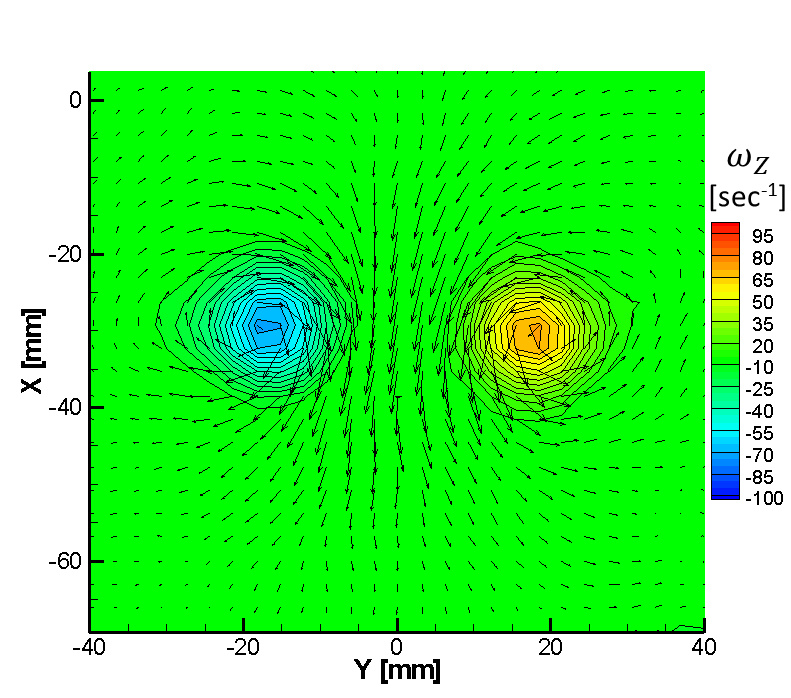}
			\caption{}
			\label{fig:CoreSep_AR100_Dyn}
		\end{subfigure}\hspace{02mm}
		\begin{subfigure}[b]{0.32\textwidth}
			\includegraphics[width=\textwidth]{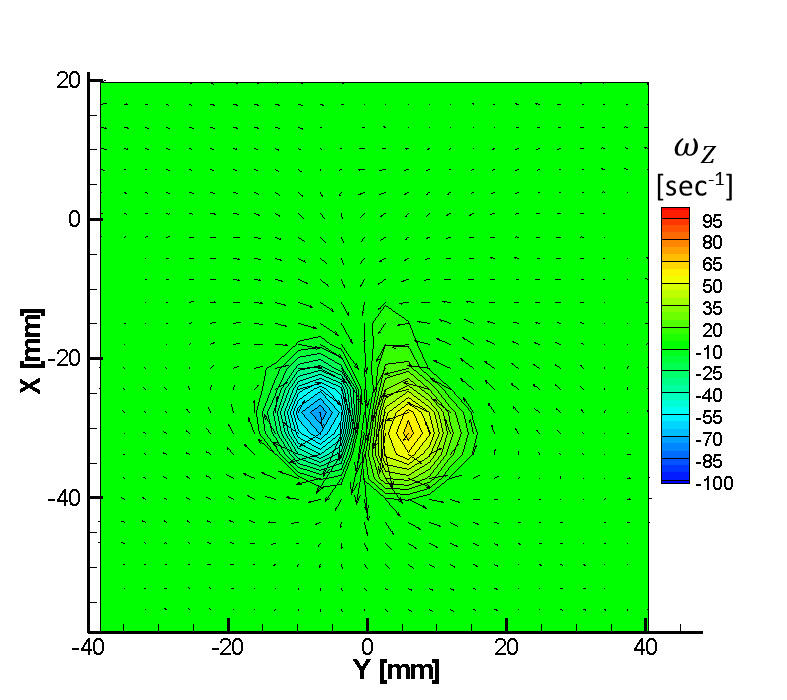}
			\caption{}
			\label{fig:CoreSep_AR200_Dyn}
		\end{subfigure}
		\caption{Vortex pairs in the XY plane (a) for the body with $d^*$= 1.5 at 300 ms, (b) for the body with $d^*$= 1.0 at 300 ms, (c) for the body with $d^*$= 0.5 at 200 ms. Other parameters: stiffness per unit depth = $Kt_1$, $M^*$=1.0, $2\theta_o$= 60 deg. }\label{fig:CoreSep_PIV}
	\end{figure} 

	\begin{figure}
		\centering
		\begin{subfigure}[b]{0.3\textwidth}
			\includegraphics[width=\textwidth]{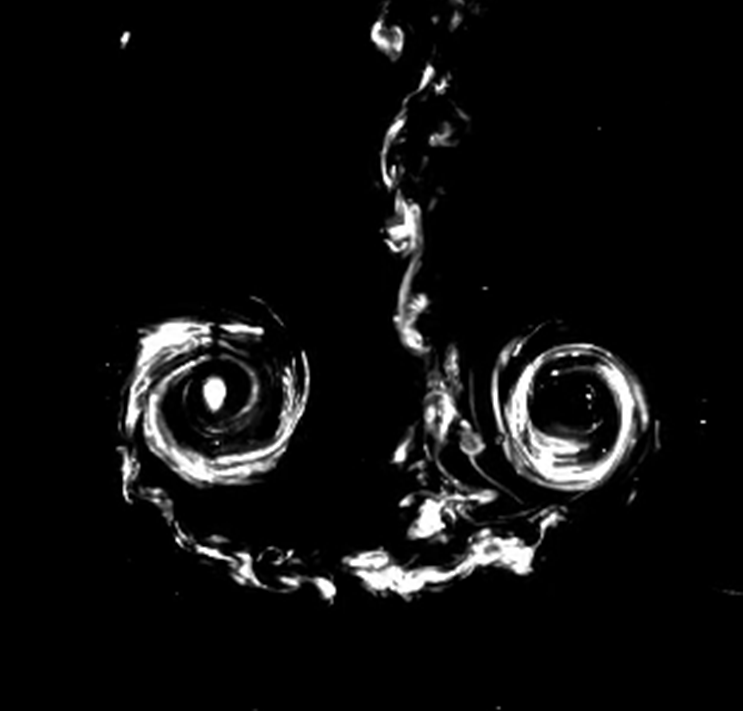}
			\caption{}
			\label{fig:Vis_AR067_Dyn}
		\end{subfigure}\hspace{05mm}
		\begin{subfigure}[b]{0.3\textwidth}
			\includegraphics[width=\textwidth]{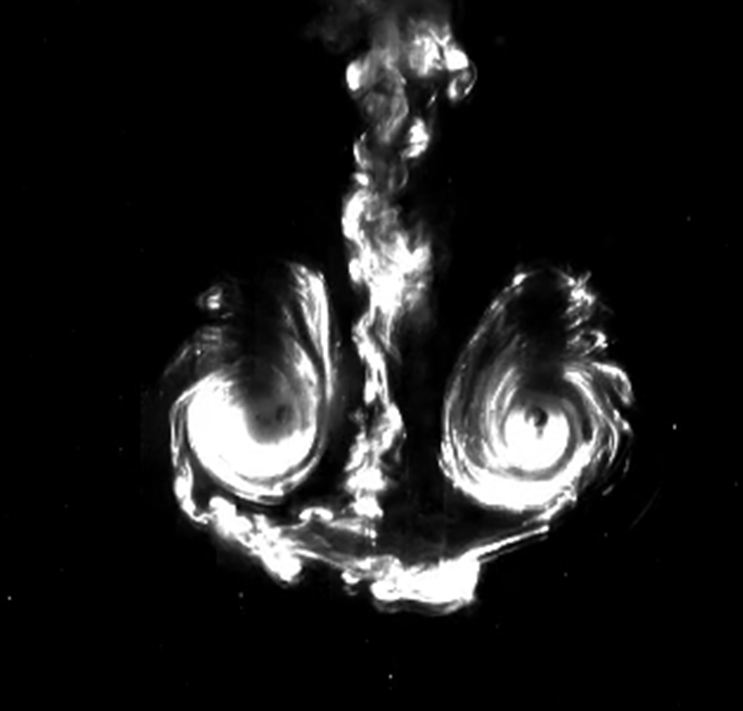}
			\caption{}
			\label{fig:Vis_AR100_Dyn}
		\end{subfigure}\hspace{05mm}
		\begin{subfigure}[b]{0.3\textwidth}
			\includegraphics[width=\textwidth]{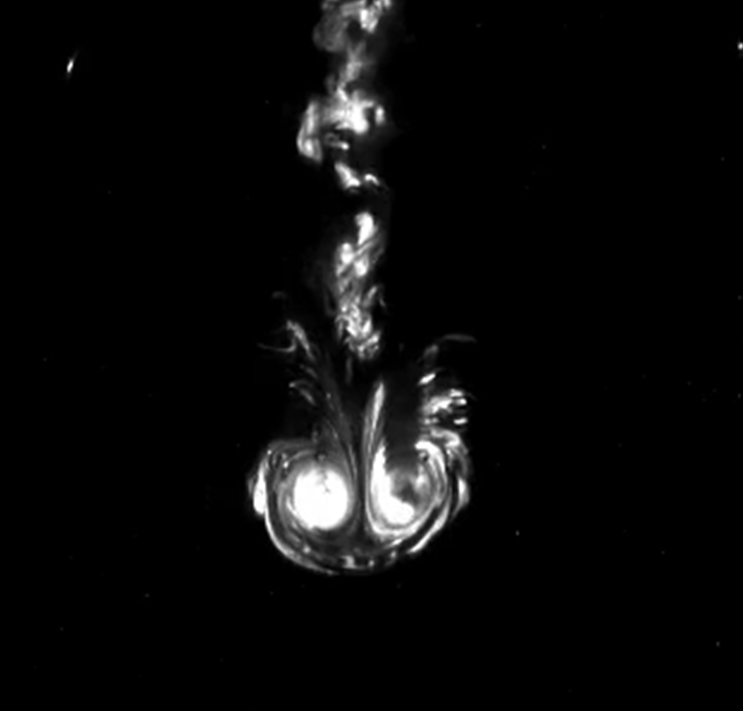}
			\caption{}
			\label{fig:Vis_AR200_Dyn}
		\end{subfigure}\vspace{07mm}
		\caption{ Dye visualization in the XY plane (a) for the body with $d^*$= 1.5 at 300 ms, (b) for the body with $d^*$= 1.0 at 300 ms, (c) for the body with $d^*$= 0.5 at 200 ms. Other parameters: stiffness per unit depth = $Kt_1$, $M^*$=1.0, $2\theta_o$= 60 deg. }\label{fig:Dye_vis_dyn}
	\end{figure}
	
	\begin{figure}
		\centering
		\begin{subfigure}[b]{0.4\textwidth}
			\includegraphics[width=\textwidth]{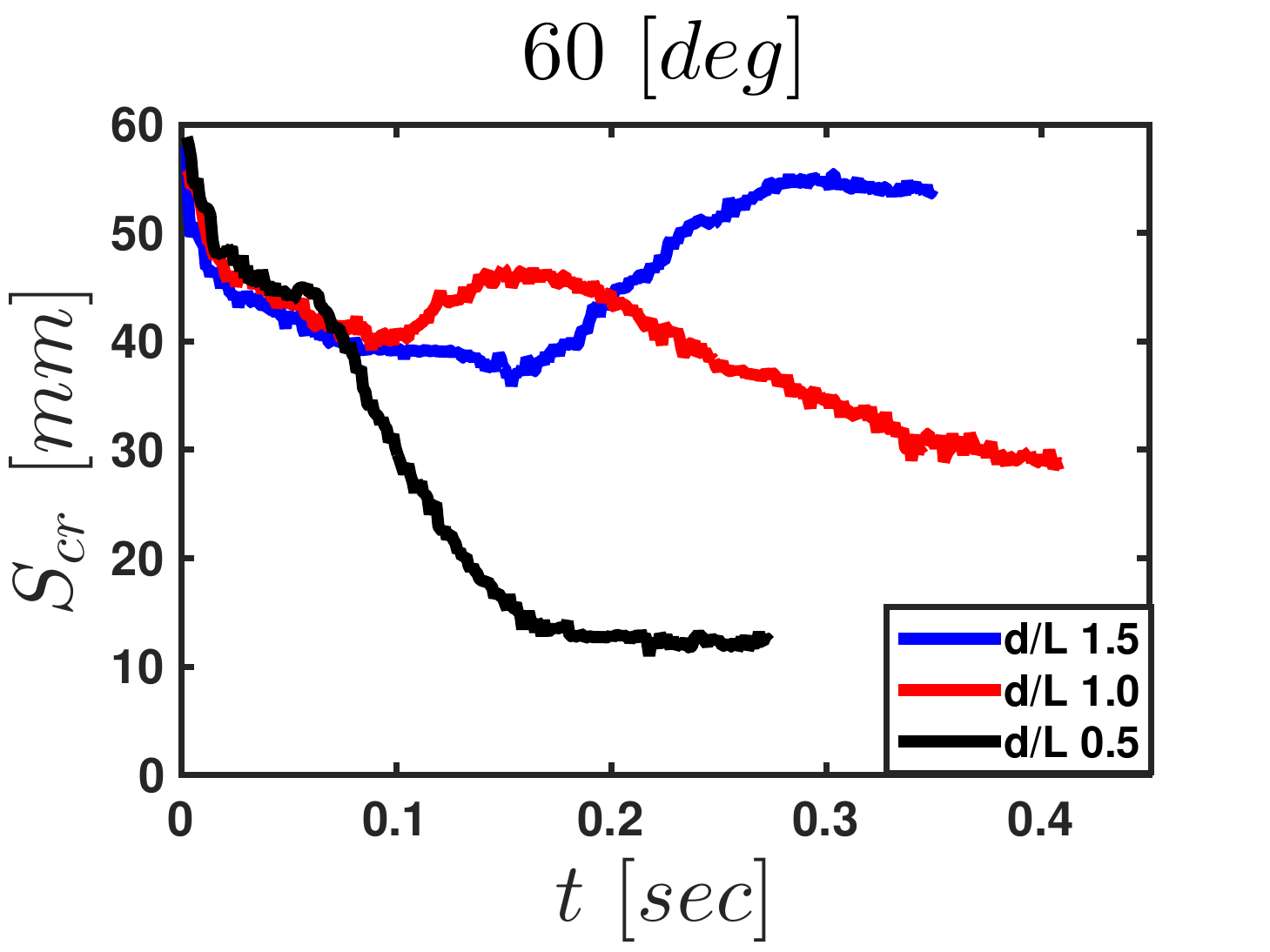}
			\caption{}
			\label{fig:CoreSepCurv_AR067_Dyn}
		\end{subfigure}\hspace{08mm}
		\begin{subfigure}[b]{0.4\textwidth}
			\includegraphics[width=\textwidth]{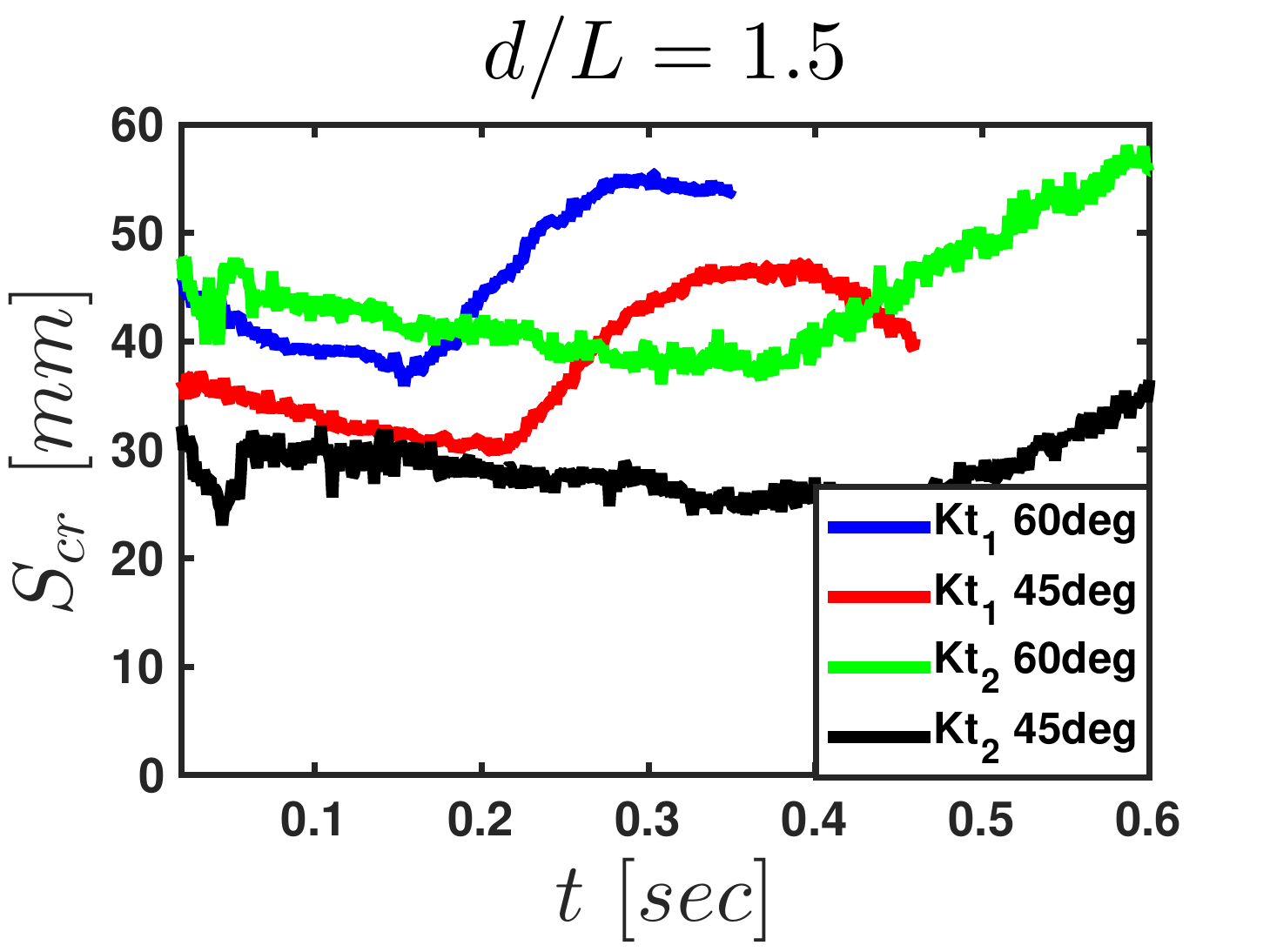}
			\caption{}
			\label{fig:CoreSepCurv_Kt_angle_Dyn}
		\end{subfigure}\vspace{05mm}
		\begin{subfigure}[b]{0.4\textwidth}
			\includegraphics[width=\textwidth]{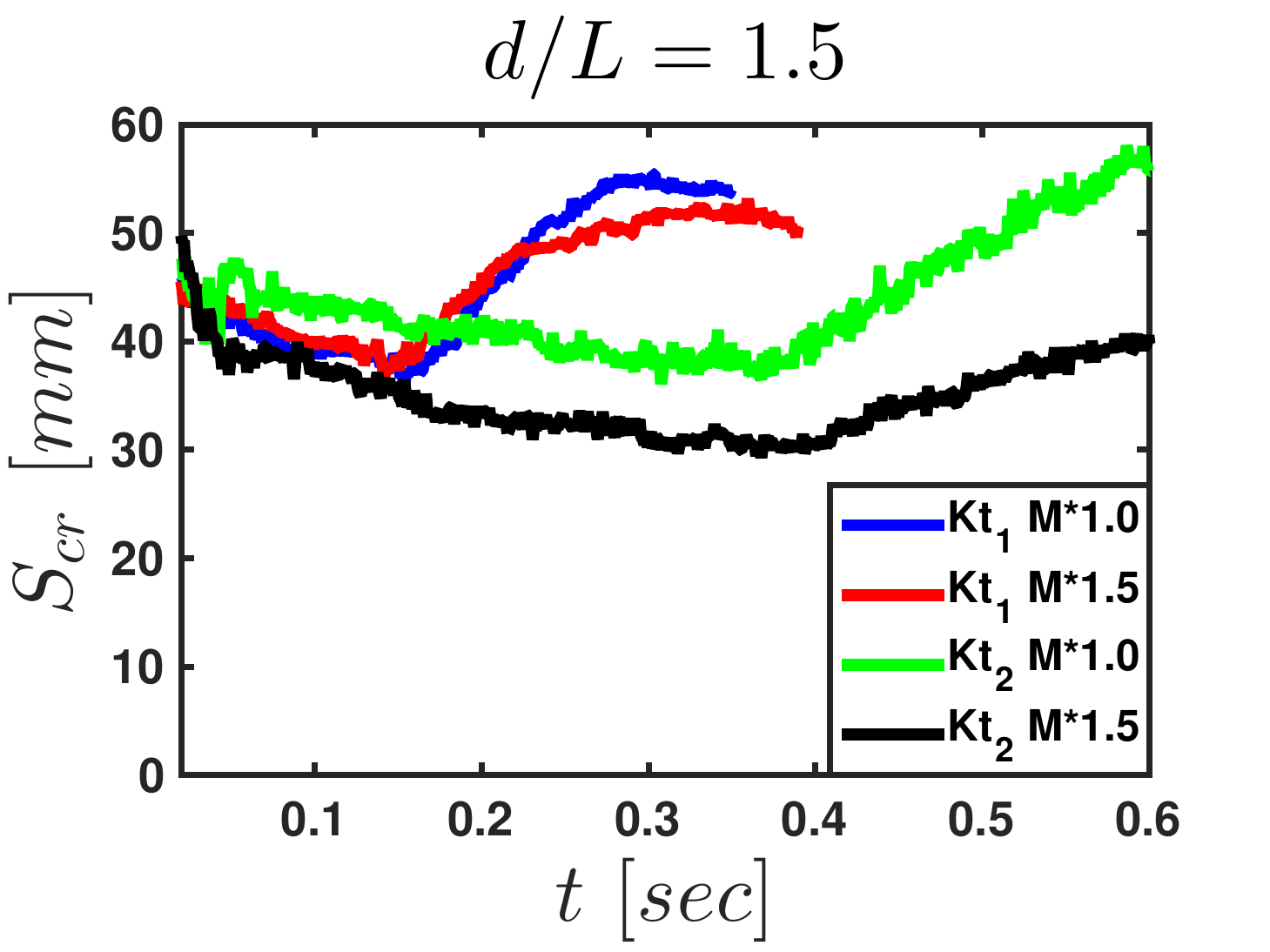}
			\caption{}
			\label{fig:CoreSepCurv_Mstr_angle_Dyn}
		\end{subfigure}\hspace{08mm}
		\begin{subfigure}[b]{0.4\textwidth}
			\includegraphics[width=\textwidth]{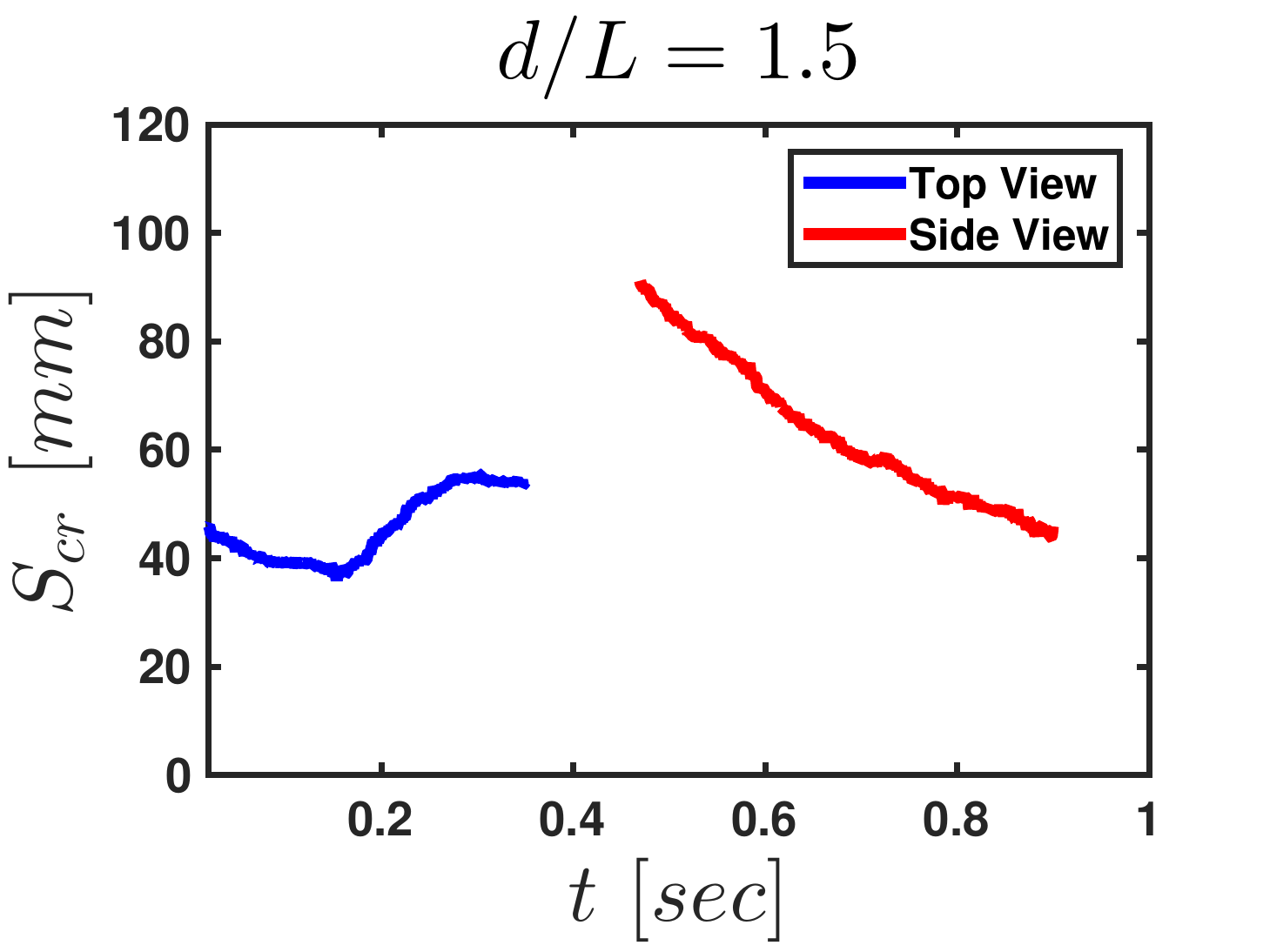}
			\caption{}
			\label{fig:CoreSep_AR067_TV_SV_Dyn}
		\end{subfigure}
		\caption{ Variation of core separation ($S_{cr}$) in the starting vortices with time, (a) for different $d^*$ values, and with stiffness per unit depth = $Kt_1$, $2 \theta_o$= 60 deg and $M^*$= 1.0. (b) for different values of $Kt$ and clapping angle $2 \theta_o$, and with $d^*$= 1.5 and $M^*$= 1.0. (c) for different values of $M^*$ and $Kt$, and with $d^*$= 1.5, $2 \theta_o$= 60 deg. For (a), (b) and (c), circulation is from flow fields measured in the XY plane. (d) The time evolution of $S_{cr}$ of the vortices in the XY (top view) and XZ (side view) planes during different time periods. Other  parameters: $Kt_1$, $2\theta_o$ = 60 deg, $M^*$ = 1.0. }\label{fig:CoreSepCurves_dyn}
	\end{figure}
	  
	\begin{figure}
	 	\centering
	 	\begin{subfigure}[b]{0.4\textwidth}
	 		\includegraphics[width=\textwidth]{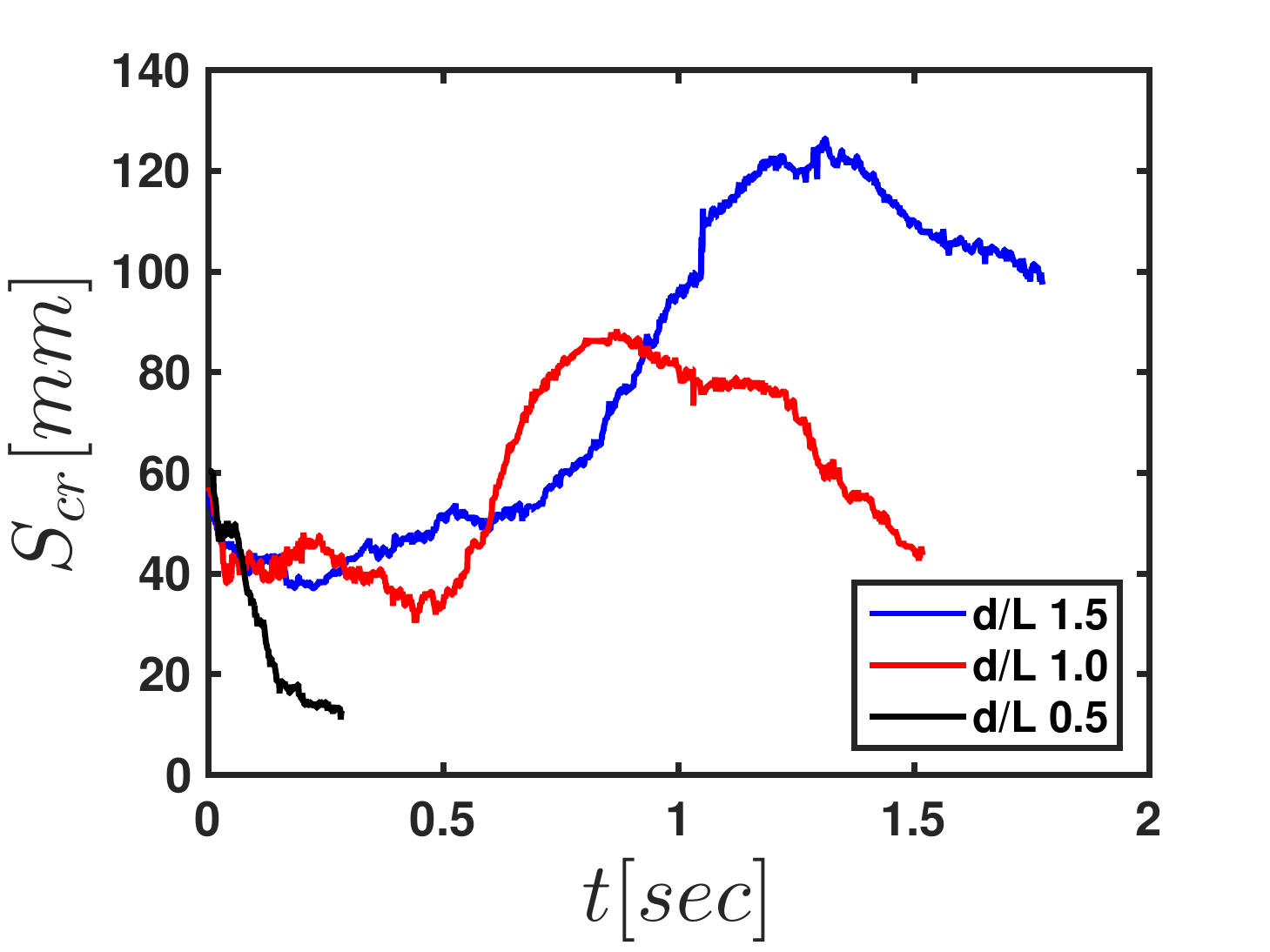}	
	 		\label{fig:Dye_Vis_Scr}
	 	\end{subfigure}
	 	\caption{Time variation of core separation $S_{cr}$ as obtained using PLIF visualization. Other parameters: stiffness per unit depth = $Kt_1$, $2\theta_o$ = 60 deg, $M^*$ = 1.0, $d^*$ = 1.5, 1.0 and 0.5.}\label{fig:CoreSepCurves_Dye_vis}
	 \end{figure} 
  
     Figure \hyperref[fig:CoreSep_AR067_TV_SV_Dyn]{24d} shows the vortex separation in the side view (XZ plane) for $d^*$= 1.5 case, with the other parameters being same as in figure figure \hyperref[fig:CoreSepCurv_AR067_Dyn]{24a} and figure \hyperref[fig:CoreSepCurves_Dye_vis]{25}. As discussed above, vortex cores are seen in the XZ plane only after the  vortex reconnection, which happens at around 300ms. In figure \hyperref[fig:CoreSepCurv_AR067_Dyn]{24d} we see that a gradual reduction in vortex sepration in the XZ plane can be observed. In the XY plane, during the vortex reconnection process $S_{cr}$ shows reduction till 100ms followed by an increase from 200 to 300ms. The plots in figures \hyperref[fig:CoreSepCurv_AR067_Dyn]{24d} and \hyperref[fig:CoreSepCurves_Dye_vis]{25} are consistent with the axis switching phenomenon depicted for $d^*$= 1.5 in figure \hyperref[fig:3DVort_AR067_Dyn_01]{17}. The major axis of the elliptical ring is initially along the depth (d=133mm), and $S_{cr}$ indicates the minor axis slightly less than the initial tip distance (figure \hyperref[fig:3DVort_AR067_Dyn_01]{17a}). At around 1300ms the major axis is along the Y direction with $S_{cr}$ =126mm (figure \hyperref[fig:CoreSepCurves_Dye_vis]{25}). During this phase, the circulation is nearly constant (figure \hyperref[fig:Circulation_AR067_TV_SV_Dyn]{21d}).  \par

\end{subsubsection}

\end{subsection}

\begin{subsection}{Wake energetics}
	\label{sec:WakeEnergetics}
	In the previous sections, the wake structure obtained in the wake of the clapping body has been discussed. Next, we look at the energy budget, i.e., the conversion of the strain energy ($SE$) initially in the spring into kinetic energy in the body and in the fluid. At the end of the clapping motion, the energy balance with the assumption of negligible viscous dissipation of energy is     
	
	\begin{gather}\label{eq:energ_bal}
	SE = KE_b + KE_f \\
	KE_b = 0.5\ (m_b + m_{add}) u_m^2  
	\end{gather}
	
	$KE_b$ in \eqref{eq:energ_bal} represents the maximum kinetic energy attained by the body, corresponding to the time when body velocity has reached its maximum value $u_m$, and $KE_f$ represents the kinetic energy in the fluid. To account for all the kinetic energy in the fluid in the wake of the clapping body, we need 3D velocimetry data, which is not available. $m_{add}$ is a mass of fluid that is moving with the body. Since, at the end of the clapping motion, both plates touch each other resulting in a thin, streamlined body configuration, we neglect $m_{add}$ contribution. The energy stored in the two steel plates for the initial angular deflection $\theta_o$ is 
	
	\begin{gather}
	\label{eq:SE_Kt}
	SE = 2\ K_{td} \ {\theta_o}^2 
	\end{gather}
	
	In the \eqref{eq:SE_Kt}, $K_{td}$ is the stiffness coefficient which correlates the strain energy to angular deflection. The $K_{td}$ is calculated experimentally for each body. The $K_{td}$ for all 12 clapping bodies is listed in \hyperref[tab:DesignData]{Table 1}. The details of $SE$ measurement are provided in \S \hyperref[sec:SE_Exp]{5}. In equation \eqref{eq:energ_bal}, $SE$ is calculated using initial clapping angle, and kinetic energy of the body is calculated using image analysis; the only unknown is $KE_f$. Further \hyperref[tab:wake_energy]{table 8} shows that approximately  80\% initial stored energy is transferred to the fluid. \par 
  %%%%%%%%%%%%% Energy analysis of clapping body %%%%%%%%%%%%%%%%%%%%%%%%%%%%%%%%%%%%%%%
	% Table generated by Excel2LaTeX from sheet 'Energy_2'
	
	\begin{table}
		\centering
		
		\begin{tabular}{cccccccc}
			\toprule
			\text{$Kt'$} & \text{M*} & \text{d*} & \text{$2\theta_o$}    & \text{$SE$} & \text{$KE_b$} & \text{$KE_f$} & \text{$KE_f$} \\
			\text{[mJ / mm.rad$^2$]} &       &       & \multicolumn{1}{p{3em}}{\text{[Deg]}} & \text{[mJ]} & \text{[mJ]} & \text{[mJ]} & \text{\%} \\
			\midrule
			\multirow{12}[12]{*}{0.8-1.1} & \multirow{6}[6]{*}{1.0} & \multirow{2}[2]{*}{1.5} & 57.9  & 54.7  & 8.8   & 45.9  & 83.9 \\
			&       &       & 45.9  & 33.9  & 6.4   & 27.5  & 81.2 \\
			\cmidrule{3-8}          &       & \multirow{2}[2]{*}{1.0} & 59.4  & 44.9  & 5.0   & 39.9  & 88.8 \\
			&       &       & 43.2  & 22.1  & 3.1   & 19.0  & 86.0 \\
			\cmidrule{3-8}          &       & \multirow{2}[2]{*}{0.5} & 62.5  & 21.0  & 2.7   & 18.3  & 87.1 \\
			&       &       & 49.0  & 12.4  & 1.7   & 10.7  & 86.0 \\
			\cmidrule{2-8}          & \multirow{6}[6]{*}{1.5} & \multirow{2}[2]{*}{1.5} & 64.2  & 69.5  & 9.3   & 60.2  & 86.6 \\
			&       &       & 44.7  & 33.5  & 3.6   & 30.0  & 89.4 \\
			\cmidrule{3-8}          &       & \multirow{2}[2]{*}{1.0} & 62.2  & 40.8  & 5.7   & 35.1  & 86.0 \\
			&       &       & 42.9  & 18.9  & 2.5   & 16.3  & 86.6 \\
			\cmidrule{3-8}          &       & \multirow{2}[2]{*}{0.5} & 63.0  & 22.4  & 2.3   & 20.1  & 89.9 \\
			&       &       & 47.8  & 11.9  & 1.4   & 10.5  & 88.4 \\
			\midrule
			\multirow{12}[12]{*}{0.3-0.6} & \multirow{6}[6]{*}{1.0} & \multirow{2}[2]{*}{1.5} & 58.8  & 18.7  & 3.1   & 15.6  & 83.4 \\
			&       &       & 40.4  & 8.7   & 1.6   & 7.1   & 81.1 \\
			\cmidrule{3-8}          &       & \multirow{2}[2]{*}{1.0} & 61.7  & 13.3  & 1.7   & 11.6  & 87.2 \\
			&       &       & 40.8  & 5.3   & 0.8   & 4.5   & 85.1 \\
			\cmidrule{3-8}          &       & \multirow{2}[2]{*}{0.5} & 58.8  & 7.1   & 0.7   & 6.4   & 90.4 \\
			&       &       & 41.4  & 2.0   & 0.3   & 1.8   & 86.4 \\
			\cmidrule{2-8}          & \multirow{6}[6]{*}{1.5} & \multirow{2}[2]{*}{1.5} & 55.2  & 18.8  & 1.5   & 17.3  & 92.0 \\
			&       &       & 39.2  & 8.2   & 0.5   & 7.7   & 93.8 \\
			\cmidrule{3-8}          &       & \multirow{2}[2]{*}{1.0} & 57.6  & 13.0  & 1.1   & 11.9  & 91.6 \\
			&       &       & 40.1  & 5.1   & 0.4   & 4.8   & 92.2 \\
			\cmidrule{3-8}          &       & \multirow{2}[2]{*}{0.5} & 58.3  & 6.7   & 0.4   & 6.3   & 93.5 \\
			&       &       & 40.4  & 1.6   & 0.1   & 1.5   & 91.4 \\
			\bottomrule
		\end{tabular}%
		\label{tab:wake_energy}%
		\caption{Energy budget of the clapping bodies}
	\end{table}%
    
	We use an approximate relation based on the fluid volume and circulation to calculate the kinetic energy of the fluid in the wake. In the case of bodies with $d^*$= 1.0 and 1.5, the wake can be modeled as a single vortex ring; further for scaling purposes we assume the ring to be circular. The vortex ring radius($R_v$) can be assumed to scale as the cube-root of volume($\forall$) of the fluid present in the interplate cavity; the volume of the cavity is the product of body depth ($d$) and the area of the triangular region($0.5\ R^2 sin(2\theta_o)$) indicated by the yellow dashed line (figure \hyperref[fig:TV_superimposed_Lines]{2e}); the approximate radius of rotation($R$) is 58mm.
		
	\begin{gather}
	\label{eq:Scale_Rv}
	R_v\ \sim \ {\forall}^{1/3}\  
	\end{gather}

	Expressions are available for impulse ($I_v$)  and kinetic energy ($KE_v$)  of thin vortex rings (eg. Sullivan et al.\cite{Sullivan08}) in terms of $R_v$, density $\rho$ and circulation $\Gamma$. For our case we write  
		
	\begin{gather}
		KE_v \sim \rho\  R_v\ \Gamma_m^2\    \\
	    I_v \sim \rho\  R_v^2\ \Gamma_m 
	\end{gather}
	
	where $\Gamma_m$ is the maximum circulation. A detailed discussion on the impulse-momentum approach for a body impulsively set in motion in an unbounded fluid is given by Epps\cite{Epps10}. Equating the momentum in the body to the momentum in the fluid (see equation \eqref{eq:linMom}), and neglecting the added mass component, we get 

	\begin{gather}
	\label{eq:ring_mom}
	m_b\ u_m = c_1\ \rho\ R_v^2\ \Gamma_{m} 
	\end{gather}
	
	In figure \hyperref[fig:C1_dyn]{26}, we plot the $m_b \ u_m$ versus $\rho \ R_v^2 \ \Gamma_m$. We see that the two are approximately linearly related to each other, with a linear fit giving a value of $c_1$ = 0.45. This suggests that the assumption of the linear momentum of the fluid being equal to that of an equivalent vortex ring is valid. The plot contains data from all the experiments. The $u_m$ and $\Gamma_{m} $ are obtained from measurements, and $R_v$ is obtained from \eqref{eq:Scale_Rv}. \par
	
	The equation for energy \eqref{eq:energ_bal} at the end of clapping motion with the assumptions of negligible added mass and negligible viscous dissipation of energy becomes  

	\begin{gather}
	\label{eq:ring_energy}
		2\ K_{td}\ {\theta_o}^2 = 0.5\ m_b\ u_m^2 +\ c_2\ \rho\  R_v\ \Gamma_m^2
	\end{gather}
	where $c_2$ is a constant.  Equations \eqref{eq:ring_mom} and \eqref{eq:ring_energy} may be solved to obtain $u_m$ and $\Gamma_m$ in terms the parameters  $Kt_d$, $\theta_o$, $m_b$ and $R_v$. 
	   
	\begin{gather}
	\label{eq:um_scale}
	u_m=\ \theta_o\sqrt{\frac{2{\ K}_{td}}{m_b}}\ \ \left[\frac{1}{2}+\ \frac{c_2}{{c_1}^2}\frac{{m_b}}{\ \rho{\ R_v}^3}\right]^{-1/2}\ 
	\end{gather}
		
	\begin{gather}
	\label{eq:gm_scale}
	{\ \mathrm{\Gamma}}_{m}\ =\ \theta_o \frac{\sqrt{2{\ K}_{td} \ m_b}}{\rho \ R_v^2}\ \ \left[\ \frac{{c_1}^2}{2}+ c_2 \frac{m_b}{\rho{\ R_v}^3} \right]^{-1/2}  
	\end{gather} 
	
	\begin{figure}
		\centering
		\begin{subfigure}[b]{0.48\textwidth}
			\includegraphics[width=\textwidth]{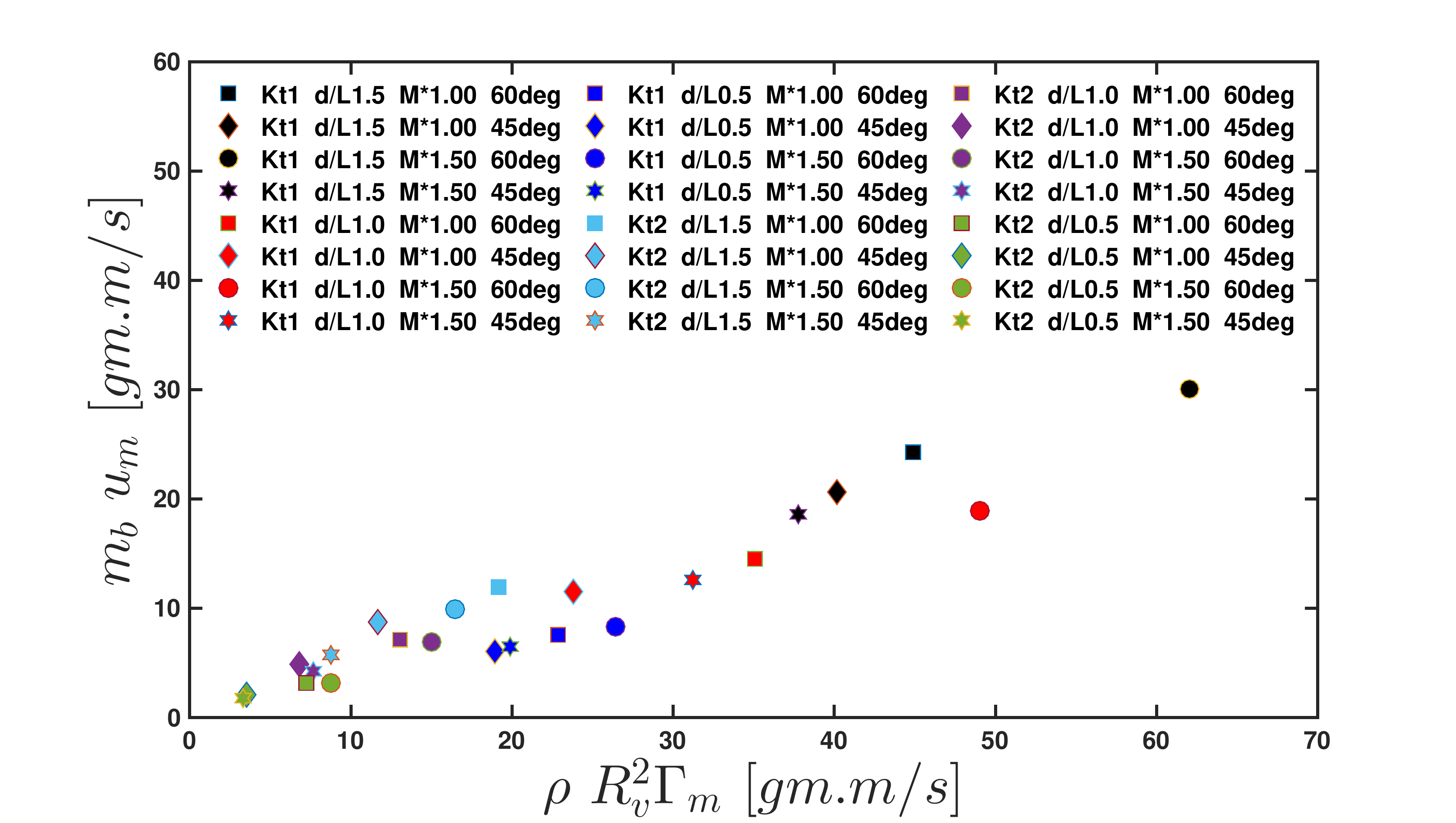}
			\label{fig:C1}
		\end{subfigure}
		\caption{Plot showing the linear dependence between body momentum and fluid momentum.}\label{fig:C1_dyn}
	\end{figure}
	
	\begin{figure}
		\centering
		\begin{subfigure}[b]{0.48\textwidth}
			\includegraphics[width=\textwidth]{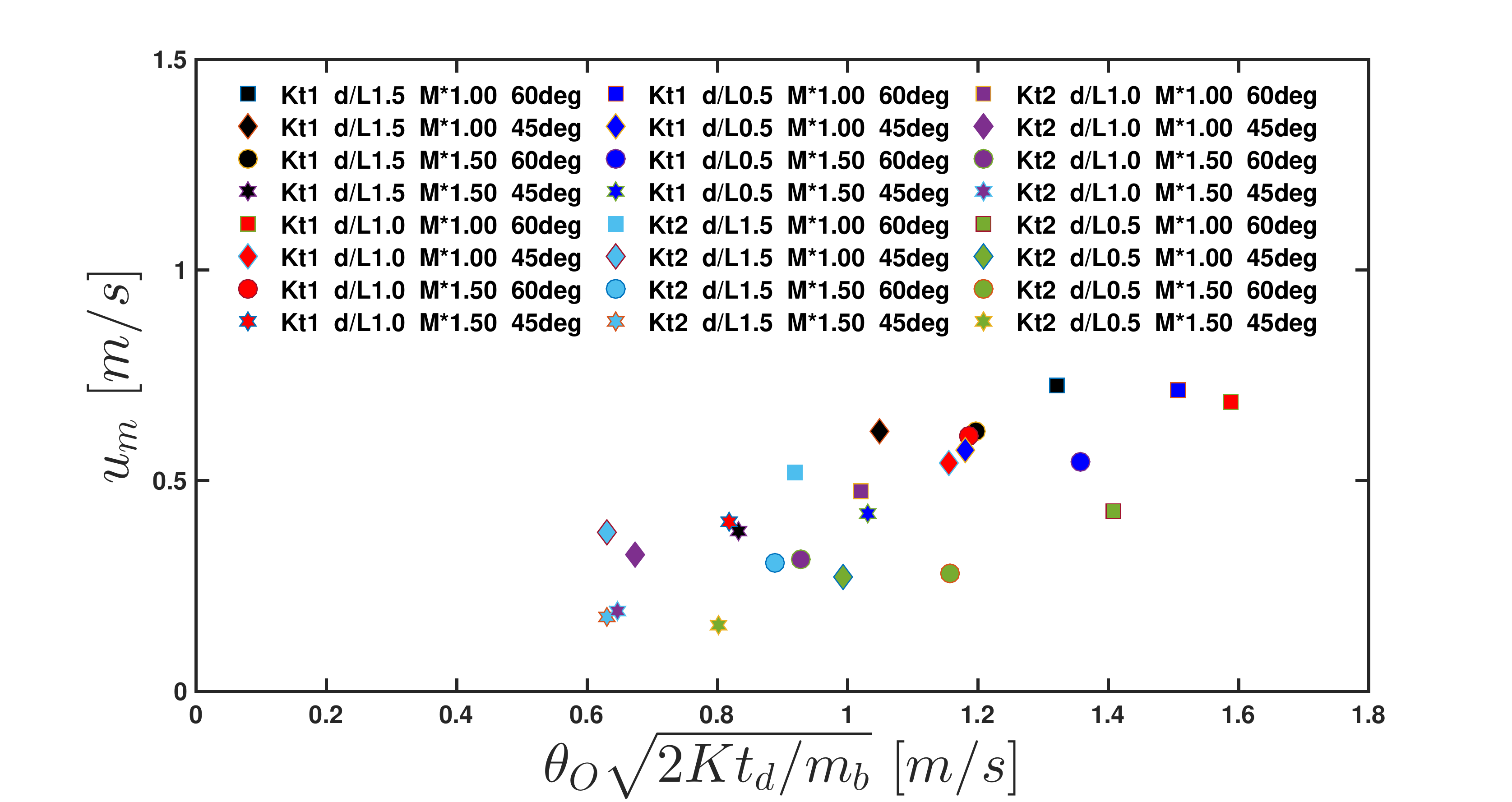}
			\caption{}
			\label{fig:um_scale}
		\end{subfigure}\hspace{02mm}
		\begin{subfigure}[b]{0.48\textwidth}
			\includegraphics[width=\textwidth]{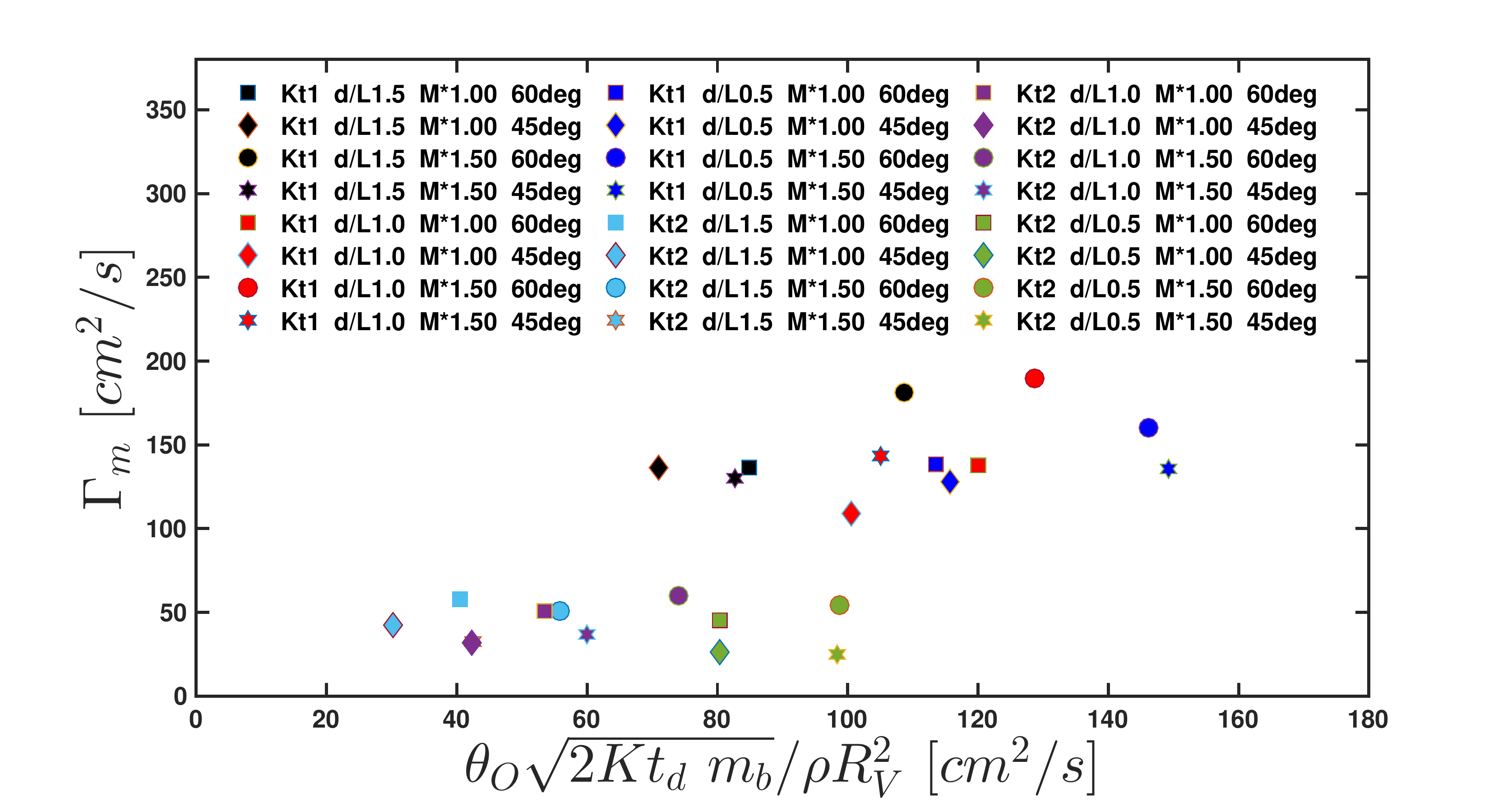}
			\caption{}
			\label{fig:Gamma_scale}
		\end{subfigure}
		\caption{(a) Measured maximum body velocity ($u_m$) plotted versus the body velocity obtained from equation \eqref{eq:um_scale}, neglecting the contribution of $m_b / (\rho{\ R_v}^3)$. (b) Measured maximum circulation ($\Gamma_m$) plotted versus the circulation obtained from equation \eqref{eq:gm_scale}, neglecting the contribution of $m_b / (\rho{\ R_v}^3)$.}\label{fig:scaling_dyn}
	\end{figure}	
	
	The above expressions may be used to obtain the velocity of a self-propelling body and circulation in the starting vortex due to the ejection of an impulsive jet in terms of the initial strain energy or work done by the body. In the limit of small body mass compared to ejected fluid mass, the 2nd terms within the parenthesis in \eqref{eq:um_scale} and \eqref{eq:gm_scale} may be neglected. Figure \hyperref[fig:um_scale]{27a} shows experimental data plotted using equation \eqref{eq:um_scale},  indicating an approximately linear relation between $\theta_o  \sqrt{2{\ K}_{td} / m_b} $ and $u_m$; similarly, a linear relation between and $\theta_o  \sqrt{2{\ K}_{td} \ m_b} / \rho R_V^2$ and $\Gamma_m$ can be observed in figure \hyperref[fig:Gamma_scale]{27b}, obtained from the experimental data plotted using the equation \eqref{eq:gm_scale}. Both figures \hyperref[fig:scaling_dyn]{27a-b} show among all the other parameters (such as $\theta_o$, $m_b$ and $R_v$) an increase in stiffness per unit depth from $Kt_2$ to $Kt_1$ resulted in a significant increase in $u_m$ and $\Gamma_m$. \par
	
	A commonly used performance metric for locomotory bodies, including underwater ones is  the cost of transport ($COT$) (Vogel\cite{Vogel88}, Videler\cite{Videler93}), defined as the ratio of work done by the body to mass times distance moved. In our case we may write
	
	\begin{gather}
	\label{eq:COT}
	COT=\frac{SE}{m_b\times \Delta S}\ 
	\end{gather}
	 
	where $SE$ is the initial energy stored in the body of mass $m_b$ and $\Delta S$ is the distance traveled.\par
	
	We may obtain through simple analysis a scaling for COT for our clapping body. The clapping body travels most of the distance in the retardation phase as the acceleration phase ends in a short time, see figures \hyperref[fig:Lin_Velo_Dyn_AR200]{4a-b}. The total distance traveled by the body is approximately the distance traveled in the retardation phase and can be obtained by integrating the body velocity with time $(=\int{u_b \ dt_r})$. Using equations \eqref{eq:ub_model} and \eqref{eq:phi_exp}, 	and subject to conditions $u_b = u_m$ as the initial time till the body reaches a small fraction($\eta$) of $u_m$ ($ub = \eta u_m$ ($\eta \ll 1 $)) we obtain 
	
	\begin{gather}
	\label{eq:S_retard}
	\Delta S\cong -{\phi }^{-1} \mathrm{log}\ \eta \
	\end{gather}
	
	Obtaining $SE$ ($ \sim 0.5m_b\ u_m^2$) using equations \eqref{eq:SE_Kt} and \eqref{eq:um_scale} with the assumption of $m_b/(\rho\ R_V^3) \sim 0$, and obtaining $\Delta S$ using \eqref{eq:S_retard}, we get the scaling for $COT$ as,
		
	\begin{gather}
	\label{eq:COT_scale}
	COT \sim \ \frac{-\rho L}{2~log\eta }~\frac{u^2_m~C_d~d}{m_b}
	\end{gather}

	 Figure \hyperref[fig:COT_scale]{28} shows the values of $COT$ obtained in the experiments using equation \eqref{eq:COT} plotted versus the scale for $COT$ given by equation \eqref{eq:COT_scale}; in the plot, $\eta$ is taken as 0.1. First, the $COT$ values for the clapping body in our study vary between 2 and 8, which lies between the $COT$ of jellyfish and the $COT$ of squid (Gemmel et al.\cite{Gemmell13}). Second, the variation in $COT$ is reasonably well captured by the scaling given by \eqref{eq:COT_scale}, which indicates the dependence on the various parameters, body mass, body dimensions, and drag coefficient. The slower-moving bodies have a lower $COT$ value. The lower $COT$ for jellyfish may be partly ascribed to their low speeds of the order of a few centimeters per second. The RHS of equation \eqref{eq:COT_scale} varies proportionately with $COT$ values obtained using equation \eqref{eq:COT}; see figure \hyperref[fig:COT_scale]{28}. It shows $COT$ for the self-propelling clapping bodies strongly depends on maximum body velocity $u_m$. The fluid density also controls the $COT$.
	 
	\begin{figure}
		\centering
		\begin{subfigure}[b]{0.48\textwidth}
			\includegraphics[width=\textwidth]{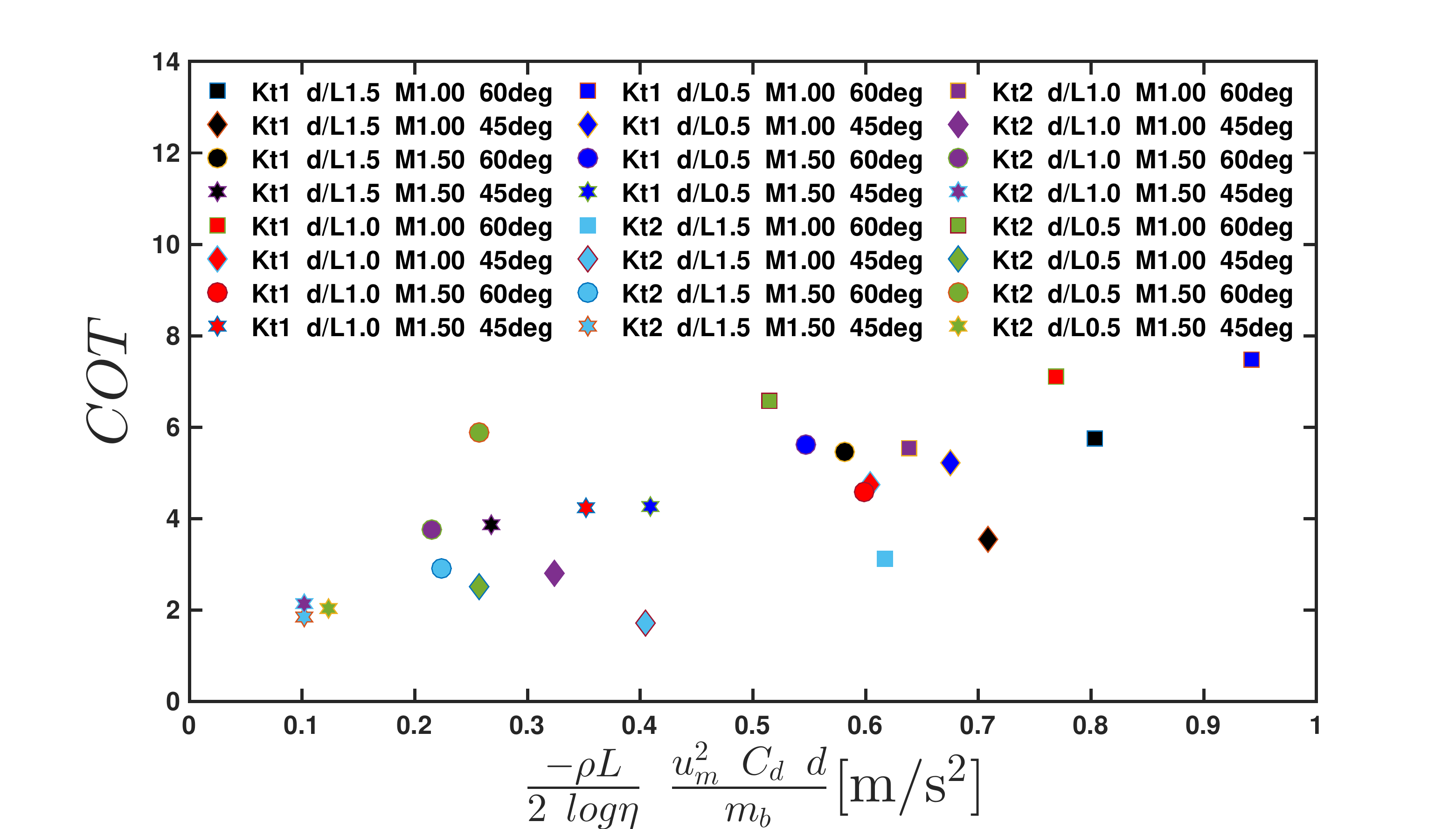}
			\label{fig:C1}
		\end{subfigure}
		\caption{Experimentally obtained $COT$ values plotted versus the analytically predicted values given by equation \eqref{eq:COT_scale}.}\label{fig:COT_scale}
	\end{figure}

\end{subsection}
\end{section}

\begin{section}{Concluding remarks}
\label{sec:Conclusion}

The hydrodynamics of a simple clapping propulsion-based self-propelling body is investigated experimentally. The self-propelling body consists of two plates pivoted together at the front with a ‘torsion’ spring. The clapping action is achieved by release of the plates that are initially pulled apart at the other end. The clapping action of the plates generates a jet that propels the body in a forward direction. A lot of effort went into making the body neutrally buoyant and fixing the centers of mass and buoyancy to ensure stable straight-line motion. Experiments were done by varying the spring stiffness per unit depth ($Kt$), body mass ($m_b$), body aspect ratio ($d^*$), and the initial clapping angle ($\theta_o$). A total of 24 cases were studied. We used high-speed imaging to obtain the body kinematics and PIV and LIF visualization to study the wake structure. \par

The body motion has two phases: a rapid linear acceleration accompanied by the rapid reduction in the clapping angle, which is followed by a relatively slow deceleration of the body till it stops. The first phase is when the forward thrust is produced due to clapping action, and in the second phase, it is the drag force on the closed body that slows it down. The drag coefficient  obtained experimentally is in the range 0.02-0.06 lying between the $C_d$ corresponding to a flat plate and a symmetric aerofoil. We found that the translational velocity of the body is nearly independent of $d^*$, though the wake structure showed large differences with change in $d^*$.  \par

The wake of the clapping body has complex vortex structures whose cross-section in the XY plane shows an isolated vortex pair that travels opposite to the body with lower translational velocity than the body. The initial circulation of the  vortices is approximately independent of $d^*$. The later evolution of the wake is strongly dependent on $d^*$: for the $d^*$ = 1.0 and 1.5 bodies, the vortex loops display axis switching (figures \hyperref[fig:3D Vortex loop_d*1.5]{17},\hyperref[fig:3D Vortex loop_d*1.0]{18}) characteristic of elliptical rings, whereas, for the shorter body ($d^*$ = 0.5), we observe multiple ringlets (figure \hyperref[fig:3D Vortex loop_d*0.5]{19}). \par

Using a simple vortex ring to model the wake, we use conservation of momentum and energy to derive expressions for body velocity \eqref{eq:um_scale} and circulation \eqref{eq:gm_scale} in the starting vortex in terms of the initial stored strain energy in the spring and the other parameters, $d^*$, $m_b$, and $\theta_o$. These relations will be useful for calculating the velocity of self-propelling bodies under pulsed jet propulsion. The data from our experiments are consistent with the analysis. Analysis of the energy budget shows that more than 80\% of initially stored energy is transferred to the fluid.
$COT$ of the clapping body varies between 2 to 8 $\mathrm{J \ kg^{-1} m^{-1}}$, and it lies between COT corresponding to squid and jellyfish (Gemmell et al.\cite{Gemmell13}). The $COT$ scaling shows its strong dependence on the maximum body velocity $u_m$ \eqref{eq:COT_scale}. It must be noted that this COT calculation accounts for only the acceleration phase of the body, though the additional energy required for opening the cavity may not be much, especially if done slowly compared to the clapping motion. \par

Most of the earlier laboratory experiments of clapping propulsion have been with bodies that are constrained from moving forward (D. Kim et al.\cite{Kim13}, Brodsky\cite{Brodsky91}). It is expected that the clapping kinematics and the hydrodynamics will be different when the body is allowed to move, and that is what we have been able to do in the present study reproducing what happens in practice. Some studies have been done on freely swimming animals. Bartol et al.\cite{Bartol2D09} measured the swimming speed of squid {\it Lolliguncula brevis} to be in the range of 2.43-22cm/s. Dabiri et al.\cite{Dabiri06} measured the swimming speeds of jellyfish   {\it Aglantha digitale} to be about 13BL/s in fast swimming. In our experiments, the maximum speed attained by the clapping body is 73cm/s (8 BL/s). Our analysis has revealed that the body speed depends on a variety of factors, including the stored strain energy. This study offers an insight into the flow dynamics and the kinematics of a freely moving clapping body, and can have direct practical utility in the design of aquatic robots based on pulsed propulsion. The present investigation is confined to one phase, the jet ejection phase. The other phase when the plates open out and fluid enters the cavity will bring in additional parameters and fluid mechanics.

\textbf{Funding.} This work was supported by Department of Science and Technology, India under 'Fund for Improvement of S\&T' (grant number: FA/DST0-16.009) and Naval Research Board, India (grant number: NRB/456/19-20). \\

\textbf{Declaration of interests.} The authors report no conflict of interest.\\

\textbf{Author ORCID} \\
Suyog Mahulkar, \href{https://orcid.org/my-orcid?orcid=0000-0001-8497-183X}{https://orcid.org/0000-0001-8497-183X}.
\end{section}

\begin{section}{Appendix}

	\label{sec:SE_Exp}
	Experiments have been conducted to measure initial strain energy ($SE$) stored in two steel plates of the clapping body. In this experiment, force $F$ is applied at the trailing edge of the clapping plate, produces moment $M$ on the steel plate of length $L_e$. The $M$ is given as a sum of the constant moment ($FL_P$) due to the rigid plastic plate of length $L_P$ and variable moment ($Fx$), refer \eqref{eq:Toatal_moment}. Due to the applied moment $M$, the steel plate shows angular deflection $\Theta$. The corresponding strain energy is given by LHS of equation \eqref{eq:SE_Mo}, where the RHS correlates $SE$ to angular deflection.
	
	\begin{gather}
	\label{eq:SE_Mo}
	K_{td} \ {\Theta}^2=\frac{1}{2EI}\int M^2\ dx; \\
	\label{eq:Toatal_moment}
	M= -FL_p -Fx  
	\end{gather}
	
	The Young modulus $E$ is experimentally measured as 212-239 Gpa for the steel plate thickness of 0.14mm (stiffness per unit depth = $Kt_1$) and 175-208 GPa for the steel plate thickness of 0.10mm (stiffness per unit depth = $Kt_2$). The $SE$ vs. $\Theta$ curve is parabolic in nature, and the second-order polynomial fit gives the values for coefficient $K_{td}$.  \par

\end{section}

\bibliographystyle{jfm}
%\bibliography{jfm2esam}

\end{document}